\documentclass[preprint]{elsarticle}

\makeatletter
\def\ps@pprintTitle{
 \let\@oddhead\@empty
 \let\@evenhead\@empty
 \def\@oddfoot{\centerline{\thepage}}
 \let\@evenfoot\@oddfoot}
\makeatother

\usepackage{hyperref}
\usepackage{amssymb,amsmath,amsthm,subcaption}
\usepackage{stmaryrd}

\theoremstyle{definition}
\newtheorem{theorem}{Theorem}[section]
\newtheorem{lemma}[theorem]{Lemma}
\newtheorem{definition}[theorem]{Definition}

\newcommand\abs[1]{\ensuremath{\lvert#1\rvert}}

\usepackage{tabularx,environ}
\makeatletter
\newcommand{\specinput}[1]{\gdef\@specinput{#1}}%
\newcommand{\specoutput}[1]{\gdef\@specoutput{#1}}
\NewEnviron{spec}{
  \BODY
  \par\addvspace{.25\baselineskip}
  \noindent
  \begin{tabularx}{\textwidth}{@{\hspace{\parindent}} l X c}
    \textbf{Input:} & \@specinput \\
    \textbf{Output:} & \@specoutput
  \end{tabularx}
  \par\addvspace{.25\baselineskip}
}
\makeatother

\usepackage{adjustbox,graphicx,tikz,xcolor,pgfplots}
\tikzset{gp2 node/.style={draw, circle, thick, minimum width=0.64cm}}
\tikzset{root node/.style={draw, circle, thick, minimum width=0.64cm, double, double distance=0.3mm}}
\pgfplotsset{compat=1.15}
\usepackage[final]{listings}
\usepackage{float}

\definecolor{gp2green}{RGB}{153, 255, 170}
\definecolor{gp2blue}{RGB}{153, 187, 255}
\definecolor{gp2red}{RGB}{233, 73, 87}
\definecolor{gp2pink}{RGB}{255, 153, 238}
\definecolor{gp2grey}{RGB}{194, 194, 194}
\definecolor{plot1}{RGB}{70, 116, 193}
\definecolor{plot2}{RGB}{235, 125, 60}
\definecolor{plot3}{RGB}{165, 165, 165}
\definecolor{plot4}{RGB}{252, 190, 45}
\definecolor{plot5}{RGB}{94, 156, 210}
\definecolor{plot6}{RGB}{113, 171, 77}
\definecolor{plot7}{RGB}{156, 72, 25}
\definecolor{plot8}{RGB}{40, 69, 117}

{\begin{trivlist}\item[]%
{\textit{Proof sketch.}}}%
{\qed\end{trivlist}}

{\begin{trivlist}\item[]%
{\textit{Proof of #1.}}}%
{\qed\end{trivlist}}

\newcommand{\ttt}{\texttt}

\newcommand{\mtt}{\mathtt}
\newcommand{\mrm}{\mathrm}

\newcommand{\failrm}{\mathrm{fail}}
\newcommand{\failtt}{\mathtt{fail}}
\newcommand{\skiptt}{\mathtt{skip}}

\newcommand{\G}{\mathcal{G}}

\newcommand{\Sem}[1]{\llbracket{#1}\rrbracket}

\newcommand{\tuple}[1]{\langle#1\rangle}

\newcommand{\ifte}[3]{\mathtt{if}\ #1\ \mathtt{then}\ #2\ \mathtt{else}\ #3}
\newcommand{\ift}[2]{\mathtt{if}\ #1\ \mathtt{then}\ #2}
\newcommand{\tryt}[2]{\mathtt{try}\ #1\ \mathtt{then}\ #2}
\newcommand{\trye}[2]{\mathtt{try}\ #1\ \mathtt{else}\ #2}
\newcommand{\tryte}[3]{\mathtt{try}\ #1\ \mathtt{then}\ #2\ \mathtt{else}\ #3}

\newcommand{\DSto}{\mathop{\to}\limits}

\newcommand{\dder}{\Rightarrow}

\newcommand{\DSdder}{\mathop{\Rightarrow}\limits}

\newenvironment{allintypewriter}{\ttfamily}{\par}

\begin{document}

\begin{frontmatter}

\title{Fast Rule-Based Graph Programs}

\author[1]{Graham Campbell\fnref{fn1}}
\ead{g.j.campbell2@newcastle.ac.uk}

\author[2]{Brian Courtehoute}
\ead{bc956@york.ac.uk}

\author[2]{Detlef Plump}
\ead{detlef.plump@york.ac.uk}

\address[1]{School of Mathematics, Statistics and Physics, Newcastle University, Newcastle upon Tyne, United Kingdom}
\address[2]{Department of Computer Science, University of York, York, United Kingdom}

\fntext[fn1]{Supported by a Doctoral Training Grant from the Engineering and Physical Sciences Research Council (EPSRC) Grant No. (2281162) in the UK.}

\begin{abstract}
Implementing graph algorithms efficiently in a rule-based language is challenging because graph pattern matching is expensive. In this paper, we present a number of linear-time implementations of graph algorithms in GP 2, an experimental programming language based on graph transformation rules which aims to facilitate program analysis and verification. We focus on two classes of rule-based graph programs: graph reduction programs which check some graph property, and programs using a depth-first search to test some property or perform an operation such as producing a 2-colouring or a topological sorting. Programs of the first type run in linear time without any constraints on input graphs while programs of the second type require input graphs of bounded degree to run in linear time. Essential for achieving the linear time complexity are so-called rooted rules in GP 2, which, in many situations, can be matched in constant time. For each of our programs, we prove both correctness and complexity, and also give empirical evidence for their run time.
\end{abstract}


\end{frontmatter}

\section{Introduction}
\label{sec:introduction}
	
Rule-based graph transformation was established as a research field in the 1970s and has since then been the subject of countless articles. While many of these contributions have a theoretical nature (see the monograph \cite{Ehrig-Ermel-Golas-Hermann15a} for a recent overview), there has also been work on languages and tools for executing and analysing graph transformation systems.

Languages based on graph transformation rules include 
AGG \cite{Runge-Ermel-Taentzer11a},
GReAT \cite{Agrawal-Karsai-Neema-Shi-Vizhanyo06a},
GROOVE \cite{Ghamarian-deMol-Rensink-Zambon-Zimakova12a},
GrGen.Net \cite{Jakumeit-Buchwald-Kroll10a},
Henshin \cite{Arendt-Biermann-Jurack-Krause-Taentzer10a} and
PORGY \cite{Fernandez-Kirchner-Mackie-Pinaud14a}.
This paper focuses on GP\,2 \cite{Plump12a}, an experimental graph programming language which aims to facilitate formal reasoning on programs. The language has a simple formal semantics and is computationally complete in that every computable function on graphs can be programmed \cite{Plump17a}. Research on graph programs has provided, for example, a Hoare-calculus for program verification \cite{Poskitt-Plump12a,Poskitt-Plump14a} and a static analysis for confluence checking \cite{Hristakiev-Plump18a}.

A challenge for the design and implementation of graph transformation languages is to narrow the performance gap between imperative and rule-based graph programming. The bottleneck for achieving fast graph transformation is the cost of graph matching. In general, matching the left-hand graph $L$ of a rule within a \emph{host graph} $G$ requires time $\mrm{size}(G)^{\mrm{size}(L)}$ which is polynomial since $L$ is fixed. (We denote by $\mrm{size}(X)$ the number of nodes and edges in a graph $X$.) As a consequence, linear-time imperative graph algorithms may be slowed down to polynomial time when they are recast as rule-based graph programs. 

To mitigate this problem, GP\,2 supports \emph{rooted} graph transformation which was first proposed by D\"orr \cite{Doerr95b}. The idea is to distinguish certain nodes as \emph{roots} and to match roots in rules with roots in host graphs. Then only the neighbourhood of \emph{host graph} roots needs to be searched for matches, allowing, under mild conditions, to match rules in constant time. In \cite{Bak-Plump12a}, \emph{fast} rules were identified as a class of rooted rules that can be applied in constant time if host graphs have a bounded node degree and contain a bounded number of roots. 

The condition of a bounded number of \emph{host graph} roots can be satisfied by requiring unrooted input graphs and using in loops only rules that do not increase the number of roots. This simply means that no such rule must have more roots in its right-hand side than in its left-hand side. (A refined condition considers the ``root balance" of all rules in a loop body simultaneously.) The condition that host graphs must have a bounded node degree depends on the application domain of a program. For example, traffic networks or digital circuits can be considered as graphs of bounded degree.

The first linear-time graph problem implemented by a GP\,2 program with fast rules was 2-colouring. In \cite{Bak-Plump12a,Bak-Plump16a} it is shown that this program colours connected graphs of bounded degree in linear time. The compiled program even matches the speed of Sedgewick's textbook C program \cite{Sedgewick02a} on grid graphs of up to 100,000 nodes. Since then, the compiler has received some major improvements, in particular relating to the runtime graph data structure used by the compiled programs \cite{Campbell-Romo-Plump20d}, which has allowed us to achieve linear time worst-case performance for a wider class of programs than was previously possible, in some cases even on input graph classes of unbounded degree.

In this paper, we continue to provide evidence that rooted graph programs can rival the time complexity of graph algorithms in conventional programming languages. We present five case studies the first three of which are based on graph reduction programs for recognising cycle graphs, trees, and binary DAGs. The other two case studies are based on programs using depth-first search to check connectedness resp.\ produce a topological sorting of an acyclic graph. Each of these problems is solvable in linear time with algorithms in imperative languages. For each problem, we present a GP\,2 program with fast rules, show its correctness, and prove its linear time complexity (on graphs of bounded node degree for the last two problems). We also give empirical evidence for the linear run time by presenting benchmark results for graphs of up to around 500,000 nodes in various graph classes.

This paper is a revised and significantly extended version of \cite{Campbell-Courtehoute-Plump19b}, which was written before the improvements to the GP\,2 run time system. We now have graph reduction programs which recognise cycle graphs, trees and binary DAGs in linear time among arbitrary input graphs, without the restriction of bounded node degrees. We present the recognition programs together with new proofs of correctness and complexity. They all run in linear time, which previously was possible only if input graphs have a bounded node degree. We also revisit the topological sorting program of \cite{Campbell-Courtehoute-Plump19b} and the 2-colouring program of Bak and Plump \cite{Bak-Plump16a}, giving more rigorous analyses. The topological sorting program has been re-worked so that it doubles as a program for checking acyclicity.  In addition, we present and analyse a program for checking connectedness using depth-first search. 

Finally, it is worth noting that rooted rules per se are not a blueprint for imitating algorithms in imperative languages. This is because GP\,2 intentionally does not provide access to the graph data structure of its implementation. 

\section{The Graph Programming Language GP\,2}
\label{sec:gp2}
	
This section briefly introduces GP\,2, a non-deterministic language based on graph-transformation rules, first defined in \cite{Plump12a}. An up-to-date version of the syntax of GP\,2 can be found in \cite{Bak15a}. The language is implemented by a compiler generating C code \cite{Bak-Plump16a,Campbell-Romo-Plump20d}, and the source code is available on GitHub\footnote{\url{https://github.com/UoYCS-plasma/GP2}}.

\subsection{Graphs, Rules and Programs}
\label{subsec:graphs_rules_programs}

GP\,2 programs transform input graphs into output graphs, where graphs are directed and may contain parallel edges and loops. Both nodes and edges are labelled with lists consisting of integers and character strings. This includes the special case of items labelled with the empty list which may be considered as ``unlabelled''. 

The principal programming construct in GP\,2 consist of conditional graph transformation rules labelled with expressions. For example, the rule \ttt{push} in Figure \ref{fig:is-tree-program} has three formal parameters of type \ttt{list}, a left-hand graph and a right-hand graph which are specified graphically, and a textual condition starting with the keyword \ttt{where}.

The small numbers attached to nodes are identifiers, all other text in the graphs consist of labels. Parameters are typed but in this paper, we only need the most general type \ttt{list} which represents lists with arbitrary values. 

Besides carrying expressions, nodes and edges can be \emph{marked} red, green or blue. In addition, nodes can be marked grey and edges can be dashed. For example, rule \ttt{push} in Figure \ref{fig:is-tree-program} contains red and grey nodes and unmarked edges. Nodes and edges in rules can also be marked magenta, which allows the mark to be bound to any mark at the point of rule instantiation (see the next paragraph). Marks are convenient, among other things, to record visited items during a graph traversal and to encode auxiliary structures in graphs. The programs in the following sections use marks extensively.

Rules operate on \emph{host graphs} which are labelled with constant values (lists containing integers and character strings). Formally, the application of a rule to a \emph{host graph} is defined as a two-stage process in which first the rule is instantiated by replacing all variables with values of the same type, and evaluating all expressions. This yields a standard rule (without expressions) in the so-called double-pushout approach with relabelling \cite{Habel-Plump02c}. In the second stage, the instantiated rule is applied to the \emph{host graph} by constructing two suitable pushouts. We refer to \cite{Bak15a} for details and only give an equivalent operational description of rule application.

Applying a rule $L \dder R$ to a \emph{host graph} $G$ works roughly as follows: (1) Replace the variables in $L$ and $R$ with constant values and evaluate the expressions in $L$ and $R$, to obtain an instantiated rule $\hat{L} \dder \hat{R}$. (2) Choose a subgraph $S$ of $G$ isomorphic to $\hat{L}$ such that the dangling condition and the rule's application condition are satisfied (see below). (3) Replace $S$ with $\hat{R}$ as follows: numbered nodes stay in place (possibly relabelled), edges and unnumbered nodes of $\hat{L}$ are deleted, and edges and unnumbered nodes of $\hat{R}$ are inserted. 

In this construction, the \emph{dangling condition} requires that nodes in $S$ corresponding to unnumbered nodes in $\hat{L}$ (which should be deleted) must not be incident with edges outside $S$. The rule's application condition is evaluated after variables have been replaced with the corresponding values of $\hat{L}$, and node identifiers of $L$ with the corresponding identifiers of $S$. For example, the condition $\mtt{\mathrm{indeg}(1)<2}$ of rule \ttt{push} in Figure \ref{fig:is-tree-program} requires that node $g(\mtt{1})$ has at most one incoming edge, where $g(\mtt{1})$ is the node in $S$ corresponding to \ttt{1}. 

A program consists of declarations of conditional rules and procedures, and exactly one declaration of a main command sequence, which is a distinct procedure named \ttt{Main}. Procedures must be non-recursive, they can be seen as macros. We describe GP\,2's main control constructs.

The call of a rule set $\{r_1,\dots,r_n\}$ non-deterministically applies one of the rules whose left-hand graph matches a subgraph of the \emph{host graph} such that the dangling condition and the rule's application condition are satisfied. The call \emph{fails} if none of the rules is applicable to the host graph. 

The command \ttt{if} $C$ \ttt{then} $P$ \ttt{else} $Q$ is executed on a \emph{host graph} $G$ by first executing $C$ on a copy of $G$. If this results in a graph, $P$ is executed on the original graph $G$; otherwise, if $C$ fails, $Q$ is executed on $G$. The \ttt{try} command has a similar effect, except that $P$ is executed on the result of $C$'s execution. 

The loop command $P!$ executes the body $P$ repeatedly until it fails. When this is the case, $P!$ terminates with the graph on which the body was entered for the last time. The \ttt{break} command inside a loop terminates that loop and transfers control to the command following the loop.

In general, the execution of a program on a \emph{host graph} may result in different graphs, fail, or diverge. This is formally defined in the next subsection. 

\subsection{Operational Semantics of GP\,2}
\label{subsec:semantics}

This subsection reviews the semantics of GP\,2, except for the definition of rule applications, in the style of structural operational semantics \cite{Plotkin04a}. In this approach, inference rules inductively define a small-step transition relation $\to$ on \emph{configurations}. In the setting of GP 2, a configuration is either a command sequence together with a host graph, just a host graph or the special element fail:
\[ \to \;\; \subseteq \; (\text{ComSeq} \times \G) \times 
      ((\text{ComSeq} \times \G) \cup \G \cup \{\failrm\}) \]
where $\G$ is the set of GP\,2 host graphs. Configurations in $\text{ComSeq} \times \G$, given by a rest program and a host graph, represent states of unfinished computations while graphs in $\G$ are final states or \emph{results} of computations. The element fail represents a failure state. A configuration $\gamma$ is said to be \emph{terminal} if there is no configuration $\delta$ such that $\gamma \to \delta$.

Figure \ref{fig:core_sos_rules} shows the inference rules for the core commands of GP\,2. The rules contain meta-variables for command sequences and graphs, where $R$ stands for a call of a rule set or of a rule, $C,P,P',Q$ stand for command sequences, and $G,H$ stand for host graphs. 
The transitive and reflexive-transitive closures of $\to$ are written $\to^+$ and $\to^*$, respectively. We write $G \dder_R H$ if $H$ results from host graph $G$ by applying the rule set $R$, while $G \not\dder_R$ means that there is no graph $H$ such that $G \dder_R H$ (application of $R$ fails). 

\begin{figure}[!ht]
\begin{center}
\scalebox{.9}{
\begin{tabular}{ll}
$\mathrm{[call_1]}$ $\frac{\displaystyle G \dder_R H}{\displaystyle\tuple{R,\,G} \to H}$ 
&
$\mathrm{[call_2]}$ $\frac{\displaystyle G \not\dder_R}{\displaystyle\tuple{R,\,G} \to \failrm}$
\\\\
$\mathrm{[seq_1]}$ $\frac{\displaystyle \tuple{P,\, G} \to \tuple{P',\, H}}{\displaystyle \tuple{P;Q,\, G} \to \tuple{P';Q,\, H}}$ 
&
$\mathrm{[seq_2]}$ $\frac{\displaystyle \tuple{P,\, G} \to H}{\displaystyle \tuple{P;Q,\, G}\to \tuple{Q,\, H}}$
\\\\
$\mathrm{[seq_3]}$ $\frac{\displaystyle \tuple{P,\, G} \to \failrm}{\displaystyle \tuple{P;Q,\, G}\to \failrm}$
\\\\
$\mathrm{[if_1]}$ $\frac{\displaystyle \tuple{C,\, G} \to^+ H}{\displaystyle \tuple{\ifte{C}{P}{Q},\, G}\to \tuple{P,\, G}}$
\\\\
$\mathrm{[if_2]}$ $\frac{\displaystyle \tuple{C,\, G} \to^+ \failrm}{\displaystyle \tuple{\ifte{C}{P}{Q},\, G} \to \tuple{Q,\, G}}$
\\\\
$\mathrm{[try_1]}$ $\frac{\displaystyle \tuple{C,\, G} \to^+ H}{\displaystyle \tuple{\tryte{C}{P}{Q},\, G}\to \tuple{P,\, H}}$
\\\\
$\mathrm{[try_2]}$ $\frac{\displaystyle \tuple{C,\, G} \to^+ \failrm}{\displaystyle \tuple{\tryte{C}{P}{Q},\, G} \to \tuple{Q,\, G}}$
\\\\
$\mathrm{[alap_1]}$ $\frac{\displaystyle \tuple{P,\, G} \to^+ H}{\displaystyle \tuple{P!,\, G} \to \tuple{P!,\, H}}$
&
$\mathrm{[alap_2]}$ $\frac{\displaystyle \tuple{P,\, G} \to^+ \failrm}{\displaystyle \tuple{P!,\, G} \to G}$
\\\\
$\mrm{[alap_3]}$ $\frac{\displaystyle \tuple{P,\, G} \to^* \tuple{\mtt{break}, H}}{\displaystyle \tuple{P!,\, G} \to H}$
&
$\mrm{[break]}$ $\tuple{\mtt{break}; P,\, G} \to \tuple{\mtt{break},\, G}$
\end{tabular}
}
\end{center}
\vspace{-0.333333em}
\caption{Inference rules for core commands}
\label{fig:core_sos_rules}
\end{figure}

The inference rules for the remaining GP\,2 commands are given in Figure \ref{fig:derived_sos_rules}. These commands are referred to as \emph{derived} commands because they can be defined by the core commands. 

\begin{figure}[!ht]
\begin{center}
\scalebox{.9}{
\begin{tabular}{llll}
$\mrm{[or_1]}$ & $\tuple{P\mspace{.5mu} \mathop{\mtt{or}}\, Q,\, G} \to \tuple{P,\, G}$ \hspace{4em} & $\mrm{[or_2]}$ & $\tuple{P\mspace{.5mu} \mathop{\mtt{or}}\, Q,\, G} \to \tuple{Q,\, G}$
\\[1.5ex]
$\mrm{[skip]}$ & $\tuple{\skiptt,\, G} \to G$ & $\mrm{[fail]}$ & $\tuple{\failtt,\, G} \to \failrm$
\\[1.5ex]
$\mrm{[if_3]}$ & \multicolumn{3}{l}{$\tuple{\ift{C}{P},\, G} \to \tuple{\ifte{C}{P}{\skiptt},\, G}$}
\\[1.5ex]
$\mrm{[try_3]}$ & \multicolumn{3}{l}{$\tuple{\tryt{C}{P},\, G} \to \tuple{\tryte{C}{P}{\skiptt},\, G}$}
\\[1.5ex]
$\mathrm{[try_4]}$ & \multicolumn{3}{l}{$\tuple{\trye{C}{P},\, G} \to \tuple{\tryte{C}{\skiptt}{P},\, G}$}
\\[1.5ex]
$\mathrm{[try_5]}$ & \multicolumn{3}{l}{$\tuple{\mtt{try}\ C,\, G} \to \tuple{\tryte{C}{\skiptt}{\skiptt},\, G}$}
\end{tabular}
}
\end{center}
\vspace{-0.333333em}
\caption{Inference rules for derived commands}
\label{fig:derived_sos_rules}
\end{figure}

The meaning of GP\,2 programs is summarised by the semantic function $\Sem{\_}$ which assigns to each command sequence $P$ the function $\Sem{P}$ mapping an input graph $G$ to the set $\Sem{P}G$ of all possible results of executing $P$ on $G$. The value fail indicates a failed program run while $\bot$ indicates a run that does not terminate or gets stuck. Program $P$ \emph{can diverge from} $G$ if there is an infinite sequence $\tuple{P,\,G} \to \tuple{P_1,\,G_1} \to \tuple{P_2,\,G_2} \to \dots$ Also, $P$ \emph{can get stuck from} $G$ if there is a terminal configuration $\tuple{Q,\,H}$ such that $\tuple{P,\,G} \to^* \tuple{Q,\,H}$.

The \emph{semantic function} $\Sem{\_}\colon \mathrm{ComSeq} \to (\G \to 2^{\G^{\oplus}})$ is defined by
   \[ \setlength{\arraycolsep}{2pt}
      \begin{array}{rcl}
       \Sem{P}G & = & 
            \{X \in (\G \cup \{\failrm\}) \mid \tuple{P,\,G}\DSto^+ X\}\ \cup\\
       && \{\bot \mid \text{$P$ can diverge or get stuck from $G$}\},
      \end{array} \]
where ComSeq is the set of GP\,2 command sequences and $\G^{\oplus} = \G \cup \{\bot,\mrm{fail}\}$. Getting stuck indicates a form of divergence that can happen with a command $\ifte{C}{P}{Q}$ or $\tryte{C}{P}{Q}$ in case $C$ can diverge from a graph $G$ and neither produce a graph nor fail from $G$, or with a loop $B!$ whose body $B$ possesses the said property.

\subsection{The GP\,2-to-C Compiler}
\label{subsec:compiler}

GP\,2's primary implementation is a GP\,2-to-C compiler, and all our complexity assumptions will be compatible with this implementation. The original GP\,2 compiler is described in detail by Bak's thesis \cite{Bak15a}, and recent modifications are described in \cite{Campbell-Romo-Plump20d}. These modifications are important for two reasons. The first is that there have been significant changes to the compiler's generated code and runtime system which enable us to write reduction programs in GP\,2 that are linear time on graph classes of unbounded degree, which has not been possible before. The second important change is to the theoretical model (and implementation) of root nodes and morphisms, requiring morphisms to reflect rootedness as well as preserving it. We will assume all morphisms are rootedness reflecting in this paper.

As shown in Figure \ref{fig:gp2-to-c-compiler}, GP\,2 programs are first compiled to C programs using the GP\,2-to-C compiler implementation, which are in turn compiled by GCC to a platform dependent executable. This executable then reads an input graph from disk and executes the program, with one of four possible outcomes:

\begin{enumerate}
    \item the program produces an output graph;
    \item the program evaluates to \ttt{fail};
    \item the program encounters a runtime error;
    \item the program does not terminate.
\end{enumerate}

\begin{figure}[!ht]
\centering
\begin{tikzpicture}
\node at (1,1.7) {GP\,2 Program};
\draw [->] (1,1.5) -- (1,1);
\draw [fill=black!10] (0,0) rectangle node{GP\,2-to-C} (2,1);
\node at (2.75,0.75) {C};
\node at (2.75,0.25) {Program};
\draw [->] (2,0.5) -- (3.5,0.5);
\draw [fill=black!10] (3.5,0) rectangle node{GCC} (5.5,1);
\node at (6.25,0.75) {Binary};
\draw [->] (5.5,0.5) -- (7,0.5);
\node at (8,1.7) {Input Graph};
\draw [->] (8,1.5) -- (8,1);
\draw [fill=black!10] (7,0) rectangle node{Execute} (9,1);
\draw [->] (9,0.5) -- (10.5,0.5);
\node at (9.75,0.75) {Output};
\node at (9.75,0.25) {Graph};
\end{tikzpicture}
\caption{The GP\,2-to-C compiler}
\label{fig:gp2-to-c-compiler}
\end{figure}
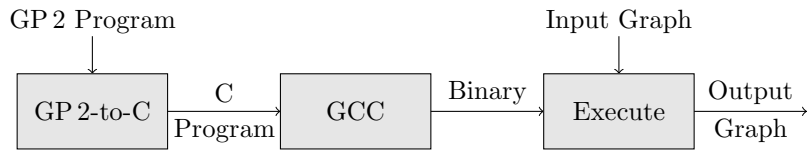

In this paper, we will ignore runtime errors such as division by zero in a label operation, or the physical memory of the machine being exhausted.

GP\,2 programs are graphically visualised throughout this paper. We now look at an example program which computes the transitive closure of a graph (Figure \ref{fig:transitive-closure-program}). The concrete syntax for the program is available on GitHub\footnote{\url{https://gist.github.com/GrahamCampbell/c429491366119c898c9d1b5ad0459ab1}}.

\begin{figure}[!ht]
\centering
\fbox{\begin{minipage}{11.78cm}
\begin{allintypewriter}
Main = link!

\setlength{\tabcolsep}{0.4cm}

\medskip
\smallskip

\begin{tabular}{ p{10.5cm} }
	
	link(a,b,x,y,z:list) \\
	
	\begin{tikzpicture}
        
		\node (a) at (0,0)       [gp2 node, fill=gp2grey] {x};
		\node (b) at (1.25,0)    [gp2 node, fill=gp2grey] {y};
		\node (c) at (2.5,0)     [gp2 node, fill=gp2grey] {z};
		
		\node (d) at (3.375,0)   {$\Rightarrow$};
		
		\node (e) at (4.25,0)    [gp2 node, fill=gp2grey] {x};
		\node (f) at (5.5,0)     [gp2 node, fill=gp2grey] {y};
		\node (g) at (6.75,0)    [gp2 node, fill=gp2grey] {z};
		
		\node (A) at (0,-.45)    {\tiny{1}};
		\node (B) at (1.25,-.45) {\tiny{2}};
		\node (C) at (2.5,-.45)  {\tiny{3}};
		\node (E) at (4.25,-.45) {\tiny{1}};
		\node (F) at (5.5,-.45)  {\tiny{2}};
		\node (G) at (6.75,-.45) {\tiny{3}};
		
		\draw (a) edge[->,thick] node[above] {a} (b)
		      (b) edge[->,thick] node[above] {b} (c)
		      (e) edge[->,thick] node[above] {a} (f)
		      (f) edge[->,thick] node[above] {b} (g)
		      (e) edge[->,thick,bend left=40] (g);
	\end{tikzpicture}
	\\
	
	\vspace{-1em}where not edge(1,3) \\
	
\end{tabular}
\end{allintypewriter}
\end{minipage}}
\caption{GP\,2 program \ttt{transitive-closure}}
\label{fig:transitive-closure-program}
\end{figure}
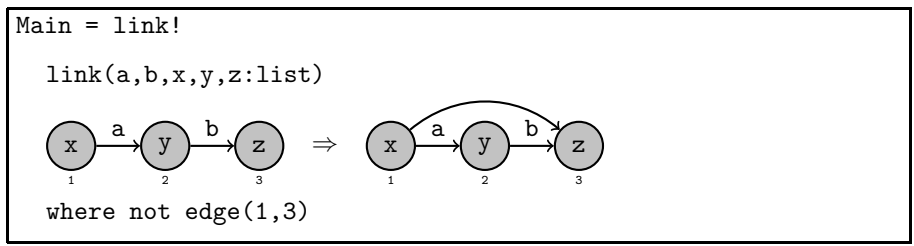

The generated C code for this GP\,2 program:

\begin{enumerate}
    \item Reads and parses the input host graph;
    \item Applies the \ttt{link} rule as long as possible;
    \item Outputs the \emph{host graph} as output.
\end{enumerate}

Looking at what it means to apply the \ttt{link} rule as long as possible, we attempt to find a \emph{match} satisfying the application condition, and if we find one, we apply the rule, adding a fresh edge. We repeat this (zero or more times) until we can no longer find a suitable match. Finding a match involves first finding a suitable node, then finding a suitable outgoing edge to a suitable node, and then another suitable outgoing edge to a suitable node, and then checking the application condition. We call this the matching algorithm. This is described in general by Bak \cite{Bak15a} in his thesis, and updates to the algorithm are defined in \cite{Campbell-Courtehoute-Plump19b}.

\subsection{Reasoning about Time Complexity}
\label{subsec:complexity}

When analysing the time complexity of programs, we assume that these are fixed. This is customary in algorithm analysis where programs are fixed and running time is measured in terms of input size \cite{Aho-Hopcroft-Ullman74a,Skiena08a}. In our setting, the input size is the \emph{size} of a host graph, which we define to be the total number of nodes and edges. Figure \ref{fig:complexity-assumptions} shows the complexity of various runtime operations, where \(n\) is the size of the current host graph.

\begin{figure}[!ht]
\begin{center}
\scalebox{.75}{
\begin{tabular}{l|l|l}
Procedure               & Description                                                    & Complexity  \\ \hline
\ttt{parseInputGraph}   & Parse and load the input graph into memory: the host graph.    & \(O(n)\)    \\
\ttt{alreadyMatched}    & Test if the given item has been matched in the host graph.     & \(O(1)\)    \\
\ttt{clearMatched}      & Clear the \ttt{is matched} flag for a given item.              & \(O(1)\)    \\
\ttt{setMatched}        & Set the \ttt{is matched} flag for a given item.                & \(O(1)\)    \\
\ttt{firstHostNode}     & Fetch the first node in the host graph.                        & \(O(1)\)    \\
\ttt{nextHostNode}      & Given a node, fetch the next node in the host graph.           & \(O(1)\)    \\
\ttt{firstHostRootNode} & Fetch the first root node in the host graph.                   & \(O(1)\)    \\
\ttt{nextHostRootNode}  & Given a root node, fetch the next root node in the host graph. & \(O(1)\)    \\
\ttt{firstInEdge}       & Given a node, fetch the first incoming edge.                   & \(O(1)\)    \\
\ttt{nextInEdge}        & Given a node and an edge, fetch the next incoming edge.        & \(O(1)\)    \\
\ttt{firstOutEdge}      & Given a node, fetch the first outgoing edge.                   & \(O(1)\)    \\
\ttt{nextOutEdge}       & Given a node and an edge, fetch the next outgoing edge.        & \(O(1)\)    \\
\ttt{printHostGraph}    & Write the current host graph state as output.                  & \(O(n)\)    \\
\end{tabular}
}
\end{center}
\caption{Runtime complexity assumptions}
\label{fig:complexity-assumptions}
\end{figure}

Throughout the paper, we provide empirical evidence of the time complexity of programs, which includes the time needed to read, parse, and load the input graph into memory, and also the time needed to write the output graph, but not the time spent compiling the program. This evidence supports our complexity proofs and also gives confidence in the accuracy of Figure \ref{fig:complexity-assumptions}. Formal complexity analysis of those operations is beyond the scope of this paper, and all our complexity results are relative to these basic complexities.

When we discuss the time complexity of programs, we will do this by reasoning about the complexity of the generated code. When considering the complexity of rule applications, it suffices to only reason about the complexity of finding a match because all the programs in this paper satisfy the assumption of the following lemma.

\begin{lemma}[Constant Time Application] \label{lem:const-time}
    Once a match has been found for a rule that does not modify labels, other than perhaps introducing a fixed label or changing marks, only constant time is needed to complete the process of applying the instantiated rule and building the result graph.
\end{lemma}

When analysing the programs of this paper, by a \emph{step} we mean a completed or failed rule application, the \ttt{break} operation, or the \ttt{fail} operation. While \ttt{break} and \ttt{fail} always finish in constant time, completed and failed rule applications can often be shown to require only constant time by the specific properties of our rules (see below) and the preconditions of some of our programs such as a bounded node degree in input graphs.
\subsection{Rooted Programs}
\label{subsec:rooted_programs}

The bottleneck for efficiently implementing algorithms in a language based on graph transformation rules is the cost of graph matching. In general, to match the left-hand graph $L$ of a rule within a \emph{host graph} $G$ requires time polynomial in the size of $L$ \cite{Bak-Plump12a,Bak-Plump16a}. As a consequence, linear-time graph algorithms in imperative languages may be slowed down to polynomial time when they are recast as rule-based programs. 

To speed up matching, GP\,2 supports \emph{rooted} graph transformation where graphs in rules and host graphs are equipped with so-called root nodes. Roots in rules must match roots in the \emph{host graph} so that matches are restricted to the neighbourhood of the host graph's roots. We draw root nodes using double circles. For example, in the rule \ttt{prune} of Figure \ref{fig:is-tree-program}, the node labelled \ttt{y} in the left-hand side and the single node in the right-hand side are roots.

A conditional rule $\tuple{L \dder R,\, c}$ is \emph{fast} if (1) each node in $L$ is undirectedly reachable from some root, (2) neither $L$ nor $R$ contain repeated occurrences of list, string or atom variables, and (3) the condition $c$ contains neither an $\mathtt{edge}$ predicate nor a test $e_1 \mtt{=} e_2$ or $e_1 \mtt{!\!=} e_2$ where both $e_1$ and $e_2$ contain a list, string or atom variable.

Conditions (2) and (3) will be satisfied by all rules occurring in the following sections; in particular, we neither use the $\mathtt{edge}$ predicate nor the equality tests. For example, the rules \ttt{prune} and \ttt{push} in Figure \ref{fig:is-tree-program} are fast rules.

\begin{theorem}[Fast Rule Matching \protect\cite{Bak-Plump12a,Bak-Plump16a}] \label{thm:rooted-matching-complexity}
Rooted graph matching can be implemented to run in constant time for fast rules, provided there are upper bounds on the maximal node degree and the number of roots in host graphs. Moreover, the GP\,2-to-C compiler produces programs that match fast rooted rules in constant time under the above conditions.
\end{theorem}

Theorem \ref{thm:rooted-matching-complexity} is used in the complexity analysis of programs with an input graph class of bounded degree. This approach is central to Section \ref{sec:dfs}. However, it is too coarse to obtain sharp bounds on complexity in general. In particular, in Section \ref{sec:reduction}, we wish to show complexity results with input graph classes not necessarily of bounded degree. We are able to do this via direct reasoning about the complexity of programs generated by the GP\,2-to-C compiler.

\section{Fast Reduction Programs}
\label{sec:reduction}

The aim of this section is to demonstrate that GP\,2 can recognise various graph classes in linear time, by means of simple reduction specifications. As discussed in the previous sections, this is made possible by GP\,2's implementation of root nodes. We do not know of any other rule-based graph programming languages that can claim linear time complexity for such tasks.

As well as providing proofs of complexity, we back-up our assumptions by timing the actual execution times of the programs produced by the GP\,2-to-C compiler on various graph classes (Figures \ref{fig:graph-class-1a}, \ref{fig:graph-class-1b}, \ref{fig:graph-class-1c}, \ref{fig:graph-class-2a}, \ref{fig:graph-class-2b}, \ref{fig:graph-class-2c}, \ref{fig:graph-class-3a} and \ref{fig:graph-class-3b}). The concrete syntax for all the programs in this section is available on GitHub\footnote{\url{https://gist.github.com/GrahamCampbell/102eb8ec101ba87f6040ec2e9f3323a2}}.

\begin{figure}[!ht]
\centering
\begin{minipage}{3.93cm}
\centering
\begin{tikzpicture}[scale=0.7]
	\node (a) at (-1.500,1.333)  [draw,circle,thick,fill=gp2grey] {\,};
	\node (b) at (0.000,1.333)   [draw,circle,thick,fill=gp2grey] {\,};
	\node (c) at (1.500,1.333)   [draw,circle,thick,fill=gp2grey] {\,};
	\node (d) at (-1.500,0.000)  [draw,circle,thick,fill=gp2grey] {\,};
	\node (e) at (0.000,0.000)   [draw,circle,thick,fill=gp2grey] {\,};
	\node (f) at (1.500,0.000)   [draw,circle,thick,fill=gp2grey] {\,};
	\node (g) at (-1.500,-1.333) [draw,circle,thick,fill=gp2grey] {\,};
	\node (h) at (0.000,-1.333)  [draw,circle,thick,fill=gp2grey] {\,};
	\node (i) at (1.500,-1.333)  [draw,circle,thick,fill=gp2grey] {\,};
\end{tikzpicture}
\captionof{figure}{Discrete graph}
\label{fig:graph-class-1a}
\end{minipage}
\begin{minipage}{3.93cm}
\centering
\begin{tikzpicture}[scale=0.7]
	\node (a) at (-1.500,1.333)  [draw,circle,thick,fill=gp2grey] {\,};
	\node (b) at (0.000,1.333)   [draw,circle,thick,fill=gp2grey] {\,};
	\node (c) at (1.500,1.333)   [draw,circle,thick,fill=gp2grey] {\,};
	\node (d) at (-1.500,0.000)  [draw,circle,thick,fill=gp2grey] {\,};
	\node (e) at (0.000,0.000)   [draw,circle,thick,fill=gp2grey] {\,};
	\node (f) at (1.500,0.000)   [draw,circle,thick,fill=gp2grey] {\,};
	\node (g) at (-1.500,-1.333) [draw,circle,thick,fill=gp2grey] {\,};
	\node (h) at (0.000,-1.333)  [draw,circle,thick,fill=gp2grey] {\,};
	\node (i) at (1.500,-1.333)  [draw,circle,thick,fill=gp2grey] {\,};
	
	\draw (a) edge[->, thick] (b)
	      (a) edge[->, thick] (d)
	      (b) edge[->, thick] (c)
	      (b) edge[->, thick] (e)
	      (c) edge[->, thick] (f)
	      (d) edge[->, thick] (e)
	      (d) edge[->, thick] (g)
	      (e) edge[->, thick] (f)
	      (e) edge[->, thick] (h)
	      (f) edge[->, thick] (i)
	      (g) edge[->, thick] (h)
	      (h) edge[->, thick] (i);
\end{tikzpicture}
\captionof{figure}{Grid graph}
\label{fig:graph-class-1b}
\end{minipage}
\begin{minipage}{3.93cm}
\centering
\begin{tikzpicture}[scale=0.7]
	\node (a) at (0.000,1.333)   [draw,circle,thick,fill=gp2grey] {\,};
	\node (b) at (1.333,0.000)   [draw,circle,thick,fill=gp2grey] {\,};
	\node (c) at (-1.333,0.000)  [draw,circle,thick,fill=gp2grey] {\,};
	\node (d) at (2.000,-1.333)  [draw,circle,thick,fill=gp2grey] {\,};
	\node (e) at (0.666,-1.333)  [draw,circle,thick,fill=gp2grey] {\,};
	\node (f) at (-0.666,-1.333) [draw,circle,thick,fill=gp2grey] {\,};
	\node (g) at (-2.000,-1.333) [draw,circle,thick,fill=gp2grey] {\,};
	
	\draw (a) edge[->, thick] (b)
	      (a) edge[->, thick] (c)
	      (b) edge[->, thick] (d)
	      (b) edge[->, thick] (e)
	      (c) edge[->, thick] (f)
	      (c) edge[->, thick] (g);
\end{tikzpicture}
\captionof{figure}{Binary tree}
\label{fig:graph-class-1c}
\end{minipage}
\end{figure}
\begin{figure}[!ht]
\centering
\begin{minipage}{3.93cm}
\centering
\begin{tikzpicture}[scale=0.7]
	\node (a) at (0.000,0.000)   [draw,circle,thick,fill=gp2grey] {\,};
	\node (b) at (0.000,1.333)   [draw,circle,thick,fill=gp2grey] {\,};
	\node (c) at (0.943,0.943)   [draw,circle,thick,fill=gp2grey] {\,};
	\node (d) at (1.333,0.000)   [draw,circle,thick,fill=gp2grey] {\,};
	\node (e) at (0.943,-0.943)  [draw,circle,thick,fill=gp2grey] {\,};
	\node (f) at (0.000,-1.333)  [draw,circle,thick,fill=gp2grey] {\,};
	\node (g) at (-0.943,-0.943) [draw,circle,thick,fill=gp2grey] {\,};
	\node (h) at (-1.333,0.000)  [draw,circle,thick,fill=gp2grey] {\,};
	\node (i) at (-0.943,0.943)  [draw,circle,thick,fill=gp2grey] {\,};
	
	\draw (a) edge[->, thick] (b)
	      (c) edge[->, thick] (a)
	      (a) edge[->, thick] (d)
	      (e) edge[->, thick] (a)
	      (a) edge[->, thick] (f)
	      (g) edge[->, thick] (a)
	      (a) edge[->, thick] (h)
	      (i) edge[->, thick] (a);
\end{tikzpicture}
\captionof{figure}{Star graph}
\label{fig:graph-class-2a}
\end{minipage}
\begin{minipage}{3.93cm}
\centering
\begin{tikzpicture}[scale=0.7]
	\node (a) at (0.0000,1.3333)   [draw,circle,thick,fill=gp2grey] {\,};
	\node (b) at (1.1545,0.6666)   [draw,circle,thick,fill=gp2grey] {\,};
	\node (c) at (1.1545,-0.6666)  [draw,circle,thick,fill=gp2grey] {\,};
	\node (d) at (0.0000,-1.3334)  [draw,circle,thick,fill=gp2grey] {\,};
	\node (e) at (-1.1545,-0.6666) [draw,circle,thick,fill=gp2grey] {\,};
	\node (f) at (-1.1545,0.6666)  [draw,circle,thick,fill=gp2grey] {\,};
	
	\draw (a) edge[->, thick] (b)
	      (b) edge[->, thick] (c)
	      (c) edge[->, thick] (d)
	      (d) edge[->, thick] (e)
	      (e) edge[->, thick] (f)
	      (f) edge[->, thick] (a);
\end{tikzpicture}
\captionof{figure}{Cycle graph}
\label{fig:graph-class-2b}
\end{minipage}
\begin{minipage}{3.93cm}
\centering
\begin{tikzpicture}[scale=0.7]
	\node (A) at (0.0000,1.4740)   [draw,circle,thick,fill=gp2grey] {\,};
	\node (a) at (0.0000,0.7370)   [draw,circle,thick,fill=gp2grey] {\,};
	\node (B) at (1.4019,0.4555)   [draw,circle,thick,fill=gp2grey] {\,};
	\node (b) at (0.7010,0.2277)   [draw,circle,thick,fill=gp2grey] {\,};
	\node (C) at (0.8665,-1.1926)  [draw,circle,thick,fill=gp2grey] {\,};
	\node (c) at (0.4332,-0.5963)  [draw,circle,thick,fill=gp2grey] {\,};
	\node (D) at (-0.8665,-1.1926) [draw,circle,thick,fill=gp2grey] {\,};
	\node (d) at (-0.4332,-0.5963) [draw,circle,thick,fill=gp2grey] {\,};
	\node (E) at (-1.4019,0.4555)  [draw,circle,thick,fill=gp2grey] {\,};
	\node (e) at (-0.7010,0.2277)  [draw,circle,thick,fill=gp2grey] {\,};
	
	\draw (a) edge[->,thick] (b)
	(a) edge[->,thick] (b)
	(b) edge[->,thick] (c)
	(c) edge[->,thick] (d)
	(d) edge[->,thick] (e)
	(e) edge[->,thick] (a)
	(A) edge[->,thick] (a)
	(B) edge[->,thick] (b)
	(C) edge[->,thick] (c)
	(D) edge[->,thick] (d)
	(E) edge[->,thick] (e);
\end{tikzpicture}
\captionof{figure}{Sun graph}
\label{fig:graph-class-2c}
\end{minipage}
\end{figure}
\begin{figure}[!ht]
\centering
\begin{minipage}{6.9cm}
\centering
\begin{tikzpicture}[scale=0.7]
	\node (a) at (0,0)             [draw,circle,thick,fill=gp2grey] {};
	\node (b) at (-.6666,-.6666)   [draw,circle,thick,fill=gp2grey] {};
	\node (c) at (.6666,-.6666)    [draw,circle,thick,fill=gp2grey] {};
	\node (d) at (-1.3333,-1.3333) [draw,circle,thick,fill=gp2grey] {};
	\node (e) at (0,-1.3333)       [draw,circle,thick,fill=gp2grey] {};
	\node (f) at (1.3333,-1.3333)  [draw,circle,thick,fill=gp2grey] {};
	\node (g) at (-.6666,-2)       [draw,circle,thick,fill=gp2grey] {};
	\node (h) at (.6666,-2)        [draw,circle,thick,fill=gp2grey] {};
	\node (i) at (0,-2.6666)       [draw,circle,thick,fill=gp2grey] {};
	
	\draw (a) edge[->,thick] (b)
	(a) edge[->,thick] (c)
	(b) edge[->,thick] (d)
	(b) edge[->,thick] (e)
	(c) edge[->,thick] (e)
	(c) edge[->,thick] (f)
	(d) edge[->,thick] (g)
	(e) edge[->,thick] (g)
	(e) edge[->,thick] (h)
	(f) edge[->,thick] (h)
	(g) edge[->,thick] (i)
	(h) edge[->,thick] (i);
    
	\node (j) at (2.6666,0)        [draw,circle,thick,fill=gp2grey] {};
	\node (k) at (2,-.6666)        [draw,circle,thick,fill=gp2grey] {};
	\node (l) at (3.3333,-.6666)   [draw,circle,thick,fill=gp2grey] {};
	\node (m) at (2.6666,-1.3333)  [draw,circle,thick,fill=gp2grey] {};
	\node (n) at (4,-1.3333)       [draw,circle,thick,fill=gp2grey] {};
	\node (o) at (2,-2)            [draw,circle,thick,fill=gp2grey] {};
	\node (p) at (3.3333,-2)       [draw,circle,thick,fill=gp2grey] {};
	\node (q) at (2.6666,-2.6666)  [draw,circle,thick,fill=gp2grey] {};
	
	\draw (j) edge[->,thick] (k)
	(j) edge[->,thick] (l)
	(k) edge[->,thick] (f)
	(k) edge[->,thick] (m)
	(l) edge[->,thick] (m)
	(l) edge[->,thick] (n)
	(f) edge[->,thick] (o)
	(m) edge[->,thick] (o)
	(m) edge[->,thick] (p)
	(n) edge[->,thick] (p)
	(o) edge[->,thick] (q)
	(p) edge[->,thick] (q);
    
	\node (r) at (5.3333,0)        [draw,circle,thick,fill=gp2grey] {};
	\node (s) at (4.6666,-.6666)   [draw,circle,thick,fill=gp2grey] {};
	\node (t) at (6,-.6666)        [draw,circle,thick,fill=gp2grey] {};
	\node (u) at (5.3333,-1.3333)  [draw,circle,thick,fill=gp2grey] {};
	\node (v) at (6.6666,-1.3333)  [draw,circle,thick,fill=gp2grey] {};
	\node (w) at (4.6666,-2)       [draw,circle,thick,fill=gp2grey] {};
	\node (x) at (6,-2)            [draw,circle,thick,fill=gp2grey] {};
	\node (y) at (5.3333,-2.666)   [draw,circle,thick,fill=gp2grey] {};
	
	\draw (r) edge[->,thick] (s)
	(r) edge[->,thick] (t)
	(s) edge[->,thick] (n)
	(s) edge[->,thick] (u)
	(t) edge[->,thick] (u)
	(t) edge[->,thick] (v)
	(n) edge[->,thick] (w)
	(u) edge[->,thick] (w)
	(u) edge[->,thick] (x)
	(v) edge[->,thick] (x)
	(w) edge[->,thick] (y)
	(x) edge[->,thick] (y);
\end{tikzpicture}
\captionof{figure}{Grid chain}
\label{fig:graph-class-3a}
\end{minipage}
\begin{minipage}{4.88cm}
\centering
\begin{tikzpicture}[scale=0.7]
	\node (a) at (0.0000,1.3333)   {\,};
	\node (b) at (-2.6666,0.0000)  [draw,circle,thick,fill=gp2grey] {\,};
	\node (c) at (-1.3333,0.0000)  [draw,circle,thick,fill=gp2grey] {\,};
	\node (d) at (0.0000,0.0000)   [draw,circle,thick,fill=gp2grey] {\,};
	\node (e) at (1.3333,0.0000)   [draw,circle,thick,fill=gp2grey] {\,};
	\node (f) at (2.6666,0.0000)   [draw,circle,thick,fill=gp2grey] {\,};
	\node (g) at (0.0000,-1.3333)  {\,};
	
	\draw (b) edge[->, thick] (c)
	      (c) edge[->, thick] (d)
	      (d) edge[->, thick] (e)
	      (e) edge[->, thick] (f);
\end{tikzpicture}
\captionof{figure}{Linked list}
\label{fig:graph-class-3b}
\end{minipage}
\end{figure}

For the purposes of Section \ref{sec:reduction} only, we define:

\begin{definition}[Input Graph]
\label{def:input-graph}
    An \emph{input graph} is an arbitrarily labelled GP\,2 \emph{host graph} such that:
    \begin{enumerate}
        \item every node is marked grey;
        \item every node is unrooted;
        \item every edge is unmarked.
    \end{enumerate}
\end{definition}

\subsection{Recognising Cycle Graphs}
\label{subsec:cycle_graph_rec}

The class of cycle graphs, up to labelling, consists of the graph containing one node and one edge, and graphs containing \(n\) nodes, connected by a directed cycle, for all \(n \geq 2\). It is reasonably straightforward to specify this class using reduction rules, yielding a GP\,2 program (Figure \ref{fig:is-cycle-slow-program}) which, given an \emph{input graph} \(G\), fails if and only if \(G\) it is not a cycle graph.

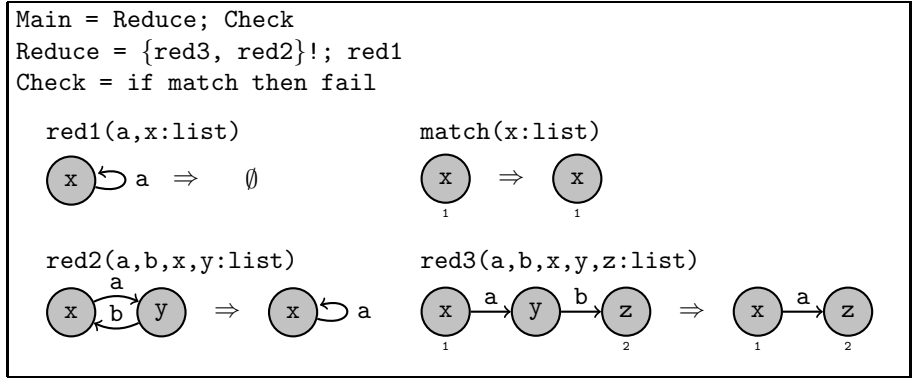
\begin{figure}[!ht]
\centering
\fbox{\begin{minipage}{11.78cm}
\begin{allintypewriter}
Main = Reduce; Check

Reduce = \{red3, red2\}!; red1

Check = if match then fail

\setlength{\tabcolsep}{0.4cm}

\medskip
\smallskip

\begin{tabular}{ p{4.18cm} p{6.00cm} }
	
	red1(a,x:list) & match(x:list) \\
							
	\begin{tikzpicture}
		\node (a) at (0,0)      [gp2 node, fill=gp2grey] {x};
		
		\node (b) at (1.525,0)  {$\Rightarrow$};
		
		\node (c) at (2.4,0.02) {$\emptyset$};
        
        \node (X) at (0.0,-.48) {\tiny{\,}};
		
		\draw (a) edge[->,black,in=15,out=-15,loop,thick] node[right] {a} (a);
	\end{tikzpicture}
	&
	
	\begin{tikzpicture}
        \node (a) at (0.0,0) [gp2 node, fill=gp2grey] {x};

        \node (b) at (.875,0) {$\Rightarrow$};

        \node (c) at (1.75,0) [gp2 node, fill=gp2grey] {x};

        \node (X) at (0.0,.28) {\tiny{\,}};
        \node (A) at (0.0,-.45) {\tiny{1}};
        \node (C) at (1.75,-.45) {\tiny{1}};
	\end{tikzpicture}
	\\

\end{tabular}

\smallskip

\begin{tabular}{ p{4.18cm} p{6.00cm} }
	
	red2(a,b,x,y:list) & red3(a,b,x,y,z:list) \\

	\vspace{-1.15em}
	\begin{tikzpicture}
		\node (a) at (0,0)     [gp2 node, fill=gp2grey] {x};
		\node (b) at (1.2,0)   [gp2 node, fill=gp2grey] {y};
		
		\node (c) at (2.075,0) {$\Rightarrow$};
		
		\node (d) at (2.95,0)  [gp2 node, fill=gp2grey] {x};
		
		\draw (a) edge[->,black,thick,bend left=25] node[above, yshift=-1pt] {a} (b)
		      (b) edge[->,black,thick,bend left=25] node[above, yshift=-1pt] {b} (a)
		      (d) edge[->,black,in=15,out=-15,loop,thick] node[right] {a} (d);
	\end{tikzpicture}					
	&

	\vspace{-0.76em}
	\begin{tikzpicture}
		\node (a) at (0,0)        [gp2 node, fill=gp2grey] {x};
		\node (b) at (1.2,0)     [gp2 node, fill=gp2grey] {y};
		\node (c) at (2.4,0)      [gp2 node, fill=gp2grey] {z};
		
		\node (d) at (3.275,0)    {$\Rightarrow$};
		
		\node (e) at (4.15,0)     [gp2 node, fill=gp2grey] {x};
		\node (f) at (5.35,0)      [gp2 node, fill=gp2grey] {z};
		
		\node (A) at (0,-.45)     {\tiny{1}};
		\node (B) at (2.4,-.45)   {\tiny{2}};
		\node (D) at (4.15,-.45)  {\tiny{1}};
		\node (E) at (5.35,-.45)   {\tiny{2}};
		
		\draw (a) edge[->,black,thick] node[above, yshift=-1pt] {a} (b)
		      (b) edge[->,black,thick] node[above, yshift=-1pt] {b} (c)
		      (e) edge[->,black,thick] node[above, yshift=-1pt] {a} (f);
	\end{tikzpicture}
	\\
\end{tabular}
\end{allintypewriter}
\end{minipage}}
\caption{GP\,2 program \ttt{is-cycle-slow}}
\label{fig:is-cycle-slow-program}
\end{figure}

Notice that, while the number of computation steps is only linear in the \emph{size} of input graph, the program need not terminate in linear time, due to the fact that the time needed to find a match for each rule application need not be only constant, as we discussed in subsection \ref{subsec:rooted_programs}. One of the novelties of GP\,2 is its implementation of root nodes. Using root nodes, we are able to direct the matching algorithm to only consider a constant \emph{size} subgraph of the input, giving us a program that is genuinely linear time (Figure \ref{fig:is-cycle-program}).

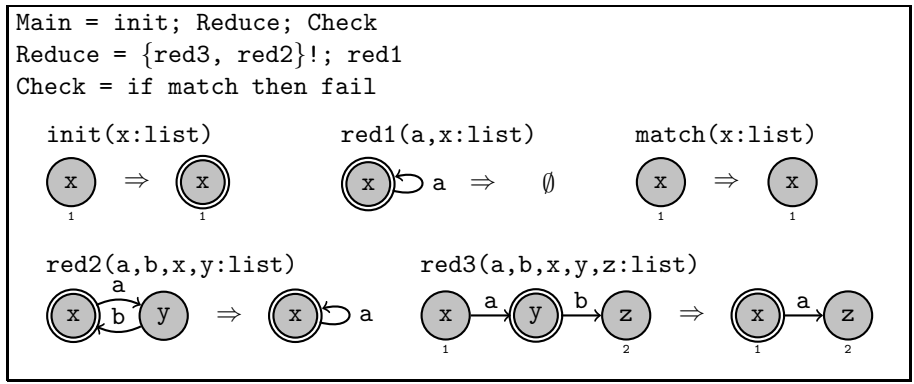
\begin{figure}[!ht]
\centering
\fbox{\begin{minipage}{11.78cm}
\begin{allintypewriter}
Main = init; Reduce; Check

Reduce = \{red3, red2\}!; red1

Check = if match then fail

\setlength{\tabcolsep}{0.4cm}

\medskip
\smallskip

\begin{tabular}{ p{3.12cm} p{3.12cm} p{3.12cm} }

    init(x:list) & red1(a,x:list) & match(x:list) \\

    \begin{tikzpicture}
        \node (a) at (0.0,0) [gp2 node, fill=gp2grey] {x};

        \node (b) at (.875,0) {$\Rightarrow$};

        \node (c) at (1.75,0) [root node, fill=gp2grey] {x};

        \node (X) at (0.0,.28) {\tiny{\,}};
        \node (A) at (0.0,-.45) {\tiny{1}};
        \node (C) at (1.75,-.45) {\tiny{1}};
    \end{tikzpicture}
    &

	\begin{tikzpicture}
		\node (a) at (0,0)      [root node, fill=gp2grey] {x};
		
		\node (b) at (1.525,0)  {$\Rightarrow$};
		
		\node (c) at (2.4,0.02) {$\emptyset$};
        
        \node (X) at (0.0,-.48) {\tiny{\,}};
		
		\draw (a) edge[->,black,in=15,out=-15,loop,thick] node[right] {a} (a);
	\end{tikzpicture}
    &

    \begin{tikzpicture}
        \node (a) at (0.0,0) [gp2 node, fill=gp2grey] {x};

        \node (b) at (.875,0) {$\Rightarrow$};

        \node (c) at (1.75,0) [gp2 node, fill=gp2grey] {x};
        
        \node (X) at (0.0,.28) {\tiny{\,}};
        \node (A) at (0.0,-.45) {\tiny{1}};
        \node (C) at (1.75,-.45) {\tiny{1}};
    \end{tikzpicture}
    \\
\end{tabular}

\smallskip

\begin{tabular}{ p{4.18cm} p{6.00cm} }
	
	red2(a,b,x,y:list) & red3(a,b,x,y,z:list) \\

	\vspace{-1.15em}
	\begin{tikzpicture}
		\node (a) at (0,0)     [root node, fill=gp2grey] {x};
		\node (b) at (1.2,0)   [gp2 node, fill=gp2grey] {y};
		
		\node (c) at (2.075,0) {$\Rightarrow$};
		
		\node (d) at (2.95,0)  [root node, fill=gp2grey] {x};
		
		\draw (a) edge[->,black,thick,bend left=25] node[above, yshift=-1pt] {a} (b)
		      (b) edge[->,black,thick,bend left=25] node[above, yshift=-1pt] {b} (a)
		      (d) edge[->,black,in=15,out=-15,loop,thick] node[right] {a} (d);
	\end{tikzpicture}
	&

	\vspace{-0.76em}
	\begin{tikzpicture}
		\node (a) at (0,0)        [gp2 node, fill=gp2grey] {x};
		\node (b) at (1.2,0)     [root node, fill=gp2grey] {y};
		\node (c) at (2.4,0)      [gp2 node, fill=gp2grey] {z};
		
		\node (d) at (3.275,0)    {$\Rightarrow$};
		
		\node (e) at (4.15,0)     [root node, fill=gp2grey] {x};
		\node (f) at (5.35,0)      [gp2 node, fill=gp2grey] {z};
		
		\node (A) at (0,-.45)     {\tiny{1}};
		\node (B) at (2.4,-.45)   {\tiny{2}};
		\node (D) at (4.15,-.45)  {\tiny{1}};
		\node (E) at (5.35,-.45)   {\tiny{2}};
		
		\draw (a) edge[->,black,thick] node[above, yshift=-1pt] {a} (b)
		      (b) edge[->,black,thick] node[above, yshift=-1pt] {b} (c)
		      (e) edge[->,black,thick] node[above, yshift=-1pt] {a} (f);
	\end{tikzpicture}
	\\
\end{tabular}
\end{allintypewriter}
\end{minipage}}
\caption{GP\,2 program \ttt{is-cycle}}
\label{fig:is-cycle-program}
\end{figure}

Even though this program is simple, we will prove its correctness and complexity now, to give a flavour of the style of such proofs, in preparation for the later proofs. By \emph{total correctness} of a program with respect to a specification, we mean that on all graphs satisfying the input description, the program terminates with output
satisfying the output description.

We must start by showing, by induction on derivation length, that the reduction rules \ttt{red2} and \ttt{red3} reduce a \emph{rooted cycle graph} to a 1-cycle, and no \emph{rooted input graphs}, where by a \emph{rooted cycle graph}, we mean a cycle graph with exactly one of the nodes a root, and similarly for a \emph{rooted input graph}.

\begin{lemma}[Cycle Reduction] \label{lem:cycle-reduction}
    Applying \ttt{red2} and \ttt{red3} to a non-empty \emph{rooted input graph} \(G\) will terminate after \(\abs{V_G} - 1\) steps, yielding a \emph{rooted 1-cycle graph} if and only if \(G\) is a \emph{rooted cycle graph}, and otherwise will yield a non-empty \emph{rooted input graph} which is not a \emph{rooted cycle graph}.
\end{lemma}

\begin{proof}
    Since both rules are size-reducing, reducing the number of nodes by \(1\), and require at least \(2\) nodes to be applicable, we have termination after \(\abs{V_G} - 1\) steps.
    
    For closedness, suppose first that \(G\) is a \emph{rooted cycle graph} with \(n\) nodes. Then the application of either one of the rules has the effect of transforming \(G\) into a \emph{rooted cycle graph} with \(n - 1\) nodes, if \(n > 1\) and if \(n = 1\) the rules aren't applicable.
    
    If \(G\) is not a \emph{rooted cycle graph}, then either the root node lies in a connected component that is a cycle graph, and the application of one of the rules has the effect of reduction that connected component only, leaving the other components alone, or there is only one connected component and it is not a cycle graph due to an additional edge present. Suppose that extra edge is a loop, then \ttt{red2} must not be applicable due to the dangling condition, and if that extra edge is a proper edge, then \ttt{red3} is either not applicable because the loop is on the root node, or \ttt{red3} is applicable and preserves the additional edge, so preserves the fact that the graph is not a \emph{rooted cycle graph}.
\end{proof}

\begin{theorem}[Correctness of \ttt{is-cycle}] \label{thm:is-cycle-correctness}
	The program \ttt{is-cycle} (Figure \ref{fig:is-cycle-program}) is totally correct with respect to the specification:
	\begin{spec}
		\specinput{An \emph{input graph}.}
		\specoutput{Fail if and only if the input is not a cycle graph.}
	\end{spec}
\end{theorem}

\begin{proof}
    Termination follows from Lemma \ref{lem:cycle-reduction}. It remains to show partial correctness. Suppose the input graph empty, then \ttt{init} will not be applicable, and the program will terminate with \ttt{fail}. It remains to analyze the case where the input graph is non-empty. Clearly, \ttt{init} will always be applicable, and will have the effect of rooting exactly one of the nodes of the graph, producing a rooted input graph which is a rooted cycle graph if and only if the original input graph was a cycle graph.
    
    What happens next is that \ttt{red3} and \ttt{red2} are applied as long as possible. We know by Lemma \ref{lem:cycle-reduction} that the result of this computation will be a rooted 1-cycle if the input graph was a cycle graph, and some non-empty non-cycle graph otherwise. In the first case, \ttt{red1} will be applied, deleting the graph, and then \ttt{match} will not be applicable, so the program succeeds (with the empty graph as output). Otherwise, either it is the case that \ttt{red1} is applicable and then \ttt{match} is too, so the program fails, or \ttt{red1} is not applicable, so the program fails.
\end{proof}

\begin{lemma}[Complexity of \ttt{init}] \label{lem:is-cycle-init-complexity}
    \ttt{init} terminates in constant time on an \emph{input graph} \(G\).
\end{lemma}

\begin{proof}
    The search plan will iterate all nodes in \(G\) looking for the first grey unrooted node. Testing if a node is unrooted and grey takes only constant time. Since every node in an \emph{input graph} is unrooted and grey, the search will stop at the first node, or fail in constant time if \(G\) is empty.
\end{proof}

\begin{lemma}[Complexity of \ttt{Reduce}] \label{lem:is-cycle-reduce-complexity}
    \ttt{Reduce} terminates in linear time with respect to the number of nodes in a \emph{rooted input graph} \(G\).
\end{lemma}

\begin{proof}
    First, we must argue that \ttt{red1}, \ttt{red2} and \ttt{red3} each take only constant time to either evaluate to \ttt{fail} or produce a result graph. Then by Lemma \ref{lem:cycle-reduction}, we know that \ttt{\{red3, red2\}!} runs for only a linear number of steps, and each takes only constant time. So \ttt{Reduce} must take only linear time, since it involves executing \ttt{\{red3, red2\}!} followed by \ttt{red1}.
    
    \ttt{red1} is the simplest to analyze. The search plan will look for a root node, and find the first one in constant time, and check that it is grey in constant time, which it is. Next, it will check the incoming and outcoming degrees are both one in constant time. If they are not, then the node is rejected and the search plan looks for another root node and determines there are no more in constant time, and matching fails. If the degrees are correct, next the search plan grabs the first outgoing edge in constant time and checks if it is an unmarked and a loop in constant time. If it is not, then matching fails as before. Otherwise, matching succeeds.
    
    For \ttt{red2}, the search plan first looks at the root nodes in exactly the same way as for \ttt{red1}, with the same degree checking and rejection handling. The first difference occurs when the search plan looks at the first outgoing edge. It instead checks the edge is proper, also in constant time. It then checks, in constant time, that the target node is grey, unrooted, and has incoming and outgoing degree one. If it is not, the root node is rejected, as before. Otherwise, the first outgoing edge of the new node is grabbed in constant time and checking that it is unmarked and has the root node as a target occurs in constant time. If it is not, then we must look for a new root node, as before, which necessarily fails, and we stop in constant time.
    
    \ttt{red3} will turn out to be constant time also, due to the important decision to make the non-interface node the root node. Correctness would not have been impacted had we made a different node in the left-hand side graph the root, but matching complexity would have been impacted! Initially, the search plan proceeds as in \ttt{red1}, again, finding a grey root with incoming and outgoing degree one. Next, the search plan grabs the first outgoing edge in constant time and checks it is unmarked and proper in constant time, and that the target node is grey and unrooted. If this fails, then we bailout, as usual, having to look for another root node which fails in constant time. If successful, we then due the dual for the incoming edge of the root node in constant time.
\end{proof}

\begin{lemma}[Complexity of \ttt{Check}] \label{lem:is-cycle-check-complexity}
    \ttt{Check} terminates in constant time on an \emph{rooted input graph} \(G\).
\end{lemma}

\begin{proof}
    The search plan will iterate all nodes in \(G\) looking for the first grey unrooted node. Testing if a node is unrooted and grey takes only constant time. The search plan will only consider at most \(2\) nodes. There are three cases to consider. The first case is that the first node we consider is what we want, and we stop. The second case is that the first node we consider is a root node and there are no other nodes, so we stop. The final case is that the first node we consider is a root, and then the second node is the one we are looking for.
\end{proof}

\begin{theorem}[Complexity of \ttt{is-cycle}] \label{thm:is-cycle-complexity}
    The program \ttt{is-cycle} (Figure \ref{fig:is-cycle-program}) terminates in linear time with respect to the \emph{size} of its input.
\end{theorem}

\begin{proof}
    Due to Lemma \ref{lem:is-cycle-init-complexity} \ttt{init} takes only constant time to either evaluate to \ttt{fail}, which stops the whole program, or evaluate to a \emph{rooted input graph}. Next, by Lemma \ref{lem:is-cycle-reduce-complexity} \ttt{Reduce} takes linear time in the number of nodes, either evaluating to \ttt{fail}, which stops the whole program, or evaluates to a \emph{rooted input graph} (Lemma \ref{lem:cycle-reduction}). Finally, \ttt{fail} takes only constant time by Lemma \ref{lem:is-cycle-check-complexity}.
\end{proof}

Finally, we have collected empirical timing results for \ttt{is-cycle}, supporting our claim that the program runs in linear time, even on graph classes that do not have bounded degree (Figure \ref{fig:is-cycle-timing}).

\begin{figure}[!ht]
\centering
\begin{tikzpicture}[scale=0.9]
    \begin{axis}[
    xlabel={Size of input graph},
    ylabel={Execution time (ms)},
    xmin=0,
    ymin=0,
    width=9.2cm,height=7.2cm,
    legend style={at={(1.5,0.82)}},
    ymajorgrids=true,
    grid style=dashed,
    ]
        \addplot[color=plot1, mark=square*] 
        coordinates {
        	(20000,52.98)
        	(60000,90.55)
        	(100000,130.82)
        	(140000,169.22)
        	(180000,206.81)
        	(220000,244.40)
        	(260000,282.45)
        	(300000,323.84)
        	(340000,363.60)
        	(380000,402.76)
        	(420000,442.13)
        	(460000,482.32)
        	(500000,521.78)
        };
        \addplot[color=plot2, mark=square*] 
        coordinates {
        	(19040,43.14)
        	(32865,54.87)
        	(50440,67.98)
        	(71765,83.19)
        	(96840,101.55)
        	(125665,118.96)
        	(158240,140.18)
        	(194565,165.48)
        	(234640,196.94)
        	(278465,224.64)
        	(326040,255.64)
        	(377365,291.32)
        	(432440,329.67)
        	(491265,372.50)
        };
        \addplot[color=plot3, mark=square*] 
        coordinates {
        	(19837,44.61)
        	(30955,53.68)
        	(45601,63.60)
        	(64261,74.91)
        	(87421,90.34)
        	(115567,111.33)
        	(149185,137.87)
        	(188761,159.28)
        	(234781,193.49)
        	(287731,226.46)
        	(348097,270.38)
        	(416365,320.28)
        	(493021,371.04)
        };
        \addplot[color=plot4, mark=square*] 
        coordinates {
        	(16381,43.69)
        	(32765,57.26)
        	(65533,78.95)
        	(131069,126.71)
        	(262141,218.22)
        	(524285,405.35)
        };
        \addplot[color=plot5, mark=square*] 
        coordinates {
        	(20000,52.10)
        	(60000,89.39)
        	(100000,126.53)
        	(140000,162.53)
        	(180000,199.15)
        	(220000,234.06)
        	(260000,268.46)
        	(300000,306.00)
        	(340000,341.08)
        	(380000,378.25)
        	(420000,414.83)
        	(460000,451.23)
        	(500000,486.79)
        };
        \addplot[color=plot6, mark=square*] 
        coordinates {
        	(20000,58.12)
        	(60000,83.78)
        	(100000,116.79)
        	(140000,145.64)
        	(180000,176.19)
        	(220000,204.11)
        	(260000,233.54)
        	(300000,263.46)
        	(340000,296.00)
        	(380000,322.80)
        	(420000,355.37)
        	(460000,385.27)
        	(500000,413.78)
        };
        \addplot[color=plot7, mark=square*] 
        coordinates {
        	(19999,53.50)
        	(59999,87.99)
        	(99999,124.22)
        	(139999,158.54)
        	(179999,192.51)
        	(219999,224.70)
        	(259999,260.09)
        	(299999,294.13)
        	(339999,329.65)
        	(379999,363.40)
        	(419999,398.76)
        	(459999,435.23)
        	(499999,470.31)
        };
        \addplot[color=plot8, mark=square*] 
        coordinates {
        	(19999,52.37)
        	(59999,89.89)
        	(99999,124.16)
        	(139999,162.55)
        	(179999,196.38)
        	(219999,230.35)
        	(259999,265.48)
        	(299999,303.13)
        	(339999,338.17)
        	(379999,372.74)
        	(419999,410.43)
        	(459999,446.84)
        	(499999,485.26)
        };
        \addlegendentry{Discrete Graph}
        \addlegendentry{Grid Graph}
        \addlegendentry{Grid Chain}
        \addlegendentry{Binary Tree}
        \addlegendentry{Cycle Graph}
        \addlegendentry{Sun Graph}
        \addlegendentry{Linked List}
        \addlegendentry{Star Graph}
    \end{axis}
\end{tikzpicture}
\caption{Measured performance of \ttt{is-cycle}}
\label{fig:is-cycle-timing}
\end{figure}
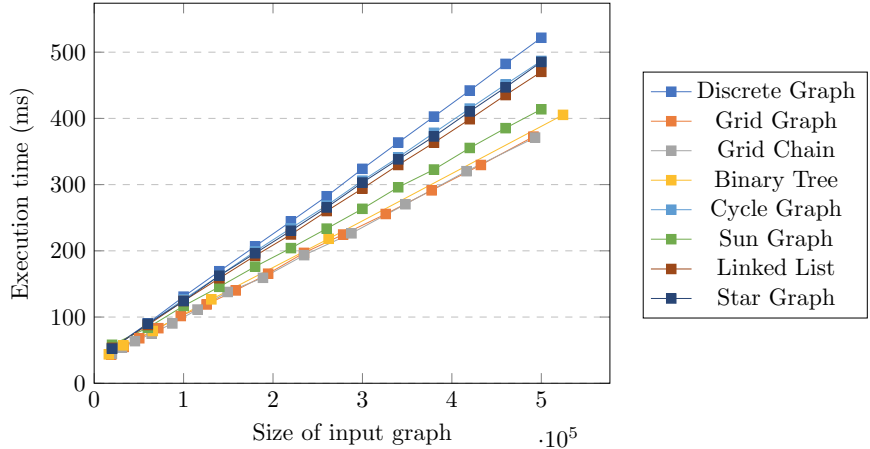

\subsection{Recognising Trees}
\label{subsec:tree_reco}

A tree is a graph containing a node from which there is a unique directed path to each node in the graph. It is easy to see that it is possible to generate the class of all unlabelled trees by inductively adding new leaf nodes to the discrete graph of \emph{size} one, thus the class can be specified by graph reduction. The question of linear time recognition of trees was partially solved by Campbell in 2019, producing a GP\,2 program to recognise trees in linear time, given the input class has only bounded degree \cite{Campbell19a}. In this subsection we improve on this result, removing the bounded degree restriction.

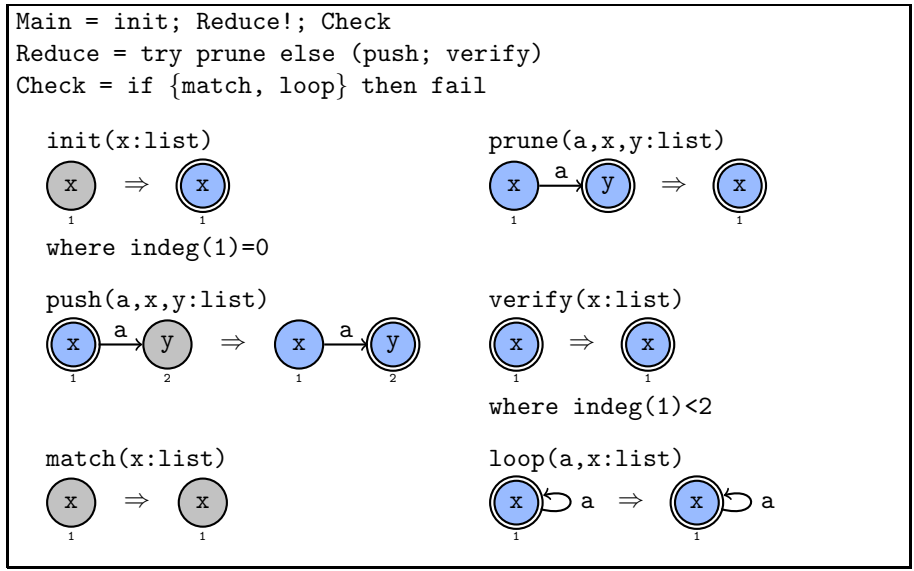
\begin{figure}[!ht]
\centering
\fbox{\begin{minipage}{11.78cm}
\begin{allintypewriter}
Main = init; Reduce!; Check

Reduce = try prune else (push; verify)

Check = if \{match, loop\} then fail

\setlength{\tabcolsep}{0.4cm}

\medskip
\smallskip

\begin{tabular}{ p{5.09cm} p{5.09cm} }
	
	init(x:list) & prune(a,x,y:list) \\
	
	\begin{tikzpicture}
		\node (a) at (0,0)        [gp2 node, fill=gp2grey] {x};
		
		\node (b) at (.875,0)     {$\Rightarrow$};
		
		\node (c) at (1.75,0)     [root node, fill=gp2blue] {x};
		
		\node (A) at (0,-.45)     {\tiny{1}};
		\node (C) at (1.75,-.45)  {\tiny{1}};
	\end{tikzpicture}
	&
	
	\begin{tikzpicture}
		\node (a) at (0,0)        [gp2 node, fill=gp2blue] {x};
		\node (b) at (1.25,0)     [root node, fill=gp2blue] {y};
		
		\node (c) at (2.125,0)    {$\Rightarrow$};
		
		\node (d) at (3,0)        [root node, fill=gp2blue] {x};
		
		\node (A) at (0,-.45)     {\tiny{1}};
		\node (D) at (3,-.45)     {\tiny{1}};
		
		\draw (a) edge[->,thick] node[above, yshift=-1pt] {a} (b);
	\end{tikzpicture}
	\\
	
	\vspace{-1.15em}where indeg(1)=0 & \\
\end{tabular}

\smallskip

\begin{tabular}{ p{5.09cm} p{5.09cm} }
	
	push(a,x,y:list) & verify(x:list) \\
	
	\begin{tikzpicture}
		\node (a) at (0,0)        [root node, fill=gp2blue] {x};
		\node (b) at (1.25,0)     [gp2 node, fill=gp2grey] {y};
		
		\node (c) at (2.125,0)    {$\Rightarrow$};
		
		\node (d) at (3,0)        [gp2 node, fill=gp2blue] {x};
		\node (e) at (4.25,0)     [root node, fill=gp2blue] {y};
		
		\node (A) at (0,-.45)     {\tiny{1}};
		\node (B) at (1.25,-.45)  {\tiny{2}};
		\node (D) at (3,-.45)     {\tiny{1}};
		\node (E) at (4.25,-.45)  {\tiny{2}};
		
		\draw (a) edge[->,thick] node[above, yshift=-1pt] {a} (b)
              (d) edge[->,thick] node[above, yshift=-1pt] {a} (e);
	\end{tikzpicture}
	&
	
	\begin{tikzpicture}
		\node (a) at (0,0)        [root node, fill=gp2blue] {x};
		
		\node (b) at (.875,0)     {$\Rightarrow$};
		
		\node (c) at (1.75,0)     [root node, fill=gp2blue] {x};
		
		\node (A) at (0,-.45)     {\tiny{1}};
		\node (C) at (1.75,-.45)  {\tiny{1}};
	\end{tikzpicture}
	\\
	
	\vspace{-1.15em}\, & \vspace{-1.15em}where indeg(1)<2 \\
\end{tabular}

\smallskip

\begin{tabular}{ p{5.09cm} p{5.09cm} }
	
	match(x:list) & loop(a,x:list) \\
	
	\begin{tikzpicture}
		\node (a) at (0,0)        [gp2 node, fill=gp2grey] {x};
		
		\node (b) at (.875,0)     {$\Rightarrow$};
		
		\node (c) at (1.75,0)     [gp2 node, fill=gp2grey] {x};
		
		\node (A) at (0,-.45)     {\tiny{1}};
		\node (C) at (1.75,-.45)  {\tiny{1}};
	\end{tikzpicture}
	&
	
	\begin{tikzpicture}
		\node (a) at (0,0)        [root node, fill=gp2blue] {x};
		
		\node (b) at (1.525,0)    {$\Rightarrow$};
		
		\node (c) at (2.4,0)      [root node, fill=gp2blue] {x};
		
		\node (A) at (0,-.45)     {\tiny{1}};
		\node (C) at (2.4,-.45)   {\tiny{1}};
		
		\draw (a) edge[->,black,in=15,out=-15,loop,thick] node[right] {a} (a)
		      (c) edge[->,black,in=15,out=-15,loop,thick] node[right] {a} (c);
	\end{tikzpicture}
	\\
\end{tabular}
\end{allintypewriter}
\end{minipage}}
\caption{GP\,2 program \ttt{is-tree}}
\label{fig:is-tree-program}
\end{figure}

\begin{figure}[!ht]
\centering
\scalebox{.65}{\begin{tikzpicture}
\node (a) at (-5.4,0)  [gp2 node, fill=gp2grey] {3};
\node (b) at (-5.9,-1) [gp2 node, fill=gp2grey] {2};
\node (c) at (-4.9,-1) [gp2 node, fill=gp2grey] {4};
\node (d) at (-5.9,-2) [gp2 node, fill=gp2grey] {1};

\draw (a) edge[->,thick] (b)
      (a) edge[->,thick] (c)
      (b) edge[->,thick] (d);

\node (t) at (-4.05,-0.9) {$\Rightarrow$};
\node (t) at (-4.05,-1.1) {\ttt{\scriptsize{init}}};

\node (a) at (-2.7,0)  [root node, fill=gp2blue] {3};
\node (b) at (-3.2,-1) [gp2 node, fill=gp2grey] {2};
\node (c) at (-2.2,-1) [gp2 node, fill=gp2grey] {4};
\node (d) at (-3.2,-2) [gp2 node, fill=gp2grey] {1};

\draw (a) edge[->,thick] (b)
      (a) edge[->,thick] (c)
      (b) edge[->,thick] (d);

\node (t) at (-1.35,-0.9) {$\Rightarrow$};
\node (t) at (-1.35,-1.1) {\ttt{\scriptsize{push}}};

\node (a) at (0,0)     [gp2 node, fill=gp2blue] {3};
\node (b) at (-0.5,-1) [root node, fill=gp2blue] {2};
\node (c) at (0.5,-1)  [gp2 node, fill=gp2grey] {4};
\node (d) at (-0.5,-2) [gp2 node, fill=gp2grey] {1};

\draw (a) edge[->,thick] (b)
      (a) edge[->,thick] (c)
      (b) edge[->,thick] (d);

\node (t) at (1.35,-0.9) {$\Rightarrow$};
\node (t) at (1.35,-1.1) {\ttt{\scriptsize{push}}};

\node (a) at (2.7,0)  [gp2 node, fill=gp2blue] {3};
\node (b) at (2.2,-1) [gp2 node, fill=gp2blue] {2};
\node (c) at (3.2,-1) [gp2 node, fill=gp2grey] {4};
\node (d) at (2.2,-2) [root node, fill=gp2blue] {1};

\draw (a) edge[->,thick] (b)
      (a) edge[->,thick] (c)
      (b) edge[->,thick] (d);

\node (t) at (4.05,-0.9) {$\Rightarrow$};
\node (t) at (4.05,-1.1) {\ttt{\scriptsize{prune}}};

\node (a) at (5.4,0)  [gp2 node, fill=gp2blue] {3};
\node (b) at (4.9,-1) [root node, fill=gp2blue] {2};
\node (c) at (5.9,-1) [gp2 node, fill=gp2grey] {4};

\draw (a) edge[->,thick] (b)
      (a) edge[->,thick] (c);

\node (t) at (6.75,-0.9) {$\Rightarrow$};
\node (t) at (6.75,-1.1) {\ttt{\scriptsize{prune}}};

\node (a) at (7.6,0)  [root node, fill=gp2blue] {3};
\node (c) at (7.6,-1) [gp2 node, fill=gp2grey] {4};

\draw (a) edge[->,thick] (c);

\node (t) at (8.45,-0.9) {$\Rightarrow$};
\node (t) at (8.45,-1.1) {\ttt{\scriptsize{push}}};

\node (a) at (9.3,0)  [gp2 node,fill=gp2blue] {3};
\node (c) at (9.3,-1) [root node,fill=gp2blue] {4};

\draw (a) edge[->,thick] (c);

\node (t) at (10.15,-0.9) {$\Rightarrow$};
\node (t) at (10.15,-1.1) {\ttt{\scriptsize{push}}};

\node (a) at (11.0,0)  [root node,fill=gp2blue] {3};

\end{tikzpicture}}
\caption{Example tree reduction}
\label{fig:is-tree-example-1}
\end{figure}
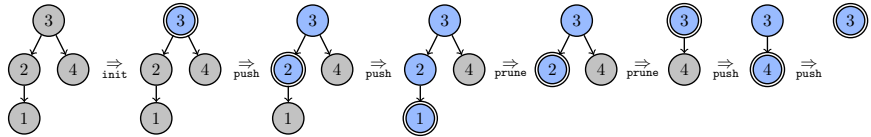

\begin{figure}[!ht]
\centering
\begin{minipage}{6.58cm}
\centering
\scalebox{.65}{\begin{tikzpicture}

\node (a) at (0.0,0)   [gp2 node, fill=gp2grey] {1};
\node (b) at (1.0,0)   [gp2 node, fill=gp2grey] {3};
\node (c) at (0.0,-1)  [gp2 node, fill=gp2grey] {2};
\node (d) at (1.0,-1)  [gp2 node, fill=gp2grey] {4};

\draw (a) edge[->,thick] (c)
      (b) edge[->,thick] (d);

\node (t) at (1.75,-0.4) {$\Rightarrow$};
\node (t) at (1.75,-0.6) {\ttt{\scriptsize{init}}};

\node (a) at (2.5,0)   [root node, fill=gp2blue] {1};
\node (b) at (3.5,0)   [gp2 node, fill=gp2grey] {3};
\node (c) at (2.5,-1)  [gp2 node, fill=gp2grey] {2};
\node (d) at (3.5,-1)  [gp2 node, fill=gp2grey] {4};

\draw (a) edge[->,thick] (c)
      (b) edge[->,thick] (d);

\node (t) at (4.25,-0.4) {$\Rightarrow$};
\node (t) at (4.25,-0.6) {\ttt{\scriptsize{push}}};

\node (a) at (5.0,0)  [gp2 node, fill=gp2blue] {1};
\node (b) at (6.0,0)  [gp2 node, fill=gp2grey] {3};
\node (c) at (5.0,-1) [root node, fill=gp2blue] {2};
\node (d) at (6.0,-1) [gp2 node, fill=gp2grey] {4};

\draw (a) edge[->,thick] (c)
      (b) edge[->,thick] (d);

\node (t) at (6.75,-0.4) {$\Rightarrow$};
\node (t) at (6.75,-0.6) {\ttt{\scriptsize{prune}}};

\node (a) at (7.5,0)  [root node, fill=gp2blue] {1};
\node (b) at (8.5,0)  [gp2 node, fill=gp2grey] {3};
\node (d) at (8.5,-1) [gp2 node, fill=gp2grey] {4};

\draw (b) edge[->,thick] (d);

\end{tikzpicture}}
\captionof{figure}{Example forest reduction}
\label{fig:is-tree-example-2a}
\end{minipage}
\begin{minipage}{5.20cm}
\centering
\scalebox{.65}{\begin{tikzpicture}[every node/.style={inner sep=0pt, text width=6.5mm, align=center}]

\node (a) at (0.0,0)   [gp2 node, fill=gp2grey] {1};
\node (b) at (0.0,-1)  [gp2 node, fill=gp2grey] {2};
\node (c) at (1.0,-1)  [gp2 node, fill=gp2grey] {3};

\draw (a) edge[->,thick] (b)
      (b) edge[->,thick, bend left=15] (c)
      (c) edge[->,thick, bend left=15] (b);

\node (t) at (1.75,-0.4) {$\Rightarrow$};
\node (t) at (1.75,-0.6) {\ttt{\scriptsize{init}}};

\node (a) at (2.5,0)   [root node, fill=gp2blue] {1};
\node (b) at (2.5,-1)  [gp2 node, fill=gp2grey] {2};
\node (c) at (3.5,-1)  [gp2 node, fill=gp2grey] {3};

\draw (a) edge[->,thick] (b)
      (b) edge[->,thick, bend left=15] (c)
      (c) edge[->,thick, bend left=15] (b);

\node (t) at (4.25,-0.4) {$\Rightarrow$};
\node (t) at (4.25,-0.6) {\ttt{\scriptsize{push}}};

\node (a) at (5.0,0)  [gp2 node, fill=gp2blue] {1};
\node (b) at (5.0,-1) [root node, fill=gp2blue] {2};
\node (c) at (6.0,-1) [gp2 node, fill=gp2grey] {3};

\draw (a) edge[->,thick] (b)
      (b) edge[->,thick, bend left=15] (c)
      (c) edge[->,thick, bend left=15] (b);

\end{tikzpicture}}
\captionof{figure}{Example reduction with cycle}
\label{fig:is-tree-example-2b}
\end{minipage}
\end{figure}

Intuitively, our new program (Figure \ref{fig:is-tree-program}) works by pushing a special \emph{root} node to the bottom of a branch, and then prunes, repeating the process as long as possible. If we start with a tree and run this until we cannot do it anymore, we must be left with a single root node (Figures \ref{fig:is-tree-example-1} and \ref{fig:is-tree-example-2a}).

Notice that we have used both grey and blue node colours. This is necessary in order to ensure termination. Consider an \emph{input graph} that is not a tree, say it contains a 2-cycle (Figure \ref{fig:is-tree-example-2b}). Then, we would have otherwise been able to push the root node round and round forever! Leaving a trail of blue nodes behind us as we push prevents this from happening.

In order to prove correctness and complexity, we must first show some intermediate results. Using the fact that an equivalent characterisation of a tree is a non-empty connected graph without undirected cycles such that every node has at most one incoming edge, we can show that our reduction procedure preserves the property of either being a tree or not being a tree. The additional \ttt{verify} rule is needed to make sure that we stay within linear time complexity, even on graphs of unbounded degree.

\begin{definition}[Live Graph]
    A \emph{live graph} (for the purposes of this subsection) is any graph that can be derived from an \emph{input graph} by applying \ttt{init} followed by zero or more applications of \ttt{push} or \ttt{prune} in any order, where these rules are given in Figure \ref{fig:is-tree-program}.
\end{definition}

\begin{lemma}[Live Graph Properties] \label{lem:live-graphs}
    Live graphs have exactly one blue root node, every other node either grey or blue, and all edges unmarked.
\end{lemma}

\begin{proof}
    The only rule that can change the number of root nodes is \ttt{init}. Now, \ttt{init} is always applied to a graph with no root nodes, so necessarily must increase the number of root nodes by exactly one. No rules have marked edges in their RHSs, and similarly, have only grey and blue nodes in their RHSs.
\end{proof}

\begin{lemma}[Tree Reduction] \label{lem:tree-reduction}
    \ttt{Reduce} is a refinement of \ttt{\{push, prune\}}. That is, if \(G \Rightarrow_{\ttt{Reduce}} H\), then there must either be a direct derivation \(G \Rightarrow_{\ttt{push}} H\) or \(G \Rightarrow_{\ttt{prune}} H\). Moreover, G is a tree if and only if \(H\) is.
\end{lemma}

\begin{proof}
	For the first part, the observation to make is that the verify rule does not modify graphs. Its only purpose is to possibly evaluate to fail. So, the result then followings from the definition of \ttt{try}.
	
	For the second part, it suffices to show that \ttt{prune} and \ttt{push} preserve the property of being a (non-)tree. Clearly \ttt{push} does not modify the structure of graphs, so we just need to analyse \ttt{prune}, which has the effect of deleting an edge and its target only when the target has no other incident edges. If \(G\) is a tree, then \(H\) is also a tree, since all we have done is deleted a leaf node. If \(G\) is not a tree, then \(H\) must also not be a tree, since if \(G\) was disconnected, \(H\) must be too, and if \(G\) has any undirected cycles, then they cannot have passed through the deleted node due to the condition on other incident edges, so \(H\) has the same undirected cycles as \(G\), and if \(G\) node with incoming degree at least \(2\), then so must \(H\) since the incoming degree of all the remaining nodes in \(H\) is the same as in \(G\) and the deleted node had only incoming degree \(1\).
\end{proof}

\begin{definition}[Blue Paths and Cycles]
    A \emph{blue} \emph{path} (\emph{cycle}) is a \emph{path} (\emph{cycle}) in a graph with all nodes blue. The \emph{length} of a \emph{blue} \emph{path} (\emph{cycle}) is the number of edges.
\end{definition}

\begin{lemma}[Live Graph Blue Paths] \label{lem:live-blue-paths}
    Given a \emph{live graph}, then:
    
    \begin{itemize}
        \item[(P1)] There is a unique blue node with incoming degree zero, and all other blue nodes have incoming degree one;
        \item[(P2)] There are no blue cycles, only blue paths;
        \item[(P3)] All maximal blue paths start from the incoming degree zero node and end at the root node.
    \end{itemize}
\end{lemma}

\begin{proof}
	First, note together with Lemma \ref{lem:live-graphs}, this means there is exactly one maximal blue path in each \emph{live graph} (the maximum blue path), otherwise, there would either be a blue cycle or a node with incoming degree of at least \(2\). Now, to show the properties, we will proceed by induction on derivation length.
	
	Let \(I\) be an arbitrary \emph{input graph} and \(I \Rightarrow_{\ttt{init}} G\). \(I\) and \(G\) are the two base cases we must check. By definition, \(I\) satisfies the properties. To see that \(G\) does, note that all nodes in \(G\) must be grey and unrooted apart from one blue root node with incoming degree \(0\), due to the application condition of \ttt{init}.
	
	Suppose now that we have \(I \Rightarrow_{\ttt{init}} G \Rightarrow_{Reduce}^n\) with \(H\) satisfying the properties by inductive hypothesis. We now consider the two possible cases for the successor graph to \(H\). By Lemma \ref{lem:tree-reduction}, we have either \(H \Rightarrow_{\ttt{prune}} M\) or \(H \Rightarrow_{\ttt{push}} M\).
	
	In the first case, \(M\) is exactly \(H\) but with the bottom of the maximal path deleted and the new bottom node rooted. There is necessarily no blue child of the new root in \(M\), since this would contradict the uniqueness of the original maximal path in \(H\). Since the incoming degree of every remaining node is left unchanged, we conclude that all of the properties hold, as required.
	
	In the second case, \(M\) is exactly \(H\) but with the bottom of the maximal path extended along an existing edge to an existing previously grey node, becoming the unique blue root in \(M\). Suppose by contradiction that this root had a blue child, then once again this contradicts the uniqueness of the maximal path ion \(H\). Since the incoming degree of every node is unchanged and the application condition of \ttt{verify} ensures our replacement root has incoming degree at most \(1\), then we conclude that all of the properties hold, as required.
\end{proof}

\begin{lemma}[Live Tree Reduction] \label{lem:live-tree-reduction}
    Given a \emph{live tree} \(G\), then either \(G\) is a single blue root node, or \ttt{Reduce} is applicable.
\end{lemma}

\begin{proof}
	Suppose \(G\) is a live tree. Then by Lemma \ref{lem:live-graphs}, it has a blue root node. If this is the only node \(G\), then \(G\) must necessarily be discrete: \(G\) is a single isolated blue root node. Alternatively, if \(G\) has more than one node, then either this root node has outgoing degree \(0\) or not. We must analyze these two cases.
	
	If the root has no outgoing edges, then it must have incoming degree \(1\) since \(G\) is a tree with at least \(2\) nodes. But then by Lemma \ref{lem:live-blue-paths}, there must be a blue node above it, so \ttt{prune} must be applicable. Thus, \ttt{Reduce} is applicable, as required.
	
	If the root has at least one outgoing edge, then by Lemma \ref{lem:live-blue-paths}, all of the nodes below it are grey, so we \ttt{push} must be applicable. Then, since the result is a tree, due to Lemma \ref{lem:tree-reduction}, we know that \ttt{verify} is always applicable, since all nodes in a tree of incoming degree at most \(1\). Thus, \ttt{Reduce} is applicable, as required.
\end{proof}

\begin{lemma}[Intermediate Complexity Result] \label{lem:is-tree-intermediate-complexity}
    Call the entire application of \ttt{init} or \ttt{Reduce} a single computation step, let \(G\) be an \emph{input graph} with \(n\) nodes, and define \ttt{P} to be the program \ttt{init; Reduce!}. Then \ttt{P} terminates after at most \(\textrm{max}(1, 2(n-1))\) steps on \(G\).
\end{lemma}

\begin{proof}
	Define the weight \(w(G)\) of \(G\) to be \(2g+b\) where \(g\) is the number of grey nodes in \(G\) and \(b\) the number of blue nodes in \(G\).
	
	From an \emph{input graph} \(G\), we either have \(G \Rightarrow_{\ttt{init}} \ttt{fail}\), or \(G \Rightarrow_{\ttt{init}} H\), where \(H\) is a \emph{live graph}. In the first case, the program has terminated after \(1\) step, and we're done. In the second case, we have performed one step to derive \(H\), and next we can proceeded to apply \ttt{Reduce} as long as possible. By Lemma \ref{lem:tree-reduction}, \ttt{Reduce} has the same effect as applying either \ttt{prune} or \ttt{push}. Both of these decrease the weight of the graph by exactly \(1\), and can only be applied to graphs of at least weight \(2\), so \ttt{Reduce} can be applied at most \(2n - 3\) times.
	
	Thus \ttt{P}, terminates after \(1\) step on \emph{input graph}s of weight less then \(2\), and otherwise terminates in \(1 + (2n-3) = 2(n-1)\) steps, as required.
\end{proof}

\begin{lemma}[Correctness of \ttt{Check}] \label{lem:is-tree-check-correctness}
    \ttt{Check} evaluates to \ttt{fail} on any graph \(G\) if and only if \(G\) has \emph{size} strictly greater than one.
\end{lemma}

\begin{proof}
	\ttt{\{match, loop\}} passes if and only if \(G\) has at least two nodes, or a loop, which happens exactly when the \emph{size} of \(G\) is at least two. The result then follows from the fact that \ttt{Check} fails if and only if \ttt{\{match, loop\}} passes.
\end{proof}

\begin{theorem}[Correctness of \ttt{is-tree}] \label{thm:is-tree-correctness}
	The program \ttt{is-tree} (Figure \ref{fig:is-tree-program}) is totally correct with respect to the specification:
	\begin{spec}
		\specinput{An \emph{input graph}.}
		\specoutput{Fail if and only if the input is not a tree.}
	\end{spec}
\end{theorem}

\begin{proof}
	Termination follows from Lemma \ref{lem:is-tree-intermediate-complexity}. To see partial correctness, let \(G\) be an \emph{input graph}. We break down our proof into two cases.
	
	First, suppose \(G\) is empty. Then \ttt{init} evaluates to \ttt{fail}, and thus \ttt{Main} evaluates to \ttt{fail}, as required.
	
	Next, suppose \(G\) is non-empty and not a tree. Then, if \ttt{init} evaluates to \ttt{fail} (such as because \(G\) is empty), then \ttt{Main} evaluates to \ttt{fail}, as required. Otherwise, \ttt{init} must have evaluated to a \emph{live graph}, then by Lemma \ref{lem:tree-reduction}, \ttt{Reduce!} will evaluate to a non-tree, which is non-empty by Lemma \ref{lem:live-graphs}. Finally, by Lemma \ref{lem:is-tree-check-correctness}, the subprogram \ttt{if Check then fail} must evaluate to \ttt{fail}, since it was fed a graph of \emph{size} at least \(2\).
	
	Finally, suppose \(G\) is a tree. Then \ttt{init} is necessarily applicable, producing a live tree. By Lemmas \ref{lem:tree-reduction} and \ref{lem:live-tree-reduction}, \ttt{Reduce!} produces a graph of \emph{size} \(1\), so by Lemma \ref{lem:is-tree-check-correctness}, the subprogram \ttt{if Check then fail} must not evaluate to \ttt{fail}, as required.
\end{proof}

Recall from Lemma \ref{lem:const-time} that all rules that do not modify labels have constant time application, once a match has been found. All rules of \ttt{is-tree} satisfy this condition, so in our remaining proofs, we omit reasoning about application complexity, only discussing matching time complexity.

\begin{lemma}[Complexity of \ttt{init}] \label{lem:is-tree-init-complexity}
    \ttt{init} terminates in linear time with respect to the number of nodes in an \emph{input graph} \(G\).
\end{lemma}

\begin{proof}
	The search plan will iterate all the nodes in \(G\) looking for the first unrooted grey node with incoming degree \(0\). Testing if a node is unrooted, grey and has incoming degree \(0\) is only constant time, so matching takes linear time in the worst case, to either find a match satisfying the application conditions or determine there is no such match.
\end{proof}

\begin{lemma}[Complexity of \ttt{Reduce}] \label{lem:is-tree-reduce-complexity}
    \ttt{Reduce} terminates in constant time on a \emph{live graph} \(G\).
\end{lemma}

\begin{proof}
	First, the program will try to find a match for \ttt{prune}. If it succeeds, there is no more time needed for matching. Otherwise, it will next try to find a match for \ttt{push}. If it succeeds, the only other matching code to run is that of \ttt{verify} on the result \(H\) of applying \ttt{push} to \(G\).
	
	When trying to find a match for \ttt{prune} in \(G\), first we identify the root in the rule LHS with the root in \(G\) (which exists by Lemma \ref{lem:live-graphs}), in constant time. We then check the root is blue in constant time, which it will be by Lemma \ref{lem:live-graphs}. Next, we check that the root has outgoing degree \(0\) and incoming degree \(1\) in constant time. If it doesn't, then we must look for a second root node in \(G\), which doesn't exist by Lemma \ref{lem:live-graphs}. We detect this in constant time, then \ttt{fail}. Otherwise, the root necessarily has a proper incoming edge. The program will check the edge is unmarked, in constant time (and it will be by Lemma \ref{lem:live-graphs}) and checks that the source node is blue, in constant time (which it will be by Lemma \ref{lem:live-blue-paths}. We have now found a match (or not) in constant time.
	
	When trying to find a match for \ttt{push} in \(G\), first we identify the root in the rule LHS with the root in \(G\) (which exists by Lemma \ref{lem:live-graphs}), in constant time. We then check the root is blue, in constant time, which it will be by Lemma \ref{lem:live-graphs}. Next, we check the root has outgoing degree a least \(1\), in constant time. If it doesn't, then we must look for a second root node in \(G\), which doesn't exist by Lemma \ref{lem:live-graphs}. We detect this in constant time, then \ttt{fail}. Otherwise, by Lemma \ref{lem:live-blue-paths}, the first outgoing edge we consider must be proper and unmarked. After verifying this, in constant time, we then look at the target node, and check it is grey, in constant time. We have now found a match (or not) in constant time.
	
	Finally, when trying to find a match for \ttt{verify} in \(H\). we identify the root in the LHS of the rule with the root in \(H\) which exists and is blue by the definition of \ttt{push} and the fact that \(H\) was the result of applying \ttt{push} to a \emph{live graph} which had exactly one root by Lemma \ref{lem:live-graphs}. We can now check the root is blue and has incoming degree at most \(1\) in constant time. If this fails, then we look for another root, and once again, don't find one, in constant time.
\end{proof}

\begin{lemma}[Complexity of \ttt{Check}] \label{lem:is-tree-check-complexity}
    \ttt{Check} terminates in constant time on a \emph{live graph} \(G\).
\end{lemma}

\begin{proof}
	The matching time of \ttt{\{match, loop\}} is the sum of the matching time of \ttt{match} and \ttt{match} given that \ttt{match} was not applicable, which tells us there are no proper edges.
	
	For \ttt{match}, we just match the first non-root of \(G\), in constant time. If there are fewer than two nodes, then we fail, in constant time.
	
	For \ttt{loop}, matching only ever happens if there is at most one node in \(G\), and so clearly, we can check for the presence of a looped edge in constant time.
	
	Thus, \ttt{Check} terminates in constant time, as required.
\end{proof}

\begin{theorem}[Complexity of \ttt{is-tree}] \label{thm:is-tree-complexity}
    The program \ttt{is-tree} (Figure \ref{fig:is-tree-program}) terminates in linear time with respect to the \emph{size} of its input.
\end{theorem}

\begin{proof}
	Let \(G\) be the input graph. Due to Lemma \ref{lem:is-tree-init-complexity} we have that \ttt{init} takes only linear time with respect to the number of nodes in \(G\), and it is only applied once, from the definition of the program. By Lemma \ref{lem:is-tree-reduce-complexity} we have that \ttt{Reduce} takes only constant time and by Lemma \ref{lem:is-tree-intermediate-complexity} is applied only a linear number of times with respect to the number of nodes in \(G\). Finally, by Lemma \ref{lem:is-tree-check-complexity} we have that \ttt{Check} takes only constant time.
	
	So, the program's \ttt{Main} procedure actually has worst-case complexity, a constant function of the number of nodes in the \emph{input graph} \(G\). However, in order to start executing \ttt{Main}, the \emph{input graph} must first be loaded into memory, which cannot be done any faster than a constant function of the \emph{size} of \(G\).
\end{proof}

Just as with the \ttt{is-cycle} program, we have collected empirical timing results for \ttt{is-tree}, supporting our claim that the program runs in linear time, even on graph classes that do not have bounded degree (Figure \ref{fig:is-tree-timing}).

\begin{figure}[!ht]
\centering
\begin{tikzpicture}[scale=0.9]
    \begin{axis}[
    xlabel={Size of input graph},
    ylabel={Execution time (ms)},
    xmin=0,
    ymin=0,
    width=9.2cm,height=7.2cm,
    legend style={at={(1.5,0.82)}},
    ymajorgrids=true,
    grid style=dashed,
    ]
        \addplot[color=plot1, mark=square*] 
        coordinates {
        	(20000,54.66)
        	(60000,90.55)
        	(100000,130.64)
        	(140000,169.61)
        	(180000,204.42)
        	(220000,242.34)
        	(260000,282.28)
        	(300000,322.39)
        	(340000,361.71)
        	(380000,399.80)
        	(420000,440.40)
        	(460000,481.56)
        	(500000,520.44)
        };
        \addplot[color=plot2, mark=square*] 
        coordinates {
        	(19040,45.04)
        	(32865,56.47)
        	(50440,70.94)
        	(71765,85.52)
        	(96840,101.15)
        	(125665,123.57)
        	(158240,145.25)
        	(194565,169.02)
        	(234640,202.29)
        	(278465,232.76)
        	(326040,264.43)
        	(377365,301.70)
        	(432440,341.77)
        	(491265,382.85)
        };
        \addplot[color=plot3, mark=square*] 
        coordinates {
        	(19837,50.71)
        	(30955,54.29)
        	(45601,67.09)
        	(64261,80.60)
        	(87421,95.50)
        	(115567,115.60)
        	(149185,137.39)
        	(188761,165.41)
        	(234781,201.63)
        	(287731,236.28)
        	(348097,277.70)
        	(416365,326.71)
        	(493021,381.38)
        };
        \addplot[color=plot4, mark=square*] 
        coordinates {
        	(16381,44.14)
        	(32765,60.35)
        	(65533,87.16)
        	(131069,146.40)
        	(262141,259.65)
        	(524285,488.53)
        };
        \addplot[color=plot5, mark=square*] 
        coordinates {
        	(20000,54.59)
        	(60000,88.62)
        	(100000,120.62)
        	(140000,157.92)
        	(180000,191.40)
        	(220000,222.07)
        	(260000,256.26)
        	(300000,291.34)
        	(340000,324.65)
        	(380000,357.05)
        	(420000,391.58)
        	(460000,425.75)
        	(500000,459.71)
        };
        \addplot[color=plot6, mark=square*] 
        coordinates {
        	(20000,55.34)
        	(60000,83.83)
        	(100000,118.14)
        	(140000,149.45)
        	(180000,179.58)
        	(220000,208.92)
        	(260000,239.12)
        	(300000,269.75)
        	(340000,300.52)
        	(380000,331.22)
        	(420000,363.94)
        	(460000,392.72)
        	(500000,424.46)
        };
        \addplot[color=plot7, mark=square*] 
        coordinates {
        	(19999,56.26)
        	(59999,98.84)
        	(99999,139.95)
        	(139999,182.87)
        	(179999,222.63)
        	(219999,265.59)
        	(259999,307.55)
        	(299999,350.42)
        	(339999,392.01)
        	(379999,434.58)
        	(419999,479.10)
        	(459999,522.38)
        	(499999,566.42)
        };
        \addplot[color=plot8, mark=square*] 
        coordinates {
        	(19999,53.18)
        	(59999,89.77)
        	(99999,124.67)
        	(139999,164.01)
        	(179999,199.78)
        	(219999,231.45)
        	(259999,268.80)
        	(299999,302.91)
        	(339999,340.55)
        	(379999,375.88)
        	(419999,413.26)
        	(459999,451.23)
        	(499999,489.12)
        };
        \addlegendentry{Discrete Graph}
        \addlegendentry{Grid Graph}
        \addlegendentry{Grid Chain}
        \addlegendentry{Binary Tree}
        \addlegendentry{Cycle Graph}
        \addlegendentry{Sun Graph}
        \addlegendentry{Linked List}
        \addlegendentry{Star Graph}
    \end{axis}
\end{tikzpicture}
\caption{Measured performance of \ttt{is-tree}}
\label{fig:is-tree-timing}
\end{figure}
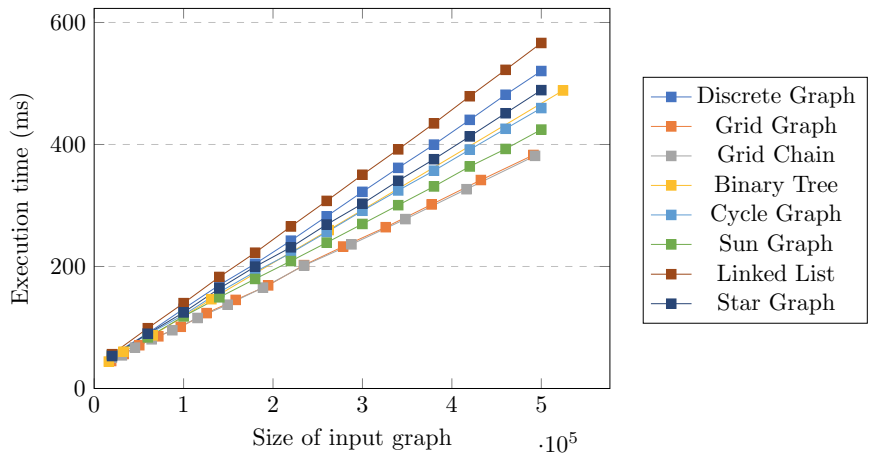

Finally, we conjecture it is possible to implement this program without needing to use blue nodes, since termination follows from the application conditions preventing the root node from entering a cycle in a graph, which is the only way to instantiate non-termination. We have omitted this program due to the correctness argument being much more fiddly, and the time complexity (and runtime performance) no better.

\subsection{Recognising Binary DAGs}
\label{subsec:binary_dag_rec}

Recall that a \emph{directed acyclic graph} (DAG) is a graph containing no directed cycles. A DAG is \emph{binary} if each of its nodes has an outgoing degree of at most two. In this subsection, we present a GP\,2 program (Figure \ref{fig:is-bin-dag-program}) that can recognise binary DAGs in linear time.

The program works by finding an incoming degree zero node, then removing it and its edges, if it has the outgoing edges one would expect. This process can be repeated until the entire graph has been deleted, if we have a binary DAG as input (Figure \ref{fig:is-bin-dag-example-1}). Otherwise, the program necessarily encounters a situation in which reduction cannot continue, and evaluates to fail (Figure \ref{fig:is-bin-dag-example-2}). Termination is ensured by dashing edges we have visited when searching for an incoming degree zero node.

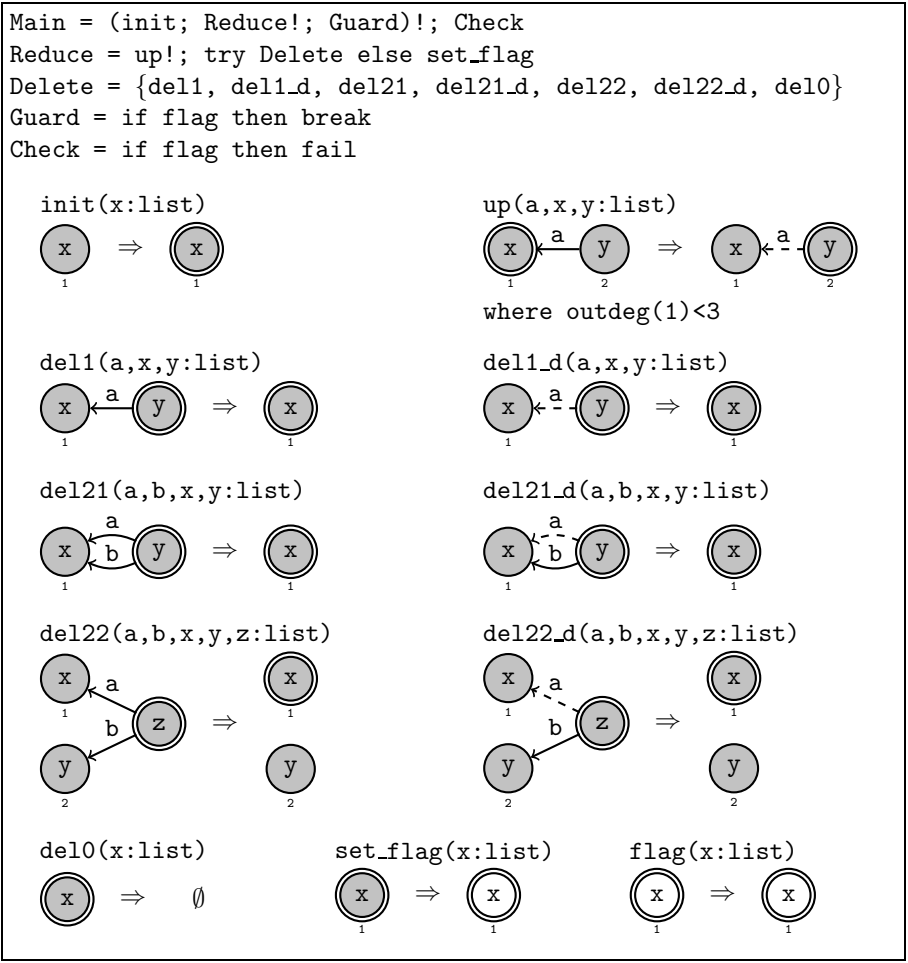
\begin{figure}[!ht]
\fbox{\begin{minipage}{11.78cm}
\begin{allintypewriter}
Main = (init; Reduce!; Guard)!; Check

Reduce = up!; try Delete else set\_flag

Delete = \{del1, del1\_d, del21, del21\_d, del22, del22\_d, del0\}

Guard = if flag then break

Check = if flag then fail

\setlength{\tabcolsep}{0.4cm}

\medskip
\smallskip

\begin{tabular}{ p{5.09cm}  p{5.09cm} }
    
    init(x:list) & up(a,x,y:list) \\

    \begin{tikzpicture}
        \node (a) at (0.0,0)     [gp2 node, fill=gp2grey] {x};

        \node (b) at (.875,0)    {$\Rightarrow$};
        
        \node (c) at (1.75,0)    [root node, fill=gp2grey] {x};
        
        \node (A) at (0.0,-.45)  {\tiny{1}};
        \node (C) at (1.75,-.45) {\tiny{1}};
    \end{tikzpicture}
    &

    \begin{tikzpicture}
        \node (a) at (0.0,0)     [root node, fill=gp2grey] {x};
        \node (b) at (1.25,0)    [gp2 node, fill=gp2grey] {y};
        
        \node (c) at (2.125,0)   {$\Rightarrow$};
        
        \node (d) at (3,0)       [gp2 node, fill=gp2grey] {x};
        \node (e) at (4.25,0)    [root node, fill=gp2grey] {y};
        
        \node (A) at (0.0,-.45)  {\tiny{1}};
        \node (B) at (1.25,-.45) {\tiny{2}};
        \node (D) at (3,-.45)    {\tiny{1}};
        \node (E) at (4.25,-.45) {\tiny{2}};
        
        \draw (b) edge[->,thick] node[above, yshift=-1pt] {a} (a)
              (e) edge[->,thick,dashed] node[above, yshift=-1pt] {a} (d);
    \end{tikzpicture}
	\\
	
	\vspace{-1.15em}\, & \vspace{-1.15em}where outdeg(1)<3 \\
\end{tabular}

\smallskip

\begin{tabular}{ p{5.09cm} p{5.09cm} }

    del1(a,x,y:list) & del1\_d(a,x,y:list) \\

    \begin{tikzpicture}
        \node (a) at (0.0,0)     [gp2 node, fill=gp2grey] {x};
        \node (b) at (1.25,0)    [root node, fill=gp2grey] {y};

        \node (c) at (2.125,0)   {$\Rightarrow$};

        \node (d) at (3,0)       [root node, fill=gp2grey] {x};

        \node (A) at (0.0,-.45)  {\tiny{1}};
        \node (D) at (3,-.45)    {\tiny{1}};

        \draw (b) edge[->,thick] node[above, yshift=-1pt] {a} (a);
    \end{tikzpicture}
    &

    \begin{tikzpicture}
        \node (a) at (0.0,0)     [gp2 node, fill=gp2grey] {x};
        \node (b) at (1.25,0)    [root node, fill=gp2grey] {y};

        \node (c) at (2.125,0)   {$\Rightarrow$};

        \node (d) at (3,0)       [root node, fill=gp2grey] {x};

        \node (A) at (0.0,-.45)  {\tiny{1}};
        \node (D) at (3,-.45)    {\tiny{1}};

        \draw (b) edge[->,thick,dashed] node[above, yshift=-1pt] {a} (a);
    \end{tikzpicture}
    \\
\end{tabular}

\smallskip

\begin{tabular}{ p{5.09cm} p{5.09cm} }

    del21(a,b,x,y:list) & del21\_d(a,b,x,y:list) \\

    \begin{tikzpicture}
        \node (a) at (0.0,0)     [gp2 node, fill=gp2grey] {x};
        \node (b) at (1.25,0)    [root node, fill=gp2grey] {y};

        \node (c) at (2.125,0)   {$\Rightarrow$};

        \node (d) at (3,0)       [root node, fill=gp2grey] {x};

        \node (A) at (0.0,-.45)  {\tiny{1}};
        \node (C) at (3,-.45)    {\tiny{1}};

        \draw (b) edge[->,thick,bend left=-25] node[above, yshift=-1pt] {a} (a)
              (b) edge[->,thick,bend left=25] node[above, yshift=-1pt] {b} (a);
    \end{tikzpicture}
    &

    \begin{tikzpicture}
        \node (a) at (0.0,0)     [gp2 node, fill=gp2grey] {x};
        \node (b) at (1.25,0)    [root node, fill=gp2grey] {y};

        \node (c) at (2.125,0)   {$\Rightarrow$};

        \node (d) at (3,0)       [root node, fill=gp2grey] {x};

        \node (A) at (0.0,-.45)  {\tiny{1}};
        \node (C) at (3,-.45)    {\tiny{1}};

        \draw (b) edge[->,thick,dashed,bend left=-25] node[above, yshift=-1pt] {a} (a)
              (b) edge[->,thick,bend left=25] node[above, yshift=-1pt] {b} (a);
    \end{tikzpicture}
    \\
\end{tabular}

\smallskip

\begin{tabular}{ p{5.09cm} p{5.09cm} }

    del22(a,b,x,y,z:list) & del22\_d(a,b,x,y,z:list) \\

    \begin{tikzpicture}
        \node (a) at (0.0,-0.0)  [gp2 node, fill=gp2grey] {x};
        \node (b) at (0.0,-1.2)  [gp2 node, fill=gp2grey] {y};
        \node (c) at (1.25,-0.6) [root node, fill=gp2grey] {z};

        \node (d) at (2.125,-.6) {$\Rightarrow$};

        \node (e) at (3,-0.0)    [root node, fill=gp2grey] {x};
        \node (f) at (3,-1.2)    [gp2 node, fill=gp2grey] {y};

        \node (A) at (0.0,-0.45) {\tiny{1}};
        \node (B) at (0.0,-1.65) {\tiny{2}};
        \node (E) at (3,-0.45)   {\tiny{1}};
        \node (F) at (3,-1.65)   {\tiny{2}};

        \draw (c) edge[->,thick] node[above] {a} (a)
              (c) edge[->,thick] node[above] {b} (b);
    \end{tikzpicture}
    &

    \begin{tikzpicture}
        \node (a) at (0.0,-0.00) [gp2 node, fill=gp2grey] {x};
        \node (b) at (0.0,-1.2)  [gp2 node, fill=gp2grey] {y};
        \node (c) at (1.25,-0.6) [root node, fill=gp2grey] {z};

        \node (d) at (2.125,-.6) {$\Rightarrow$};

        \node (e) at (3,-0.0)    [root node, fill=gp2grey] {x};
        \node (f) at (3,-1.2)    [gp2 node, fill=gp2grey] {y};

        \node (A) at (0.0,-0.45) {\tiny{1}};
        \node (B) at (0.0,-1.66) {\tiny{2}};
        \node (E) at (3,-0.45)   {\tiny{1}};
        \node (F) at (3,-1.65)   {\tiny{2}};

        \draw (c) edge[->,thick,dashed] node[above] {a} (a)
              (c) edge[->,thick] node[above] {b} (b);
    \end{tikzpicture}
    \\
\end{tabular}

\smallskip

\begin{tabular}{ p{3.12cm} p{3.12cm} p{3.12cm} }

    del0(x:list) & set\_flag(x:list) & flag(x:list) \\

    \begin{tikzpicture}
        \node (a) at (0.0,0)     [root node, fill=gp2grey] {x};

        \node (b) at (.875,0)    {$\Rightarrow$};
        
        \node (c) at (1.75,0.02) {$\emptyset$};
        
        \node (A) at (0.0,-.45)  {\tiny{\,}};
    \end{tikzpicture}
    &

    \begin{tikzpicture}
        \node (a) at (0.0,0)     [root node, fill=gp2grey] {x};

        \node (b) at (.875,0)    {$\Rightarrow$};

        \node (c) at (1.75,0)    [root node] {x};

        \node (A) at (0.0,-.45)  {\tiny{1}};
        \node (C) at (1.75,-.45) {\tiny{1}};
    \end{tikzpicture}
    &

    \begin{tikzpicture}
        \node (a) at (0.0,0)     [root node] {x};

        \node (b) at (.875,0)    {$\Rightarrow$};

        \node (c) at (1.75,0)    [root node] {x};

        \node (A) at (0.0,-.45)  {\tiny{1}};
        \node (C) at (1.75,-.45) {\tiny{1}};
    \end{tikzpicture}
    \\
\end{tabular}
\end{allintypewriter}
\end{minipage}}
\caption{GP\,2 program \ttt{is-bin-dag}}
\label{fig:is-bin-dag-program}
\end{figure}

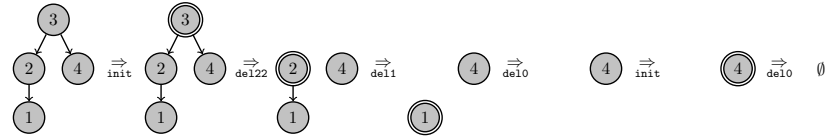
\begin{figure}[!ht]
\centering
\scalebox{.65}{\begin{tikzpicture}
\node (a) at (-5.4,0)  [gp2 node, fill=gp2grey] {3};
\node (b) at (-5.9,-1) [gp2 node, fill=gp2grey] {2};
\node (c) at (-4.9,-1) [gp2 node, fill=gp2grey] {4};
\node (d) at (-5.9,-2) [gp2 node, fill=gp2grey] {1};

\draw (a) edge[->,thick] (b)
      (a) edge[->,thick] (c)
      (b) edge[->,thick] (d);

\node (t) at (-4.05,-0.9) {$\Rightarrow$};
\node (t) at (-4.05,-1.1) {\ttt{\scriptsize{init}}};

\node (a) at (-2.7,0)  [root node, fill=gp2grey] {3};
\node (b) at (-3.2,-1) [gp2 node, fill=gp2grey] {2};
\node (c) at (-2.2,-1) [gp2 node, fill=gp2grey] {4};
\node (d) at (-3.2,-2) [gp2 node, fill=gp2grey] {1};

\draw (a) edge[->,thick] (b)
      (a) edge[->,thick] (c)
      (b) edge[->,thick] (d);

\node (t) at (-1.35,-0.9) {$\Rightarrow$};
\node (t) at (-1.35,-1.1) {\ttt{\scriptsize{del22}}};

\node (b) at (-0.5,-1) [root node, fill=gp2grey] {2};
\node (c) at (0.5,-1)  [gp2 node, fill=gp2grey] {4};
\node (d) at (-0.5,-2) [gp2 node, fill=gp2grey] {1};

\draw (b) edge[->,thick] (d);

\node (t) at (1.35,-0.9) {$\Rightarrow$};
\node (t) at (1.35,-1.1) {\ttt{\scriptsize{del1}}};

\node (c) at (3.2,-1) [gp2 node, fill=gp2grey] {4};
\node (d) at (2.2,-2) [root node, fill=gp2grey] {1};

\node (t) at (4.05,-0.9) {$\Rightarrow$};
\node (t) at (4.05,-1.1) {\ttt{\scriptsize{del0}}};

\node (c) at (5.9,-1) [gp2 node, fill=gp2grey] {4};

\node (t) at (6.75,-0.9) {$\Rightarrow$};
\node (t) at (6.75,-1.1) {\ttt{\scriptsize{init}}};

\node (c) at (8.6,-1) [root node, fill=gp2grey] {4};

\node (t) at (9.45,-0.9) {$\Rightarrow$};
\node (t) at (9.45,-1.1) {\ttt{\scriptsize{del0}}};

\node (c) at (10.3,-1) {$\emptyset$};

\end{tikzpicture}}
\caption{Example tree reduction}
\label{fig:is-bin-dag-example-1}
\end{figure}

\begin{figure}[!ht]
\centering
\scalebox{.65}{\begin{tikzpicture}

\node (a) at (-5.9,0)  [gp2 node, fill=gp2grey] {1};
\node (b) at (-5.9,-1) [gp2 node, fill=gp2grey] {2};
\node (c) at (-4.9,-1) [gp2 node, fill=gp2grey] {3};
\node (d) at (-4.9,0)  [gp2 node, fill=gp2grey] {4};

\draw (a) edge[->,thick] (b)
      (b) edge[->,thick] (c)
      (c) edge[->,thick] (d)
      (d) edge[->,thick] (a);

\node (t) at (-4.05,-0.4) {$\Rightarrow$};
\node (t) at (-4.05,-0.6) {\ttt{\scriptsize{init}}};

\node (a) at (-3.2,0)  [root node, fill=gp2grey] {1};
\node (b) at (-3.2,-1) [gp2 node, fill=gp2grey] {2};
\node (c) at (-2.2,-1) [gp2 node, fill=gp2grey] {3};
\node (d) at (-2.2,0)  [gp2 node, fill=gp2grey] {4};

\draw (a) edge[->,thick] (b)
      (b) edge[->,thick] (c)
      (c) edge[->,thick] (d)
      (d) edge[->,thick] (a);

\node (t) at (-1.35,-0.4) {$\Rightarrow$};
\node (t) at (-1.35,-0.6) {\ttt{\scriptsize{up}}};

\node (a) at (-0.5,0)  [gp2 node, fill=gp2grey] {1};
\node (b) at (-0.5,-1) [root node, fill=gp2grey] {2};
\node (c) at (0.5,-1)  [gp2 node, fill=gp2grey] {3};
\node (d) at (0.5,0)   [gp2 node, fill=gp2grey] {4};

\draw (a) edge[->,thick,dashed] (b)
      (b) edge[->,thick] (c)
      (c) edge[->,thick] (d)
      (d) edge[->,thick] (a);

\node (t) at (1.35,-0.4) {$\Rightarrow$};
\node (t) at (1.35,-0.6) {\ttt{\scriptsize{up}}};

\node (a) at (2.2,0)  [gp2 node, fill=gp2grey] {1};
\node (b) at (2.2,-1) [gp2 node, fill=gp2grey] {2};
\node (c) at (3.2,-1) [root node, fill=gp2grey] {3};
\node (d) at (3.2,0)  [gp2 node, fill=gp2grey] {4};

\draw (a) edge[->,thick,dashed] (b)
      (b) edge[->,thick,dashed] (c)
      (c) edge[->,thick] (d)
      (d) edge[->,thick] (a);

\node (t) at (4.05,-0.4) {$\Rightarrow$};
\node (t) at (4.05,-0.6) {\ttt{\scriptsize{up}}};

\node (a) at (4.9,0)  [gp2 node, fill=gp2grey] {1};
\node (b) at (4.9,-1) [gp2 node, fill=gp2grey] {2};
\node (c) at (5.9,-1) [gp2 node, fill=gp2grey] {3};
\node (d) at (5.9,0)  [root node, fill=gp2grey] {4};

\draw (a) edge[->,thick,dashed] (b)
      (b) edge[->,thick,dashed] (c)
      (c) edge[->,thick,dashed] (d)
      (d) edge[->,thick] (a);

\node (t) at (6.75,-0.4) {$\Rightarrow$};
\node (t) at (6.75,-0.6) {\ttt{\scriptsize{up}}};

\node (a) at (7.6,0)  [root node, fill=gp2grey] {1};
\node (b) at (7.6,-1) [gp2 node, fill=gp2grey] {2};
\node (c) at (8.6,-1) [gp2 node, fill=gp2grey] {3};
\node (d) at (8.6,0)  [gp2 node, fill=gp2grey] {4};

\draw (a) edge[->,thick,dashed] (b)
      (b) edge[->,thick,dashed] (c)
      (c) edge[->,thick,dashed] (d)
      (d) edge[->,thick,dashed] (a);

\node (t) at (9.45,-0.4) {$\Rightarrow$};
\node (t) at (9.45,-0.6) {\ttt{\scriptsize{set\_}}};
\node (t) at (9.45,-0.8) {\ttt{\scriptsize{flag}}};

\node (a) at (10.3,0)  [root node] {1};
\node (b) at (10.3,-1) [gp2 node, fill=gp2grey] {2};
\node (c) at (11.3,-1) [gp2 node, fill=gp2grey] {3};
\node (d) at (11.3,0)  [gp2 node, fill=gp2grey] {4};

\draw (a) edge[->,thick,dashed] (b)
      (b) edge[->,thick,dashed] (c)
      (c) edge[->,thick,dashed] (d)
      (d) edge[->,thick,dashed] (a);
\end{tikzpicture}}
\caption{Example cycle reduction}
\label{fig:is-bin-dag-example-2}
\end{figure}
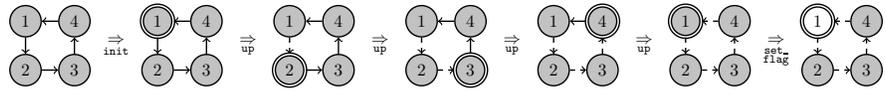

As with the tree recognition program, we show correctness and complexity, including empirical timing results (Figure \ref{fig:is-bin-dag-timing}).

\begin{definition}[Live Graph]
    A \emph{live graph} (for the purposes of this subsection) is any graph that appears as an intermediate \emph{host graph} in the process of executing \ttt{is-bin-dag} on an \emph{input graph}. Moreover, call a \emph{live graph}:
    \begin{enumerate}
        \item \emph{unrooted} if it has no root nodes;
        \item \emph{grey-rooted} if it has exactly one grey root node;
        \item \emph{white-rooted} if it has exactly one unmarked root node.
    \end{enumerate}
\end{definition}

\begin{lemma}[Live Graph Properties] \label{lem:is-bin-dag-live-graphs}
    Let \(G\) be a \emph{live graph}. Then:
    \begin{enumerate}
        \item \(G\) is either \emph{unrooted}, \emph{grey-rooted}, or \emph{white-rooted}.
        \item \(G\) has all nodes apart from possibly the root node are grey, and all edges are either unmarked or dashed.
        \item \(G\) has each node has at most two incoming dashed edges and at most one outgoing dashed edge.
    \end{enumerate}
\end{lemma}

\begin{proof}
    First, we check that any input graph \(G\) satisfies the properties. This is easy to see since \(G\) must be an unrooted graph with all nodes grey and all edges unmarked, so all of the three conditions are satisfied. After one step. Next, we analyse what happens to the procedure \ttt{init; Reduce!; Guard}. Showing that \(\ttt{init; Reduce!; Guard}(G)\) is either \ttt{fail} or an unrooted live graph if \(G\) is an unrooted live graph is sufficient to show the result by induction on the number of iterations of \ttt{(init; Reduce!; Guard)!}, since \ttt{Check} does not modify the host graph and the input graph is unrooted.
    
    It is clear that \(G_0 = \ttt{init}(G)\) is either \ttt{fail} or \(G\) but with one node turned into a grey root node. We claim that \((\ttt{Reduce!})(G_0)\) is either an unrooted live graph or a white-rooted live graph. To see this, we argue by induction on the number of iterations, showing that \(G_{i+1} = \ttt{(Reduce!})(G_i)\) is either \ttt{fail} or a graph satisfying the 3 properties of the lemma, and that if it is \ttt{fail}, then \(G_{i+1}\) must additionally be either unrooted or white-rooted. \(G'_i = (\ttt{up!})(G_i)\) must satisfy the 3 properties since \ttt{up}'s application condition prevents the 3rd property from being violated, leaving behind a grey-rooted graph. Next, \(\ttt{try Delete else set\_flag}(G'_i)\) clearly either must be the same graph only with the grey root either deleted or unmarked so the conditions are satisfied, as required. Finally, let \(H = (\ttt{init; Reduce!})(G)\). Then \(\ttt{Guard}(H)\) is either \(H\) or \ttt{fail}.
\end{proof}

\begin{lemma}[\ttt{Reduce} is Structure Preserving] \label{lem:is-bin-dag-reduce-structure}
    Given a \emph{live graph} \(G\) and \(G \Rightarrow_{\ttt{Reduce}} H\), then \(G\) is a binary DAG if and only if \(H\) is.
\end{lemma}

\begin{proof}
    We can totally ignore the rootedness and colour of nodes, and the marks of edges for the purposes of this proof, since the same program that operates ignoring everything in this way is certainly compatible, just need not terminate. Since \ttt{init}, \ttt{up}, \ttt{set\_flag} and \ttt{flag} do not modify the graph structure, then all we need to check is that the \(7\) remaining rules (ignoring the rootedness, colours, and labels) each individually preserve being a binary DAG and not being a binary DAG, and this is enough to show the result. It is clear that they preserve being a binary DAG, since all of the rules only delete nodes and edges.
    
    To see that \ttt{del0} preserves not being a binary DAG is easy, since deleting an isolated node is safe (the remaining connected component must not be a binary DAG). Looking at \ttt{del1}, if the input \(G\) is not a binary DAG and the rule is applicable, then the dangling condition ensures that the root node has no incoming edges and only one outgoing edge. Now, if \(G\) is not a binary DAG then either it has a directed cycle, or a node with outdegree at least \(3\). The matched root node must not lie on said cycle, nor does it have such outdegree, so removal of the node with its single outgoing edge must leave the directed cycle in place and also the node with outdegree at least \(3\). The argument for the remaining rules is almost identical.
\end{proof}

\begin{lemma}[Termination of \ttt{is-bin-dag}] \label{lem:is-bin-dag-termination}
    On every \emph{input graph} $G$, the program \ttt{is-bin-dag} terminates in $\mrm{O}(\mrm{size}(G))$ steps.
\end{lemma}

\begin{proof}
    Given any host graph $G$, let $\mrm{ug}(G)$ be the number of unrooted grey nodes in $G$, $\mrm{g}(G)$ be the number of grey nodes in $G$, and $\mrm{ue}(G)$ be the number of unmarked edges in $G$. Define $\#G = \mrm{ug}(G) + \mrm{g}(G) + \mrm{ue}(G)$. Then
    \begin{equation} 
     0 \leq \#G \leq 2\times|V_G| + |E_G| \leq 2\times\mrm{size}(G). 
    \end{equation}
    By inspection of the rules in \ttt{is-bin-dag}, we have 
     \begin{equation} 
     \text{$G \DSdder_r H$ implies $\#H = \#G - 1$ for $r \in \{\ttt{init},\,\ttt{up},\,\ttt{set\_flag}\}$}
    \end{equation}  
    and
    \begin{equation} 
     \text{$G \DSdder_{\mrm{Delete}} H$ implies $\#H \leq \#G - 1$}.
    \end{equation}\\
    \emph{Claim:} On every host graph $G$, the loop \ttt{Reduce!} terminates after at most $2\times\mrm{size}(G)$ successful executions of \ttt{Reduce}.\\[1ex]
    \emph{Proof:} By (2), the inner loop \ttt{up!} terminates as each application of \ttt{up} decreases the \#-value which cannot get negative. Moreover, by (2) and (3), the command \ttt{try Delete else set\_flag} either decreases the \#-value (if \ttt{Delete} or \ttt{set\_flag} is applied) or terminates the loop (if both \ttt{Delete} and \ttt{set\_flag} fail). As the \#-value cannot get negative, it follows that \ttt{Reduce!} must terminate. The number of successful executions of \ttt{Reduce} is at most $2\times\mrm{size}(G)$ because by (3) and (2), both \ttt{Delete} and \ttt{set\_flag} decrease the \#-value, and $0 \leq \#G \leq 2\times\mrm{size}(G)$ by (1). $\Box$   
    
    \vspace*{\baselineskip}                                                             
    We now show that on every live graph $G$, \ttt{(init; Reduce!; Guard)!} terminates in $\mrm{O}(\mrm{size}(G))$ steps. This finishes the proof as the execution of \ttt{Check} requires at most two steps. 
    
    By (2) and (3), each application of \ttt{init} decreases the \#-value and none of the rules in \ttt{Reduce} or \ttt{Guard} increases the \#-value. Hence each successful execution of the loop body decreases the \#-value. Together with the Claim and (1), this implies that the loop terminates after at most $2\times\mrm{size}(G)$ successful executions of its body.
    
    Hence, in total, both \ttt{init} and \ttt{flag} are applied at most $2\times\mrm{size}(G)$ times. In addition, there may be one failed application of \ttt{init} or one execution of \ttt{break}.
    
    By (2), (3) and (1), the rules \ttt{up} and \ttt{set\_flag} and the rule set \ttt{Delete} are each applied at most $2\times\mrm{size}(G)$ times in total. Each application of \ttt{Delete} involves up to six failed applications of rules from the set. Also, there are at most $2\times\mrm{size}(G)+1$ failed applications of \ttt{up} because, by the Claim, there are at most $2\times\mrm{size}(G)$ successful executions of \ttt{Reduce}. By the same argument, there are at most $2\times\mrm{size}(G)+1$ failed applications of both \ttt{set\_flag} and \ttt{Delete}. Each failed application \ttt{Delete} amounts to seven failed rule applications.
    
    Thus, altogether, the loop \ttt{(init; Reduce!; Guard)!} terminates in a number of steps that is linear in  $\mrm{size}(G))$.
\end{proof}

\begin{theorem}[Correctness of \ttt{is-bin-dag}] \label{thm:is-bin-dag-correctness}
	The program \ttt{is-bin-dag} (Figure \ref{fig:is-bin-dag-program}) is totally correct with respect to the specification:
	\begin{spec}
		\specinput{An \emph{input graph}.}
		\specoutput{Fail if and only if the input is not a binary DAG.}
	\end{spec}
\end{theorem}

\begin{proof}
    Termination follows from Lemma \ref{lem:is-bin-dag-termination}. We claim that if the input graph is a binary DAG, then the program evaluates to the empty graph, and otherwise, evaluates to fail. In particular, this means the subprogram \ttt{(init; Reduce!; Guard)!} must evaluate to the empty graph, or a \emph{white-rooted live graph}, respectively. We argue by induction of the number of iterations of the outer loop. Suppose the loop runs no times. Then \ttt{init} failed, so the input graph was empty, so the subprogram evaluates to the empty graph. Suppose we can make progress, but \ttt{Guard} causes a break. Then then \ttt{init} must be executed, followed by \ttt{Reduce} zero or more times, then \ttt{flag} must be applicable, meaning the result must be a \emph{white-rooted live graph}. By Lemma \ref{lem:is-bin-dag-reduce-structure}, this means the input graph must not have been a DAG. Finally, if we can make progress, and no break is executed, then we are ready to try to run another iteration, and again by \ref{lem:is-bin-dag-reduce-structure}, this particular iteration has preserved if the host graph was a binary DAG or not. Suppose now that the loop has run \(n\) times. If the loop cannot run again, then we repeat the argument above. Similarly, if we can make progress, then the argument above follows too.
\end{proof}

\begin{lemma}[Complexity of \ttt{init}] \label{lem:is-bin-dag-init-complexity}
    \ttt{init} terminates in constant time on an \emph{unrooted live graph} \(G\).
\end{lemma}

\begin{proof}
    Exactly the same as the proof of Lemma \ref{lem:is-cycle-init-complexity}. The search plan will iterate all nodes in \(G\) looking for the first grey unrooted node. Testing if a node is unrooted and grey takes only constant time. Since every node in an \emph{input graph} is unrooted and grey, the search will stop at the first node, or fail in constant time if \(G\) is empty.
\end{proof}

\begin{lemma}[Complexity of \ttt{up}] \label{lem:is-bin-dag-up-complexity}
    \ttt{up} terminates in constant time on a \emph{live graph} \(G\).
\end{lemma}

\begin{proof}
    The search plan will look for a root node. If there are no root nodes, matching immediately fails. Otherwise, the first root node is located in constant time, and the check that it is grey takes constant time. If it is not grey, then the matching algorithm will look for the next root node, and determines there are no more (Lemma \ref{lem:is-bin-dag-live-graphs}) in constant time, and matching fails. In the case that the matching algorithm has determined the root node is grey, it will then check that the incoming degree is at least one and the outgoing degree is at most two (our application condition), which takes only constant time. If the degree checks fail, then the search plan looks for another root node, as before.
    
    If the degrees are correct, next the search plan grabs the first incoming edge in constant time and checks if it is unmarked and proper in constant time. If it is not, then we grab the next incoming edge in constant time. Due to the fact that the outgoing degree is at most 2, there can only be at most two looped edges and due to Lemma \ref{lem:is-bin-dag-live-graphs}, there can only be at most two proper edges that are not unmarked. So, we know that the search plan will find a suitable edge after a bounded number of attempts. After finding a suitable edge, the search plan then checks the source node is grey and unrooted in constant time, which it will be.
\end{proof}

\begin{lemma}[Complexity of \ttt{Delete}] \label{lem:is-bin-dag-delete-complexity}
    \ttt{Delete} terminates in constant time on an \emph{live graph} \(G\).
\end{lemma}

\begin{proof}
    \ttt{Delete} takes at most as long as the sum of the time needed for \ttt{del1}, \ttt{del1d}, \ttt{del21}, \ttt{del21d}, \ttt{del22}, \ttt{del22d}, \ttt{del0}.
    
    We start with \ttt{del0}, the simplest. The search plan will look for a root node. If there are no root nodes, matching immediately fails. Otherwise, the first root node is located in constant time, and the check that it is grey takes constant time. If it is not grey, then the matching algorithm will look for the next root node, and determines there are no more (Lemma \ref{lem:is-bin-dag-live-graphs}) in constant time, and matching fails. In the case that the matching algorithm has determined the root node is grey, it will then check that the incoming and outgoing degrees are zero, in constant time. If they are, we're done, and if not we must look for another root node, as before.
    
    For \ttt{del1} and \ttt{del1\_d}, the search plan will look for a root node. If there are no root nodes, matching immediately fails. Otherwise, the first root node is located in constant time, and the check that it is grey takes constant time. If it is not grey, then the matching algorithm will look for the next root node, and determines there are no more (Lemma \ref{lem:is-bin-dag-live-graphs}) in constant time, and matching fails. In the case that the matching algorithm has determined the root node is grey, it will then check that incoming degree is zero and outgoing degree is one, in constant time. If they are not, we must look for another root node, which fails in constant time. Otherwise, we grab the first outgoing edge in constant time, and check that it has the appropriate mark and is proper. If it is not, we proceed to look for the next root node, as before. Otherwise, we then check the target node of the edge is grey and unrooted. If not, we return to looking for the next root, as before.
    
    The story for the remaining rules is similar. The first difference is that the initial outgoing degree check is for exactly two, and an additional edge must be matched. For \ttt{del21} and \ttt{del21\_d}, the search plan additionally checks of the second edge at the end, and for \ttt{del22} and \ttt{del22\_d} the process of checking for a second edge and target node is just an exact repeat of what happened for the first edge and target node.
\end{proof}

\begin{lemma}[Complexity of \ttt{set\_flag}] \label{lem:is-bin-dag-set-flag-complexity}
    \ttt{set\_flag} terminates in constant time on an \emph{live graph} \(G\).
\end{lemma}

\begin{proof}
    The search plan will look for a root node. If there are no root nodes, matching immediately fails. Otherwise, the first root node is located in constant time, and the check that it is grey takes constant time. If it is not grey, then the matching algorithm will look for the next root node, and determines there are no more (Lemma \ref{lem:is-bin-dag-live-graphs}) in constant time, and matching fails.
\end{proof}

\begin{lemma}[Complexity of \ttt{flag}] \label{lem:is-bin-dag-flag-complexity}
    \ttt{Guard} and \ttt{Check} both terminate in constant time on a \emph{live graph} \(G\).
\end{lemma}

\begin{proof}
    \ttt{Guard} and \ttt{Check} take only as long as \ttt{flag}. \ttt{flag}'s search plan will look for a root node, and find the first one in constant time, and check that it is unmarked in constant time. If it is unmarked, then matching succeeds. If it is not, then we look for the next root node, and determine there is no other root node in constant time (Lemma \ref{lem:is-bin-dag-live-graphs}), and fail.
\end{proof}

\begin{theorem}[Complexity of \ttt{is-bin-dag}] \label{thm:is-bin-dag-complexity}
   The program \ttt{is-bin-dag} (Figure \ref{fig:is-bin-dag-program}) terminates in linear time with respect to the \emph{size} of its input.
\end{theorem}

\begin{proof}
    Lemma \ref{lem:is-bin-dag-termination} tells us that there is only a linear number of steps and Lemma \ref{lem:is-bin-dag-live-graphs} and its proofs tells us that the conditions under which the rules can be matched in constant time (Lemmata \ref{lem:is-bin-dag-init-complexity}, \ref{lem:is-bin-dag-up-complexity}, \ref{lem:is-bin-dag-delete-complexity}, \ref{lem:is-bin-dag-set-flag-complexity}, and \ref{lem:is-bin-dag-flag-complexity}) is always preserved. Thus, we have the required result.
\end{proof}

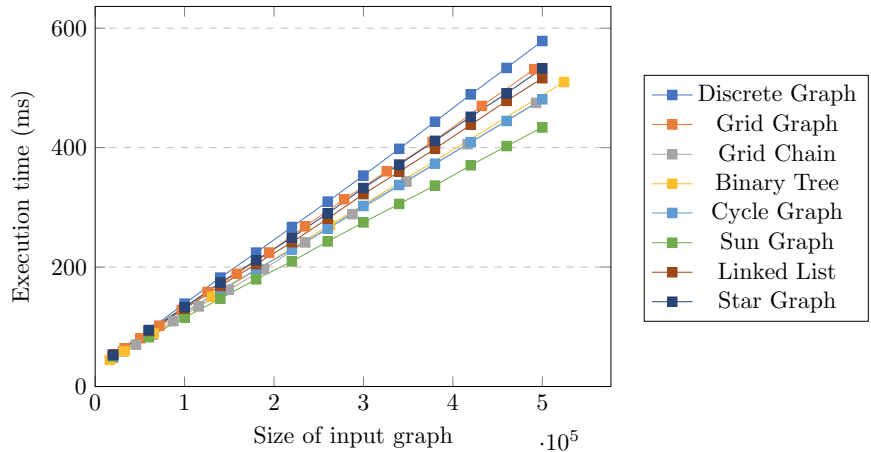
\begin{figure}[!ht]
\centering
\begin{tikzpicture}[scale=0.9]
    \begin{axis}[
    xlabel={Size of input graph},
    ylabel={Execution time (ms)},
    xmin=0,
    ymin=0,
    width=9.2cm,height=7.2cm,
    legend style={at={(1.5,0.82)}},
    ymajorgrids=true,
    grid style=dashed,
    ]
        \addplot[color=plot1, mark=square*] 
        coordinates {
        	(20000,52.26)
        	(60000,94.72)
        	(100000,138.81)
        	(140000,182.43)
        	(180000,224.27)
        	(220000,267.12)
        	(260000,309.66)
        	(300000,353.25)
        	(340000,398.06)
        	(380000,443.39)
        	(420000,489.08)
        	(460000,533.43)
        	(500000,578.53)
        };
        \addplot[color=plot2, mark=square*] 
        coordinates {
        	(19040,49.54)
        	(32865,64.01)
        	(50440,80.73)
        	(71765,101.87)
        	(96840,128.15)
        	(125665,158.18)
        	(158240,188.85)
        	(194565,224.39)
        	(234640,268.18)
        	(278465,313.53)
        	(326040,360.29)
        	(377365,409.51)
        	(432440,469.84)
        	(491265,531.44)
        };
        \addplot[color=plot3, mark=square*] 
        coordinates {
        	(19837,47.81)
        	(30955,58.68)
        	(45601,70.23)
        	(64261,86.99)
        	(87421,109.43)
        	(115567,134.11)
        	(149185,162.02)
        	(188761,196.58)
        	(234781,241.25)
        	(287731,288.55)
        	(348097,343.27)
        	(416365,405.69)
        	(493021,475.04)
        };
        \addplot[color=plot4, mark=square*] 
        coordinates {
        	(16381,44.60)
        	(32765,59.36)
        	(65533,89.31)
        	(131069,150.37)
        	(262141,269.94)
        	(524285,509.84)
        };
        \addplot[color=plot5, mark=square*] 
        coordinates {
        	(20000,52.22)
        	(60000,90.24)
        	(100000,124.39)
        	(140000,160.96)
        	(180000,195.80)
        	(220000,229.33)
        	(260000,264.55)
        	(300000,302.32)
        	(340000,337.39)
        	(380000,373.11)
        	(420000,409.10)
        	(460000,444.64)
        	(500000,480.94)
        };
        \addplot[color=plot6, mark=square*] 
        coordinates {
        	(20000,50.35)
        	(60000,82.55)
        	(100000,115.47)
        	(140000,147.23)
        	(180000,179.52)
        	(220000,209.46)
        	(260000,242.96)
        	(300000,274.80)
        	(340000,305.77)
        	(380000,336.50)
        	(420000,370.39)
        	(460000,402.62)
        	(500000,433.73)
        };
        \addplot[color=plot7, mark=square*] 
        coordinates {
        	(19999,53.87)
        	(59999,93.05)
        	(99999,130.31)
        	(139999,168.53)
        	(179999,204.82)
        	(219999,241.69)
        	(259999,281.15)
        	(299999,322.16)
        	(339999,359.62)
        	(379999,398.01)
        	(419999,438.42)
        	(459999,478.27)
        	(499999,516.23)
        };
        \addplot[color=plot8, mark=square*] 
        coordinates {
        	(19999,52.46)
        	(59999,94.14)
        	(99999,132.84)
        	(139999,174.08)
        	(179999,211.28)
        	(219999,249.47)
        	(259999,289.90)
        	(299999,332.39)
        	(339999,371.61)
        	(379999,411.40)
        	(419999,451.65)
        	(459999,490.94)
        	(499999,532.68)
        };
        \addlegendentry{Discrete Graph}
        \addlegendentry{Grid Graph}
        \addlegendentry{Grid Chain}
        \addlegendentry{Binary Tree}
        \addlegendentry{Cycle Graph}
        \addlegendentry{Sun Graph}
        \addlegendentry{Linked List}
        \addlegendentry{Star Graph}
    \end{axis}
\end{tikzpicture}
\caption{Measured performance of \ttt{is-bin-dag}}
\label{fig:is-bin-dag-timing}
\end{figure}

\section{Fast DFS-Based Programs}
\label{sec:dfs}
	
In this section we will review (undirected) DFS (depth-first search) in GP\,2, first implemented by Bak and Plump \cite{Bak-Plump12a,Bak15a,Bak-Plump16a}. We will observe, by means of an introductory example, how it can be used to recognise connected graphs. We then consider the 2-colouring problem, producing a topological sorting for a connected DAG, and how to recognise a connected DAG. The concrete syntax for all the programs in this section is available on GitHub\footnote{\url{https://gist.github.com/GrahamCampbell/79f0f62c50d7de5e7ba739ad5d4581e5}}.

\subsection{Recognising Connected Graphs using DFS}
\label{subsec:is_connected}

The program \ttt{is-connected} (Figure \ref{fig:is-connected-program}) can detect the connectedness of a graph. It fails if and only if its input graph is not connected. This can be achieved by conducting a DFS that turns grey nodes into non-grey ones. Since the DFS cannot propagate beyond the connected component it started in, the presence of a grey node indicates that the host graph is not connected.

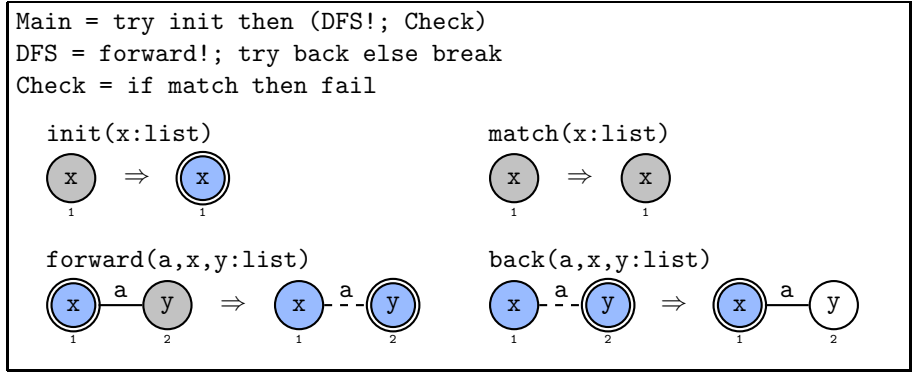
\begin{figure}[!ht]
\centering
\fbox{\begin{minipage}{11.78cm}
\begin{allintypewriter}
Main = try init then (DFS!; Check)

DFS = forward!; try back else break

Check = if match then fail

\setlength{\tabcolsep}{0.4cm}

\medskip
\smallskip

\begin{tabular}{ p{5.09cm} p{5.09cm} }
	
	init(x:list) & match(x:list) \\
	
	\begin{tikzpicture}
		\node (a) at (0,0)        [gp2 node, fill=gp2grey] {x};
		
		\node (b) at (.875,0)     {$\Rightarrow$};
		
		\node (c) at (1.75,0)     [root node, fill=gp2blue] {x};
		
		\node (A) at (0,-.45)     {\tiny{1}};
		\node (C) at (1.75,-.45)  {\tiny{1}};
	\end{tikzpicture}
	&
	
	\begin{tikzpicture}
		\node (a) at (0,0)        [gp2 node, fill=gp2grey] {x};
		
		\node (b) at (.875,0)     {$\Rightarrow$};
		
		\node (c) at (1.75,0)     [gp2 node, fill=gp2grey] {x};
		
		\node (A) at (0,-.45)     {\tiny{1}};
		\node (C) at (1.75,-.45)  {\tiny{1}};
	\end{tikzpicture}
	\\
	
\end{tabular}

\smallskip

\begin{tabular}{ p{5.09cm} p{5.09cm} }
	
	forward(a,x,y:list) & back(a,x,y:list) \\
	
	\begin{tikzpicture}
		\node (a) at (0,0)        [root node, fill=gp2blue] {x};
		\node (b) at (1.25,0)     [gp2 node, fill=gp2grey] {y};
		
		\node (c) at (2.125,0)    {$\Rightarrow$};
		
		\node (d) at (3,0)        [gp2 node, fill=gp2blue] {x};
		\node (e) at (4.25,0)     [root node, fill=gp2blue] {y};
		
		\node (A) at (0,-.45)     {\tiny{1}};
		\node (B) at (1.25,-.45)  {\tiny{2}};
		\node (D) at (3,-.45)     {\tiny{1}};
		\node (E) at (4.25,-.45)  {\tiny{2}};
		
		\draw (a) edge[thick] node[above, yshift=-1pt] {a} (b)
		      (d) edge[dashed, thick] node[above, yshift=-1pt] {a} (e);
	\end{tikzpicture}
	&
	
	\begin{tikzpicture}
		\node (a) at (0,0)        [gp2 node, fill=gp2blue] {x};
		\node (b) at (1.25,0)     [root node, fill=gp2blue] {y};
		
		\node (c) at (2.125,0)    {$\Rightarrow$};
		
		\node (d) at (3,0)        [root node, fill=gp2blue] {x};
		\node (e) at (4.25,0)     [gp2 node] {y};
		
		\node (A) at (0,-.45)     {\tiny{1}};
		\node (B) at (1.25,-.45)  {\tiny{2}};
		\node (D) at (3,-.45)     {\tiny{1}};
		\node (E) at (4.25,-.45)  {\tiny{2}};
		
		\draw (a) edge[dashed, thick] node[above, yshift=-1pt] {a} (b)
		      (d) edge[thick] node[above, yshift=-1pt] {a} (e);
	\end{tikzpicture}
	\\
	
\end{tabular}
\end{allintypewriter}
\end{minipage}}
\caption{GP\,2 program \ttt{is-connected}}
\label{fig:is-connected-program}
\end{figure}

First, we show correctness, by rigorously defining what we mean by an input graph, and then showing some intermediate results.

\begin{definition}[Input Graph]
    An \emph{input graph} (for the purposes of this subsection) is an arbitrarily labelled GP\,2 \emph{host graph} such that:
    \begin{enumerate}
        \item every node is marked grey;
        \item every node is unrooted;
        \item every edge is unmarked.
    \end{enumerate}
\end{definition}

\begin{lemma}[Invariant of \ttt{is-connected}] \label{lem:is-connected-invariant}
    Throughout the execution of the program \ttt{is-connected} on an \emph{input graph}, all non-grey nodes in the \emph{host graph} share a connected component.
\end{lemma}

\begin{proof}
    The rule \ttt{init} is only called at the start of the program, and turns a grey node into a non-grey one. Subsequently, only applications of \ttt{forward} can turn grey nodes into non-grey ones, and only if \ttt{init} was successfully applied. So let us inductively show that the invariant is satisfied.

    If \ttt{init} is applied at the start of the program, it introduces a single blue node into an \emph{input graph}, which does not violate the invariant. If \ttt{init} fails, \ttt{forward} is never called, and the invariant is trivially satisfied due to a lack of successful rule applications.
    
    Assume by induction that the invariant holds on the current host graph. An application of \ttt{forward} turns node \ttt{2} in the rule's left hand side blue. However, \ttt{forward} can only be applied if said node is adjacent to an existing blue node \ttt{1}. Hence it shares a connected component with the other non-grey nodes.
\end{proof}

\begin{lemma}[Termination of \ttt{is-connected}] \label{lem:is-connected-halt}
    On any host graph, the program \ttt{is-connected} terminates.
\end{lemma}

\begin{proof}
    Since the loop body of \ttt{forward!} consists of a single rule, \ttt{forward} either applies and reduces the number of grey nodes in the host graph, or fails to find a match and terminates the loop. At some point, since the host graph is finite, there are no grey nodes left, and \ttt{forward} cannot match, terminating the loop.

    For the termination of the loop \ttt{DFS!}, consider a couple $\#(G)$ consisting of the number of grey nodes of a host graph $G$, and the number of dashed edges of $G$ in that order. By \emph{reducing} the measure $\#$ we mean that after changing a host graph $G$ to a graph $H$, we have $\#(G)>\#(H)$ with respect to the lexicographical ordering, i.e. $\langle a,b \rangle < \langle c,d \rangle$ if either $a<c$ or both $a=c$ and $b<d$.
    
    When calling \ttt{DFS} on a host graph $G$, because of \ttt{try back else break}, either \ttt{back} is applied, or the loop terminates. When \ttt{back} is applied, the measure $\#$ is reduced. Indeed, if \ttt{forward} is applied at least once the number of grey nodes is reduced (\ttt{back} does not modify the number of grey nodes). And if \ttt{forward} is not applied, the number of grey nodes remains the same, but the number of dashed edges decreases.
    
    Due to host graphs being finite, $\#$ cannot be reduced anymore at some point, which means \ttt{back} cannot be applied. Hence \ttt{break} is invoked and the loop terminates.
\end{proof}

\begin{lemma}[Existence of a Non-Grey Connected Component]
\label{lem:is-connected-output}
    In the output graph of \ttt{try init then DFS!} executed on an \emph{input graph}, there is a connected component consisting of non-grey nodes.
\end{lemma}

\begin{proof}
    The lemma is trivially true for an empty \emph{input graph}. In the case of an \emph{input graph} consisting of a single node, \ttt{init} marks the entire graph non-grey, satisfying the lemma. So we can assume the input contains at least two nodes.

    Assume for the sake of a contradiction that all connected components have at least one grey node in the output graph. Since the input is nonempty, \ttt{init} is applied. Consider the connected component of the node \ttt{init} was applied to. Let $u$ and $v$ be non-grey and grey nodes of the output graph that are adjacent, respectively. They exist because they share a connected component that has at least one non-grey node (application of \ttt{init}) and at least one grey node (assumption). We aim to show that $u$ and $v$ are matched by \ttt{forward} during the execution, contradicting our assumption.
    
    If $u$ is unmarked, it must have been matched by \ttt{back}. Right before that happened, $u$ must have been a blue root, and \ttt{forward} cannot have been applicable to it (note that there is at most one root in the host graph at any given time). However, since $v$ is grey and adjacent to $u$, \ttt{forward} must have been applicable, which is a contradiction. So we may assume $u$ is blue.
    
    Since $u$ is blue in the output graph, it must have been matched by either \ttt{init} or \ttt{forward} at some point. Either way, after the rule application, $u$ is a blue root, and the program is executing the loop \ttt{forward!}. Since \ttt{forward} can be applied to $u$ and $v$, $u$ must have an unmarked neighbour $w$ different from $v$, otherwise $v$ is marked blue.
    
    For the next argument, let us take a look at the data structure the program creates. The dashed edges form a path of blue nodes, where a node at an end is rooted. This can be seen as a stack of blue nodes, where the root represents the top. Indeed, \ttt{init} initialises the stack, \ttt{forward} implements the \emph{push} operation, and \ttt{back} the \emph{pop} operation. Note however that \ttt{back} leaves the popped node with a blue mark, meaning it cannot be pushed again. This prevents the path from becoming a cycle, and also means that throughout the execution of \ttt{DFS!}, the number of blue nodes is reduced. Since the \emph{host graph} has finitely many nodes, \ttt{forward} is not applicable anymore at some point, due to unmarked nodes reachable by \ttt{forward} being gone. So eventually, only \ttt{back} can be applied, popping the top of the stack until the loop terminates.
    
    Coming back to $u$, $v$, and $w$, this reasoning can be applied to the loop \ttt{DFS!} from the point where $u$ is first rooted onward, in the subgraph of the nodes reachable from $w$ without going through $u$. By that reasoning, at some point, the top of the stack is popped until $u$ is the top again. When this happens, \ttt{back} is applied and the loop \ttt{DFS!} enters its next iteration, which starts with \ttt{forward!}. This leaves us in the same situation as previously, where we must assume $u$ has another unmarked neighbour, distinct from $v$ and $w$. However at some point, there will be no unmarked neighbours to apply the previous argument to (since host graphs are finite), so $v$ will have to matched by \ttt{forward}, contradicting our assumption.
\end{proof}

Finally, we show correctness and complexity of \ttt{is-connected}.

\begin{theorem}[Correctness of \ttt{is-connected}] \label{thm:is-connected-correctness}
	The program \ttt{is-connected} (Figure \ref{fig:is-connected-program}) is totally correct with respect to the specification:
	\begin{spec}
		\specinput{An \emph{input graph}.}
		\specoutput{Fail if and only if the input consists of more than one connected component.}
	\end{spec}
\end{theorem}

\begin{proof}
    Termination follows from Lemma \ref{lem:is-connected-halt}. For correctness, first assume \emph{input graph} $G$ has no connected components, i.e. $G$ is the empty graph. Then \ttt{init} cannot be applied, and the procedure \ttt{Check} is called. The rule \ttt{match} cannot be applied either, so the program terminates without failing.
    
    Assume $G$ has exactly one connected component. We know by Lemma \ref{lem:is-connected-halt} that \ttt{DFS!} terminates. Furthermore, by Lemma \ref{lem:is-connected-output}, the output $H$ of \ttt{try init then DFS!} has a connected component whose nodes are blue. Since no rule adds or deletes nodes or edges, $H$ is isomorphic to $G$ ignoring marks and roots. Hence the blue connected component must be the entirety of $H$. The procedure \ttt{Check} is called, and \ttt{match} cannot find a match in a graph containing only blue nodes. Hence \ttt{is-connected} does not fail.
    
    Assume $G$ has more than one connected component. The loop \ttt{DFS!} still terminates by Lemma \ref{lem:is-connected-halt}. Furthermore, by Lemma \ref{lem:is-connected-output}, the output $H$ of \ttt{try init then DFS!} has a connected component $C$ with blue nodes. Since by Lemma \ref{lem:is-connected-invariant}, all blue nodes share the same connected component, and since $H$ consists of more than one connected component, there is a grey node in $H-C$. Hence the rule \ttt{match} matches and the program fails.
\end{proof}

Let us now examine the complexity of \ttt{is-connected}. To do this, we consider the following measures of complexity. Let \ttt{r} be a rule. Let $s(\ttt{r})$ be an upper bound on the number of steps involving \ttt{r}, i.e. how many times \ttt{r} is called, whether it is successfully applied or not. Note that the number of steps depends on the program it is called in and the class of input graphs the program is run on. Let $t(\ttt{r})$ be an upper bound on the number of possible matches of \ttt{r} the program considers during its execution according to a GP\,2 implementation satisfying Theorem \ref{thm:rooted-matching-complexity}. We define $K(\ttt{r}) = s(\ttt{r}) \cdot t(\ttt{r})$ to serve as a measure of complexity of a rule. Consider the complexity measure $K(\ttt{p})$ of a program \ttt{p} defined as the sum of terms $K(\ttt{r})$, where \ttt{r} ranges over the rules called by \ttt{p}. Note that $K(\ttt{p})$ is an upper bound, and hence not unique.

\begin{theorem}[Complexity of \ttt{is-connected}]\label{thm:is-connected-complexity}
    On a class of \emph{bounded degree} \emph{input graphs}, the program \ttt{is-connected} (Figure \ref{fig:is-connected-program}) terminates in linear time with respect to the \emph{size} of its input.
\end{theorem}

\begin{proof}
    Since none of the rules add or delete edges, linearity of the \emph{input graph} is equivalent to linearity of all host graphs during the program's execution. So in this proof, we shall call a number that is linear in the number of nodes of the input simply \emph{linear}.
    
    Furthermore, in order for Theorem \ref{thm:rooted-matching-complexity} to be applicable, there can only be a constant number of roots in the host graph. This is indeed the case. The only rule that does not preserve the number of roots is \ttt{init}, which is only called once. So the host graph can have at most one root at any time.
    
    Let us show the linearity of $K(\ttt{Main})$ by showing that $K$ of all the rules is linear.

    The rule \ttt{init} is only called once. In the worst case, there is no match for it, and every node of the \emph{host graph} has to be considered for a match. Hence $K(\ttt{init})$ is linear.
    
    Similarly, the rule \ttt{match} is only called once, and the program has to consider each node of the \emph{host graph} for a possible match in the worst case, making $K(\ttt{match})$ linear as well.
    
    By Theorem \ref{thm:rooted-matching-complexity}, \ttt{back} matches in constant time on bounded degree graphs, hence $t(\ttt{back})$ is constant. Let us now examine how many times the rule \ttt{back} is called. Since the loop \ttt{DFS!} terminates after \ttt{back} is successfully matched (see Lemma \ref{lem:is-connected-halt}), we know that \ttt{back} succeeds at each call except for the final one, i.e. it has a constant number of unsuccessful applications. So it is enough to show the linearity of the number of successful applications of the rule. It is easy to see that \ttt{back} increases the number of unmarked nodes, while all other rules preserve it. Since \ttt{back} cannot match an unmarked node, it can only be applied a linear number of times. Hence $K(\ttt{back})$ is linear.
    
    The rule \ttt{forward} is also matched in constant time by Theorem \ref{thm:rooted-matching-complexity}, meaning $t(\ttt{back})$ is constant. Furthermore, by the previous paragraph, \ttt{back} is called a linear number of times, meaning the loop \ttt{DFS!} has a linear number of iterations. During each of these iterations, \ttt{forward} fails exactly once (termination of \ttt{forward!} by Lemma \ref{lem:is-connected-halt}), meaning the linearity of its calls is equivalent to the linearity of its successful applications. The rule \ttt{forward} decreases the number of grey nodes, while the other rules called in \ttt{DFS!} preserve it. Since it needs a grey node to match successfully, it can only do so a linear number of times. Hence $K(\ttt{forward})$ is linear.
\end{proof}

Finally, we have collected empirical timing results, supporting our claim that the program runs in linear time on graph classes of bounded degree, but not necessarily on those that do not have bounded degree (Figures \ref{fig:is-connected-timing-1} and \ref{fig:is-connected-timing-2}).

\begin{figure}[!ht]
\centering
\begin{tikzpicture}[scale=0.9]
    \begin{axis}[
    xlabel={Size of input graph},
    ylabel={Execution time (ms)},
    xmin=0,
    ymin=0,
    width=9.2cm,height=7.2cm,
    legend style={at={(1.5,0.82)}},
    ymajorgrids=true,
    grid style=dashed,
    ]
        \addplot[color=plot1, mark=square*] 
        coordinates {
        	(20000,59.28)
        	(60000,96.14)
        	(100000,135.66)
        	(140000,173.93)
        	(180000,210.73)
        	(220000,250.07)
        	(260000,289.46)
        	(300000,328.49)
        	(340000,368.17)
        	(380000,406.21)
        	(420000,447.08)
        	(460000,487.60)
        	(500000,526.23)
        };
        \addplot[color=plot2, mark=square*] 
        coordinates {
        	(19040,55.88)
        	(32865,72.95)
        	(50440,96.27)
        	(71765,127.20)
        	(96840,162.13)
        	(125665,200.51)
        	(158240,240.54)
        	(194565,292.66)
        	(234640,347.57)
        	(278465,401.33)
        	(326040,466.75)
        	(377365,533.38)
        	(432440,607.55)
        	(491265,694.51)
        };
        \addplot[color=plot3, mark=square*] 
        coordinates {
        	(19837,55.69)
        	(30955,69.13)
        	(45601,90.19)
        	(64261,115.36)
        	(87421,146.09)
        	(115567,185.23)
        	(149185,227.31)
        	(188761,276.72)
        	(234781,338.10)
        	(287731,405.46)
        	(348097,491.49)
        	(416365,578.57)
        	(493021,676.62)
        };
        \addplot[color=plot4, mark=square*] 
        coordinates {
        	(16381,51.42)
        	(32765,69.90)
        	(65533,110.89)
        	(131069,192.27)
        	(262141,356.21)
        	(524285,679.00)
        };
        \addplot[color=plot5, mark=square*] 
        coordinates {
        	(20000,63.84)
        	(60000,120.21)
        	(100000,174.29)
        	(140000,232.43)
        	(180000,281.18)
        	(220000,336.47)
        	(260000,392.64)
        	(300000,450.30)
        	(340000,508.05)
        	(380000,556.91)
        	(420000,618.18)
        	(460000,667.62)
        	(500000,728.45)
        };
        \addplot[color=plot6, mark=square*] 
        coordinates {
        	(20000,61.85)
        	(60000,112.94)
        	(100000,167.09)
        	(140000,222.93)
        	(180000,269.92)
        	(220000,323.34)
        	(260000,377.66)
        	(300000,425.65)
        	(340000,478.87)
        	(380000,530.84)
        	(420000,586.30)
        	(460000,637.63)
        	(500000,691.04)
        };
        \addplot[color=plot7, mark=square*] 
        coordinates {
        	(19999,63.73)
        	(59999,120.88)
        	(99999,178.87)
        	(139999,241.24)
        	(179999,293.36)
        	(219999,353.03)
        	(259999,416.57)
        	(299999,473.99)
        	(339999,528.13)
        	(379999,585.42)
        	(419999,647.27)
        	(459999,705.47)
        	(499999,766.68)
        };
        \addlegendentry{Discrete Graph}
        \addlegendentry{Grid Graph}
        \addlegendentry{Grid Chain}
        \addlegendentry{Binary Tree}
        \addlegendentry{Cycle Graph}
        \addlegendentry{Sun Graph}
        \addlegendentry{Linked List}
    \end{axis}
\end{tikzpicture}
\caption{Measured performance of \ttt{is-connected}}
\label{fig:is-connected-timing-1}
\end{figure}
\begin{figure}[!ht]
\centering
\begin{tikzpicture}[scale=0.9]
    \begin{axis}[
    xlabel={Size of input graph},
    ylabel={Execution time (ms)},
    xmin=0,
    ymin=0,
    width=9.2cm,height=7.2cm,
    legend style={at={(1.5,0.82)}},
    ymajorgrids=true,
    grid style=dashed,
    ]
        \addplot[color=plot7, mark=square*] 
        coordinates {
        	(1999,19.69)
        	(5999,24.52)
        	(9999,29.42)
        	(13999,35.62)
        	(17999,41.24)
        	(21999,46.14)
        	(25999,52.03)
        	(29999,57.60)
        	(33999,64.32)
        	(37999,69.07)
        	(41999,74.72)
        	(45999,80.32)
        	(49999,85.73)
        };
        \addplot[color=plot8, mark=square*] 
        coordinates {
        	(1999,25.03)
        	(5999,93.04)
        	(9999,214.41)
        	(13999,396.79)
        	(17999,669.18)
        	(21999,1095.28)
        	(25999,1622.61)
        	(29999,2253.07)
        	(33999,2927.70)
        	(37999,3712.18)
        	(41999,4646.96)
        	(45999,5707.05)
        	(49999,6817.02)
        };
        \addlegendentry{Linked List}
        \addlegendentry{Star Graph}
    \end{axis}
\end{tikzpicture}
\caption{Measured performance of \ttt{is-connected}}
\label{fig:is-connected-timing-2}
\end{figure}

\subsection{The 2-Colouring Problem}
\label{subsec:2_colouring}

Vertex colouring has many applications \cite{Skiena08a} and is among the most frequently considered graph problems. In 2016 Bak and Plump investigated the possibility of an efficient rule-based algorithm for the 2-colouring problem \cite{Bak-Plump16a}. We recall this important case study, and provide further empirical evidence for the linear time complexity on graph classes of bounded degree of the compiled program generated by the latest version of the GP\,2-to-C compiler.

In order to rigorously define what a 2-colouring program is, in the context of GP\,2, we first define what our notion of an input graph should be, and what a 2-colouring is.

\begin{definition}[Input Graph]
    An \emph{input graph} (for the purposes of this subsection) is an arbitrarily labelled, \emph{connected} GP\,2 \emph{host graph} such that:
    \begin{enumerate}
        \item every node is marked grey;
        \item every node is unrooted;
        \item every edge is unmarked.
    \end{enumerate}
\end{definition}

\begin{definition}[A 2-Colouring]
    A \emph{2-colouring} \(H\) of an \emph{input graph} \(G\) is obtained by colouring each of the nodes either red or blue such that no red (blue) node is adjacent to a red (blue) node, respectively.
\end{definition}

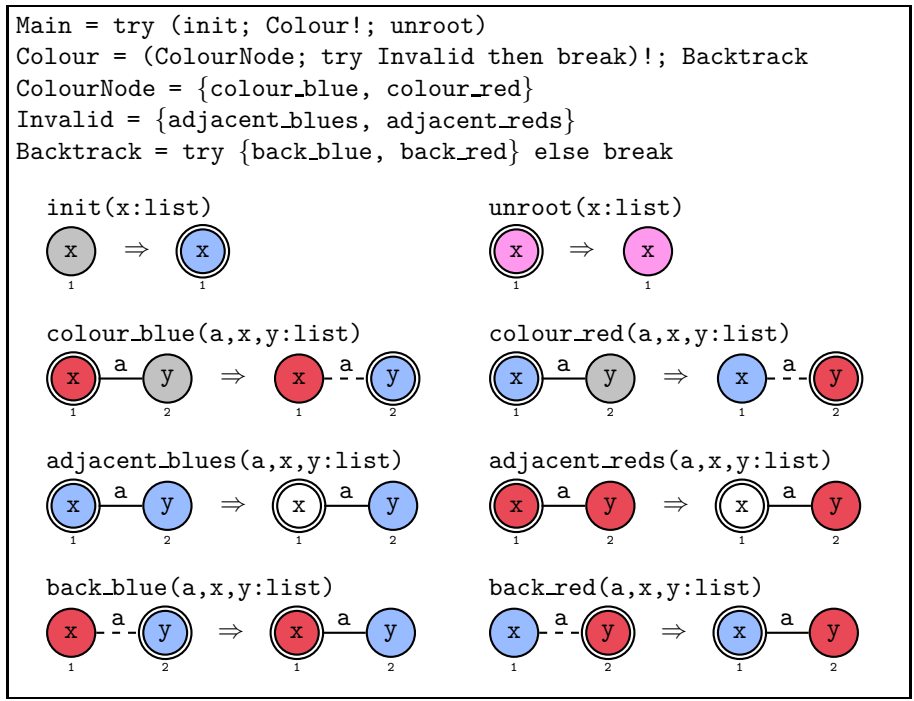
\begin{figure}[!ht]
\centering
\fbox{\begin{minipage}{11.78cm}
\begin{allintypewriter}
Main = try (init; Colour!; unroot)

Colour = (ColourNode; try Invalid then break)!; Backtrack

ColourNode = \{colour\_blue, colour\_red\}

Invalid = \{adjacent\_blues, adjacent\_reds\}

Backtrack = try \{back\_blue, back\_red\} else break

\setlength{\tabcolsep}{0.4cm}

\medskip
\smallskip

\begin{tabular}{ p{5.09cm} p{5.09cm} }
	
	init(x:list) & unroot(x:list) \\
	
	\begin{tikzpicture}
		\node (a) at (0,0)        [gp2 node, fill=gp2grey] {x};
		
		\node (b) at (.875,0)     {$\Rightarrow$};
		
		\node (c) at (1.75,0)     [root node, fill=gp2blue] {x};
		
		\node (A) at (0,-.45)     {\tiny{1}};
		\node (C) at (1.75,-.45)  {\tiny{1}};
	\end{tikzpicture}
	&
	
	\begin{tikzpicture}
		\node (a) at (0,0)        [root node, fill=gp2pink] {x};
		
		\node (b) at (.875,0)     {$\Rightarrow$};
		
		\node (c) at (1.75,0)     [gp2 node, fill=gp2pink] {x};
		
		\node (A) at (0,-.45)     {\tiny{1}};
		\node (C) at (1.75,-.45)  {\tiny{1}};
	\end{tikzpicture}
	\\
	
\end{tabular}

\smallskip

\begin{tabular}{ p{5.09cm} p{5.09cm} }
	
	colour\_blue(a,x,y:list) & colour\_red(a,x,y:list) \\
	
	\begin{tikzpicture}
		\node (a) at (0,0)        [root node,fill=gp2red] {x};
		\node (b) at (1.25,0)     [gp2 node, fill=gp2grey] {y};
		
		\node (c) at (2.125,0)    {$\Rightarrow$};
		
		\node (d) at (3,0)        [gp2 node, fill=gp2red] {x};
		\node (e) at (4.25,0)     [root node, fill=gp2blue] {y};
		
		\node (A) at (0,-.45)     {\tiny{1}};
		\node (B) at (1.25,-.45)  {\tiny{2}};
		\node (D) at (3,-.45)     {\tiny{1}};
		\node (E) at (4.25,-.45)  {\tiny{2}};
		
		\draw (a) edge[thick] node[above, yshift=-1pt] {a} (b)
		      (d) edge[dashed, thick] node[above, yshift=-1pt] {a} (e);
	\end{tikzpicture}
	&
	
	\begin{tikzpicture}
		\node (a) at (0,0)        [root node, fill=gp2blue] {x};
		\node (b) at (1.25,0)     [gp2 node, fill=gp2grey] {y};
		
		\node (c) at (2.125,0)    {$\Rightarrow$};
		
		\node (d) at (3,0)        [gp2 node, fill=gp2blue] {x};
		\node (e) at (4.25,0)     [root node, fill=gp2red] {y};
		
		\node (A) at (0,-.45)     {\tiny{1}};
		\node (B) at (1.25,-.45)  {\tiny{2}};
		\node (D) at (3,-.45)     {\tiny{1}};
		\node (E) at (4.25,-.45)  {\tiny{2}};
		
		\draw (a) edge[thick] node[above, yshift=-1pt] {a} (b)
		      (d) edge[dashed, thick] node[above, yshift=-1pt] {a} (e);
	\end{tikzpicture}
	\\
	
\end{tabular}

\smallskip

\begin{tabular}{ p{5.09cm} p{5.09cm} }
	
	adjacent\_blues(a,x,y:list) & adjacent\_reds(a,x,y:list) \\
	
	\begin{tikzpicture}
		\node (a) at (0,0)        [root node, fill=gp2blue] {x};
		\node (b) at (1.25,0)     [gp2 node, fill=gp2blue] {y};
		
		\node (c) at (2.125,0)    {$\Rightarrow$};
		
		\node (d) at (3,0)        [root node] {x};
		\node (e) at (4.25,0)     [gp2 node, fill=gp2blue] {y};
		
		\node (A) at (0,-.45)     {\tiny{1}};
		\node (B) at (1.25,-.45)  {\tiny{2}};
		\node (D) at (3,-.45)     {\tiny{1}};
		\node (E) at (4.25,-.45)  {\tiny{2}};
		
		\draw (a) edge[thick] node[above, yshift=-1pt] {a} (b)
		      (d) edge[thick] node[above, yshift=-1pt] {a} (e);
	\end{tikzpicture}
	&
	
	\begin{tikzpicture}
		\node (a) at (0,0)        [root node, fill=gp2red] {x};
		\node (b) at (1.25,0)     [gp2 node, fill=gp2red] {y};
		
		\node (c) at (2.125,0)    {$\Rightarrow$};
		
		\node (d) at (3,0)        [root node] {x};
		\node (e) at (4.25,0)     [gp2 node, fill=gp2red] {y};
		
		\node (A) at (0,-.45)     {\tiny{1}};
		\node (B) at (1.25,-.45)  {\tiny{2}};
		\node (D) at (3,-.45)     {\tiny{1}};
		\node (E) at (4.25,-.45)  {\tiny{2}};
		
		\draw (a) edge[thick] node[above, yshift=-1pt] {a} (b)
		      (d) edge[thick] node[above, yshift=-1pt] {a} (e);
	\end{tikzpicture}
	\\
	
\end{tabular}

\smallskip

\begin{tabular}{ p{5.09cm} p{5.09cm} }
	
	back\_blue(a,x,y:list) & back\_red(a,x,y:list) \\
	
	\begin{tikzpicture}
		\node (a) at (0,0)        [gp2 node, fill=gp2red] {x};
		\node (b) at (1.25,0)     [root node, fill=gp2blue] {y};
		
		\node (c) at (2.125,0)    {$\Rightarrow$};
		
		\node (d) at (3,0)        [root node, fill=gp2red] {x};
		\node (e) at (4.25,0)     [gp2 node, fill=gp2blue] {y};
		
		\node (A) at (0,-.45)     {\tiny{1}};
		\node (B) at (1.25,-.45)  {\tiny{2}};
		\node (D) at (3,-.45)     {\tiny{1}};
		\node (E) at (4.25,-.45)  {\tiny{2}};
		
		\draw (a) edge[dashed, thick] node[above, yshift=-1pt] {a} (b)
		      (d) edge[thick] node[above, yshift=-1pt] {a} (e);
	\end{tikzpicture}
	&
	
	\begin{tikzpicture}
		\node (a) at (0,0)        [gp2 node, fill=gp2blue] {x};
		\node (b) at (1.25,0)     [root node, fill=gp2red] {y};
		
		\node (c) at (2.125,0)    {$\Rightarrow$};
		
		\node (d) at (3,0)        [root node, fill=gp2blue] {x};
		\node (e) at (4.25,0)     [gp2 node, fill=gp2red] {y};
		
		\node (A) at (0,-.45)     {\tiny{1}};
		\node (B) at (1.25,-.45)  {\tiny{2}};
		\node (D) at (3,-.45)     {\tiny{1}};
		\node (E) at (4.25,-.45)  {\tiny{2}};
		
		\draw (a) edge[dashed, thick] node[above, yshift=-1pt] {a} (b)
		      (d) edge[thick] node[above, yshift=-1pt] {a} (e);
	\end{tikzpicture}
	\\
	
\end{tabular}
\end{allintypewriter}
\end{minipage}}
\caption{GP\,2 program \ttt{2-colour}}
\label{fig:2-colour-program}
\end{figure}

\begin{theorem}[Correctness of \ttt{2-colour} \cite{Bak15a}] \label{thm:2-colour-correctness}
	The program \ttt{2-colour} (Figure \ref{fig:2-colour-program}) is totally correct with respect to the specification:
	\begin{spec}
		\specinput{An \emph{input graph}.}
		\specoutput{Output a \emph{2-colouring} of the input if one exists, otherwise output the input graph unmodified (up to isomorphism).}
	\end{spec}
\end{theorem}

\begin{theorem}[Complexity of \ttt{2-colour} \cite{Bak15a}]\label{thm:2-colour-complexity}
     On a class of \emph{bounded degree} \emph{input graphs}, the program \ttt{2-colour} (Figure \ref{fig:2-colour-program}) terminates in linear time with respect to the \emph{size} of its input.
\end{theorem}

Note that while the program is only correct on connected graphs, it can be modified to work on arbitrary graphs too, but at a cost. Not only does the program become more complex, but the linear time complexity result fails also, due to there being no way to iterate all the connected components in linear time. Finally, we have collected empirical timing results, supporting our claim that the program runs in linear time on graph classes of bounded degree, but not necessarily on those that do not have bounded degree (Figures \ref{fig:2-colour-timing-1} and \ref{fig:2-colour-timing-2}).

\begin{figure}[!ht]
\centering
\begin{tikzpicture}[scale=0.9]
    \begin{axis}[
    xlabel={Size of input graph},
    ylabel={Execution time (ms)},
    xmin=0,
    ymin=0,
    width=9.2cm,height=7.2cm,
    legend style={at={(1.5,0.82)}},
    ymajorgrids=true,
    grid style=dashed,
    ]
        \addplot[color=plot1, mark=square*] 
        coordinates {
        	(20000,58.44)
        	(60000,112.07)
        	(100000,167.51)
        	(140000,220.94)
        	(180000,275.03)
        	(220000,329.83)
        	(260000,381.81)
        	(300000,438.47)
        	(340000,494.68)
        	(380000,545.62)
        	(420000,604.30)
        	(460000,657.34)
        	(500000,712.41)
        };
        \addplot[color=plot2, mark=square*] 
        coordinates {
        	(19040,54.13)
        	(32865,73.16)
        	(50440,98.21)
        	(71765,129.80)
        	(96840,166.02)
        	(125665,207.68)
        	(158240,250.63)
        	(194565,302.73)
        	(234640,358.92)
        	(278465,418.23)
        	(326040,491.75)
        	(377365,560.95)
        	(432440,639.36)
        	(491265,725.62)
        };
        \addplot[color=plot3, mark=square*] 
        coordinates {
        	(19837,57.44)
        	(30955,73.15)
        	(45601,93.75)
        	(64261,122.42)
        	(87421,155.87)
        	(115567,195.41)
        	(149185,241.27)
        	(188761,298.25)
        	(234781,362.10)
        	(287731,431.10)
        	(348097,516.28)
        	(416365,611.30)
        	(493021,714.28)
        };
        \addplot[color=plot4, mark=square*] 
        coordinates {
        	(16381,51.16)
        	(32765,73.75)
        	(65533,117.84)
        	(131069,212.17)
        	(262141,391.17)
        	(524285,754.68)
        };
        \addplot[color=plot5, mark=square*] 
        coordinates {
        	(20000,61.14)
        	(60000,122.21)
        	(100000,181.35)
        	(140000,241.94)
        	(180000,298.71)
        	(220000,359.15)
        	(260000,416.01)
        	(300000,474.63)
        	(340000,531.58)
        	(380000,587.74)
        	(420000,651.97)
        	(460000,709.67)
        	(500000,765.27)
        };
        \addplot[color=plot6, mark=square*] 
        coordinates {
        	(20000,60.12)
        	(60000,118.23)
        	(100000,175.96)
        	(140000,235.07)
        	(180000,293.44)
        	(220000,350.40)
        	(260000,408.65)
        	(300000,468.90)
        	(340000,524.40)
        	(380000,580.81)
        	(420000,645.84)
        	(460000,700.88)
        	(500000,764.96)
        };
        \addplot[color=plot7, mark=square*] 
        coordinates {
        	(19999,64.11)
        	(59999,125.65)
        	(99999,186.86)
        	(139999,249.55)
        	(179999,308.54)
        	(219999,371.48)
        	(259999,432.37)
        	(299999,494.44)
        	(339999,553.33)
        	(379999,615.55)
        	(419999,678.82)
        	(459999,738.68)
        	(499999,800.95)
        };
        \addlegendentry{Discrete Graph}
        \addlegendentry{Grid Graph}
        \addlegendentry{Grid Chain}
        \addlegendentry{Binary Tree}
        \addlegendentry{Cycle Graph}
        \addlegendentry{Sun Graph}
        \addlegendentry{Linked List}
    \end{axis}
\end{tikzpicture}
\caption{Measured performance of \ttt{2-colour}}
\label{fig:2-colour-timing-1}
\end{figure}
\begin{figure}[!ht]
\centering
\begin{tikzpicture}[scale=0.9]
    \begin{axis}[
    xlabel={Size of input graph},
    ylabel={Execution time (ms)},
    xmin=0,
    ymin=0,
    width=9.2cm,height=7.2cm,
    legend style={at={(1.5,0.82)}},
    ymajorgrids=true,
    grid style=dashed,
    ]
        \addplot[color=plot7, mark=square*] 
        coordinates {
        	(1999,22.24)
        	(5999,25.99)
        	(9999,31.79)
        	(13999,38.03)
        	(17999,43.56)
        	(21999,49.89)
        	(25999,56.40)
        	(29999,62.63)
        	(33999,68.65)
        	(37999,74.24)
        	(41999,80.21)
        	(45999,85.86)
        	(49999,92.40)
        };
        \addplot[color=plot8, mark=square*] 
        coordinates {
        	(1999,26.48)
        	(5999,94.44)
        	(9999,219.93)
        	(13999,401.70)
        	(17999,682.82)
        	(21999,1103.86)
        	(25999,1632.88)
        	(29999,2256.96)
        	(33999,2949.07)
        	(37999,3692.37)
        	(41999,4624.34)
        	(45999,5621.17)
        	(49999,6812.49)
        };
        \addlegendentry{Linked List}
        \addlegendentry{Star Graph}
    \end{axis}
\end{tikzpicture}
\caption{Measured performance of \ttt{2-colour}}
\label{fig:2-colour-timing-2}
\end{figure}

\subsection{Topological Sorting and Recognising DAGs}
\label{subsec:top_sort}

The GP\,2 program \ttt{top-sort} (Figures \ref{fig:top-sort-program-1}, \ref{fig:top-sort-program-2}, \ref{fig:top-sort-program-3}) presented in this section has two purposes: recognising whether its connected input graph is a DAG (directed acyclic graph) and if it is, producing a topological sorting of said graph.

The class of \emph{DAGs (directed acyclic graphs)} consists of all graphs that do not contain a directed cycle as a subgraph. A \emph{topological sorting} of a DAG $G$ is a total order (an antisymmetric, transitive, and connex binary relation) $\leq$ on the set of nodes of $G$, such that for each edge of source $u$ and target $v$, $u \leq v$ (\emph{topological property}). Topological sortings cannot exist for graphs containing directed cycles, since there is no way to define a total order on the nodes of a cycle such that the topological property is satisfied.

The program uses depth-first search to traverse the host graph in linear time while testing whether it is a DAG and constructing a path of blue edges that define a topological sorting. Moreover, it terminates in linear time on inputs of bounded node degree.

Example executions of \ttt{StackNodes!} and \ttt{LoopNodes!} can be found in Figures \ref{fig:top-sort-example-1} and \ref{fig:top-sort-example-2} respectively.

Let us first define what an input graph is for the purposes of this subsection, and then show the program terminates.

\begin{definition}[Input Graph]
    An \emph{input graph} (for the purposes of this subsection) is an arbitrarily labelled, \emph{connected} GP\,2 \emph{host graph} such that:
    \begin{enumerate}
        \item every node is marked grey;
        \item every node is unrooted;
        \item every edge is unmarked.
    \end{enumerate}
\end{definition}

\begin{figure}[!ht]
\centering
\fbox{\begin{minipage}{11.78cm}
\begin{allintypewriter}

Main =

\leavevmode\hphantom{..}try init then (

\leavevmode\hphantom{....}StackNodes!;

\leavevmode\phantom{....}unroot;

\leavevmode\phantom{....}LoopNodes!;

\leavevmode\phantom{....}if flag then fail

\leavevmode\phantom{..})

\smallskip
StackNodes =

\leavevmode\phantom{..}\{forward1, forward2\}!;

\leavevmode\phantom{..}try back else break

\medskip
\smallskip

\begin{tabular}{ p{5.09cm} p{5.09cm} }
	
	flag(x:list) & init(x:list) \\ 
	
	\adjustbox{valign=t}{\begin{tikzpicture}
		\node (a) at (0,0) 	 [draw,circle,thick,double,double distance=0.3mm] {x};
		
		\node (b) at (.75,0) {$\Rightarrow$};
		
		\node (c) at (1.5,0) [draw,circle,thick,double,double distance=0.3mm] {x};
		\node (d) at (0,-.42)   {\tiny{1}};
		\node (e) at (1.5,-.42) {\tiny{1}};
		\end{tikzpicture}}
	
	&
	
	\adjustbox{valign=t}{\begin{tikzpicture}
		\node (a) at (0,0)   [draw,circle,thick,fill=gp2grey] {x};
		
		\node (b) at (.75,0) {$\Rightarrow$};
		
		\node (c) at (1.5,0) [draw,circle,thick,fill=gp2red,double,double distance=0.3mm] {x};
		\node (f) at (2.5,0) [draw,circle,thick,fill=gp2green,double,double distance=0.3mm] {\phantom{x}};
		
		\node (A) at (0,-.42)   {\tiny{1}};
		\node (C) at (1.5,-.42) {\tiny{1}};
		
		\draw (f) edge[->,gp2red,thick] (c);
		\end{tikzpicture}}
	
	\\

	forward1(a,x,y:list) & forward2(a,x,y,z:list) \\  
	
	\adjustbox{valign=t}{\begin{tikzpicture}
		\node (a) at (0,0)     [draw, circle, fill=gp2red, thick, double, double distance=0.3mm] {x};
		\node (b) at (1,0)     [draw, circle, thick,fill=gp2grey] {y};
		\node (f) at (0.5,-1.1)     [draw, circle, fill=gp2green, thick, double, double distance=0.3mm] {\phantom{x}};
		
		\node (c) at (1.75,-.5)     {$\Rightarrow$};
		
		\node (d) at (2.5,0)     [draw, circle, fill=gp2red, thick] {x};
		\node (e) at (3.5,0)     [draw, circle, fill=gp2red, thick, double, double distance=0.3mm] {y};
		\node (h) at (3,-1.1)     [draw, circle, fill=gp2green, thick, double, double distance=0.3mm] {\phantom{x}};
		
		\node (A) at (0,-.42)   {\tiny{1}};
		\node (B) at (1,-.42)   {\tiny{2}};
		\node (D) at (2.5,-.42)   {\tiny{1}};
		\node (E) at (3.6,-.42)   {\tiny{2}};
		\node (F) at (0.5, -1.52)   {\tiny{3}};
		\node (H) at (3,-1.52)   {\tiny{3}};
		
		\draw (a) edge[thick] node[above] {a} (b)
		(d) edge[dashed , thick] node[above] {a} (e)
		(h) edge[->, thick, gp2red] (e)
		(e) edge[->, thick, gp2red, bend left] (d)
		(f) edge[->, thick, gp2red] (a);
		\end{tikzpicture}}
	
	& 
	
	\adjustbox{valign=t}{\begin{tikzpicture}
		\node (a) at (0,0)     [draw, circle, fill=gp2red, thick, double, double distance=0.3mm] {x};
		\node (b) at (1,0)     [draw, circle, thick,fill=gp2grey] {y};
		\node (f) at (0,-1)     [draw, circle, fill=gp2red, thick] {z};
		\node (g) at (1,-1)     [draw, circle, fill=gp2green, thick, double, double distance=0.3mm] {\phantom{x}};
		
		\node (c) at (1.75,-.5)     {$\Rightarrow$};
		
		\node (d) at (2.5,0)     [draw, circle, fill=gp2red, thick] {x};
		\node (e) at (3.5,0)     [draw, circle, fill=gp2red, thick, double, double distance=0.3mm] {y};
		\node (h) at (2.5,-1)     [draw, circle, fill=gp2red, thick] {z};
		\node (i) at (3.5,-1)     [draw, circle, fill=gp2green, thick, double, double distance=0.3mm] {\phantom{x}};
		
		\node (A) at (0,-.42)   {\tiny{1}};
		\node (B) at (1,-.42)   {\tiny{2}};
		\node (D) at (2.5,-.42)   {\tiny{1}};
		\node (E) at (3.6,-.42)   {\tiny{2}};
		\node (F) at (0, -1.42)   {\tiny{3}};
		\node (G) at (1,-1.42)   {\tiny{4}};
		\node (H) at (2.5,-1.42)   {\tiny{3}};
		\node (I) at (3.5,-1.42)   {\tiny{4}};
		
		\draw (a) edge[thick] node[above] {a} (b)
		(d) edge[thick, dashed] node[above] {a} (e)
		(g) edge[->, thick, gp2red] (f)
		(e) edge[->, thick, gp2red] (h)
		(i) edge[->, thick, gp2red] (e);
		\end{tikzpicture}}
	\\
	back(a,x,y:list) & unroot(x:list) \\   
	
	\adjustbox{valign=t}{\begin{tikzpicture}
		\node (a) at (0,0)     [draw, circle, fill=gp2red, thick] {x};
		\node (b) at (1,0)     [draw, circle, fill=gp2red, thick, double, double distance=0.3mm] {y};
		
		\node (c) at (1.75,0)     {$\Rightarrow$};
		
		\node (d) at (2.5,0)     [draw, circle, fill=gp2red, thick, double, double distance=0.3mm] {x};
		\node (e) at (3.5,0)     [draw, circle, fill=gp2red, thick] {y};
		
		\node (A) at (0,-.42)   {\tiny{1}};
		\node (B) at (1,-.42)   {\tiny{2}};
		\node (D) at (2.5,-.42)   {\tiny{1}};
		\node (E) at (3.5,-.42)   {\tiny{2}};
		
		\draw (a) edge[dashed, thick] node[above] {a} (b)
		(d) edge[thick] node[above] {a} (e);
		\end{tikzpicture}}
	
	& 
	
	\adjustbox{valign=t}{\begin{tikzpicture}
		\node (a) at (0,0) 	 [draw,circle,fill=gp2red,thick,double,double distance=0.3mm] {x};
		
		\node (b) at (.75,0) {$\Rightarrow$};
		
		\node (c) at (1.5,0) [draw,circle,fill=gp2red,thick] {x};
		
		\node (d) at (0,-.42)   {\tiny{1}};
		\node (e) at (1.5,-.42) {\tiny{1}};
	\end{tikzpicture}}
\end{tabular}
\end{allintypewriter}
\end{minipage}}
\caption{GP\,2 program \ttt{top-sort} and procedure \ttt{StackNodes}}
\label{fig:top-sort-program-1}
\end{figure}
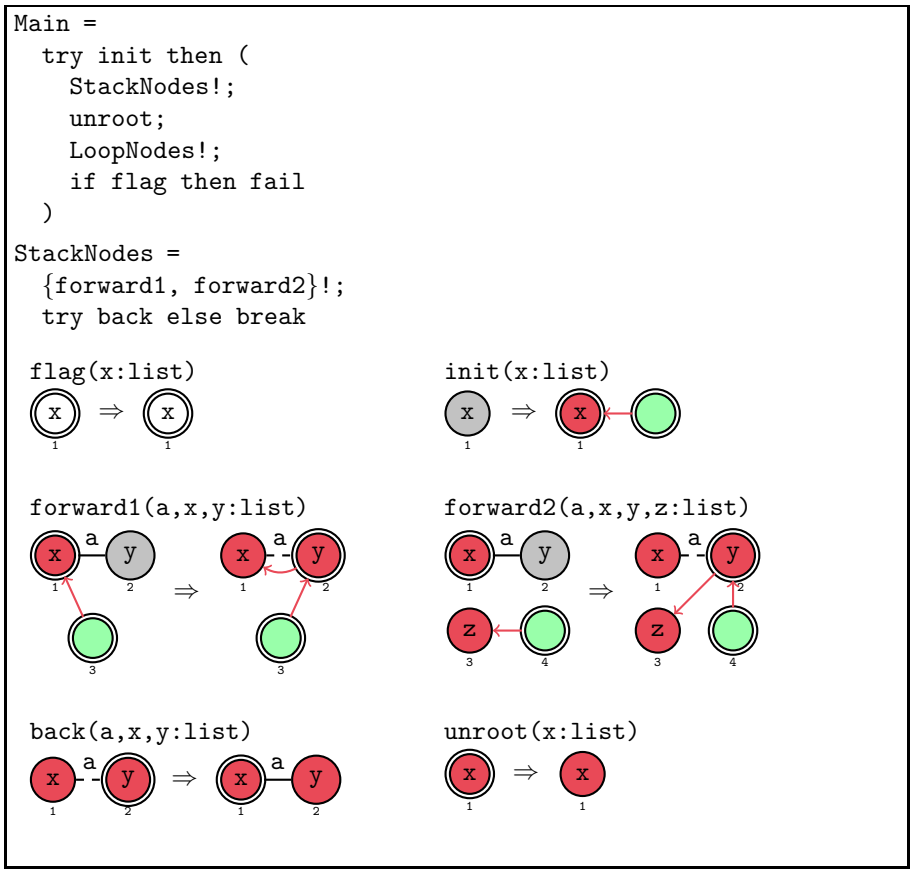
\begin{figure}[!ht]
\centering
\fbox{\begin{minipage}{11.78cm}
\begin{allintypewriter}
LoopNodes =

\leavevmode\phantom{..}if flag then break;

\leavevmode\phantom{..}try skip1 else (

\leavevmode\phantom{....}try skip2 else (

\leavevmode\phantom{......}try init1 then (

\leavevmode\phantom{........}SortNodes!

\leavevmode\phantom{......}) else (

\leavevmode\phantom{........}try init2 then (

\leavevmode\phantom{..........}SortNodes!

\leavevmode\phantom{........}) else (

\leavevmode\phantom{..........}break

\leavevmode\phantom{........})

\leavevmode\phantom{......})

\leavevmode\phantom{....})

\leavevmode\phantom{..})

\medskip
\smallskip

\begin{tabular}{ p{5.09cm} p{5.09cm} }
	
	skip1(x,y,z:list) & skip2(x,y:list) \\  
	
	\adjustbox{valign=t}{\begin{tikzpicture}
		\node (a) at (0,0)     [draw, circle, fill=gp2blue, thick] {x};
		\node (b) at (1,0)     [draw, circle, fill=gp2pink, thick] {y};
		\node (f) at (0.5,-1.1)     [draw, circle, fill=gp2green, thick, double, double distance=0.3mm] {z};
		
		\node (c) at (1.75,-.5)     {$\Rightarrow$};
		
		\node (d) at (2.5,0)     [draw, circle, fill=gp2blue, thick] {x};
		\node (e) at (3.5,0)     [draw, circle, fill=gp2pink, thick] {y};
		\node (h) at (3,-1.1)     [draw, circle, fill=gp2green, thick, double, double distance=0.3mm] {z};
		
		\node (A) at (0,-.42)   {\tiny{1}};
		\node (B) at (1,-.42)   {\tiny{2}};
		\node (D) at (2.5,-.42)   {\tiny{1}};
		\node (E) at (3.5,-.42)   {\tiny{2}};
		\node (F) at (0.5, -1.52)   {\tiny{3}};
		\node (H) at (3,-1.52)   {\tiny{3}};
		
		\draw (a) edge[->, thick, gp2red] (b)
		(f) edge[->, thick, gp2red] (a)
		(h) edge[->, thick, gp2red] (e);
		\end{tikzpicture}}
	
	& 
	
	\adjustbox{valign=t}{\begin{tikzpicture}
		\node (a) at (0,0)     [draw, circle, fill=gp2blue, thick] {x};
		\node (f) at (0,-1)     [draw, circle, fill=gp2green, thick, double, double distance=0.3mm] {y};
		
		\node (c) at (0.75,-.5)     {$\Rightarrow$};
		
		\node (d) at (1.5,0)     [draw, circle, fill=gp2blue, thick] {x};
		\node (h) at (1.5,-1)     [draw, circle, fill=gp2green, thick, double, double distance=0.3mm] {y};
		
		\node (A) at (0.1,-.42)   {\tiny{1}};
		\node (D) at (1.5,-.42)   {\tiny{1}};
		\node (F) at (0, -1.42)   {\tiny{2}};
		\node (H) at (1.5,-1.42)   {\tiny{2}};
		
		\draw (f) edge[->, thick, gp2red] (a);
		\end{tikzpicture}}
	\\
	
	init1(x,y,z:list) & init2(x,y:list) \\  
	
	\adjustbox{valign=t}{\begin{tikzpicture}
		\node (a) at (0,0)     [draw, circle, fill=gp2red, thick] {x};
		\node (b) at (1,0)     [draw, circle, fill=gp2pink, thick] {y};
		\node (f) at (0.5,-1.1)     [draw, circle, fill=gp2green, thick, double, double distance=0.3mm] {z};
		
		\node (c) at (1.75,-.5)     {$\Rightarrow$};
		
		\node (d) at (2.5,0)     [draw, circle, fill=gp2grey, thick, double, double distance=0.3mm] {x};
		\node (e) at (3.5,0)     [draw, circle, fill=gp2pink, thick] {y};
		\node (h) at (3,-1.1)     [draw, circle, fill=gp2green, thick, double, double distance=0.3mm] {z};
		
		\node (A) at (0,-.42)   {\tiny{1}};
		\node (B) at (1,-.42)   {\tiny{2}};
		\node (D) at (2.5,-.42)   {\tiny{1}};
		\node (E) at (3.5,-.42)   {\tiny{2}};
		\node (F) at (0.5, -1.52)   {\tiny{3}};
		\node (H) at (3,-1.52)   {\tiny{3}};
		
		\draw (a) edge[->, thick, gp2red] (b)
		(f) edge[->, thick, gp2red] (a)
		(h) edge[->, thick, gp2red] (e);
		\end{tikzpicture}}
	
	& 
	
	\adjustbox{valign=t}{\begin{tikzpicture}
		\node (a) at (0,0)     [draw, circle, fill=gp2red, thick] {x};
		\node (f) at (0,-1)     [draw, circle, fill=gp2green, thick, double, double distance=0.3mm] {y};
		
		\node (c) at (0.75,-.5)     {$\Rightarrow$};
		
		\node (d) at (1.5,0)     [draw, circle, fill=gp2grey, thick, double, double distance=0.3mm] {x};
		\node (h) at (1.5,-1)     [draw, circle, fill=gp2green, thick, double, double distance=0.3mm] {y};
		
		\node (A) at (0.1,-.42)   {\tiny{1}};
		\node (D) at (1.5,-.42)   {\tiny{1}};
		\node (F) at (0, -1.42)   {\tiny{2}};
		\node (H) at (1.5,-1.42)   {\tiny{2}};
		
		\draw (f) edge[->, thick, gp2red] (a);
		\end{tikzpicture}}

\end{tabular}
\end{allintypewriter}
\end{minipage}}
\caption{GP\,2 procedure \ttt{LoopNodes}}
\label{fig:top-sort-program-2}
\end{figure}
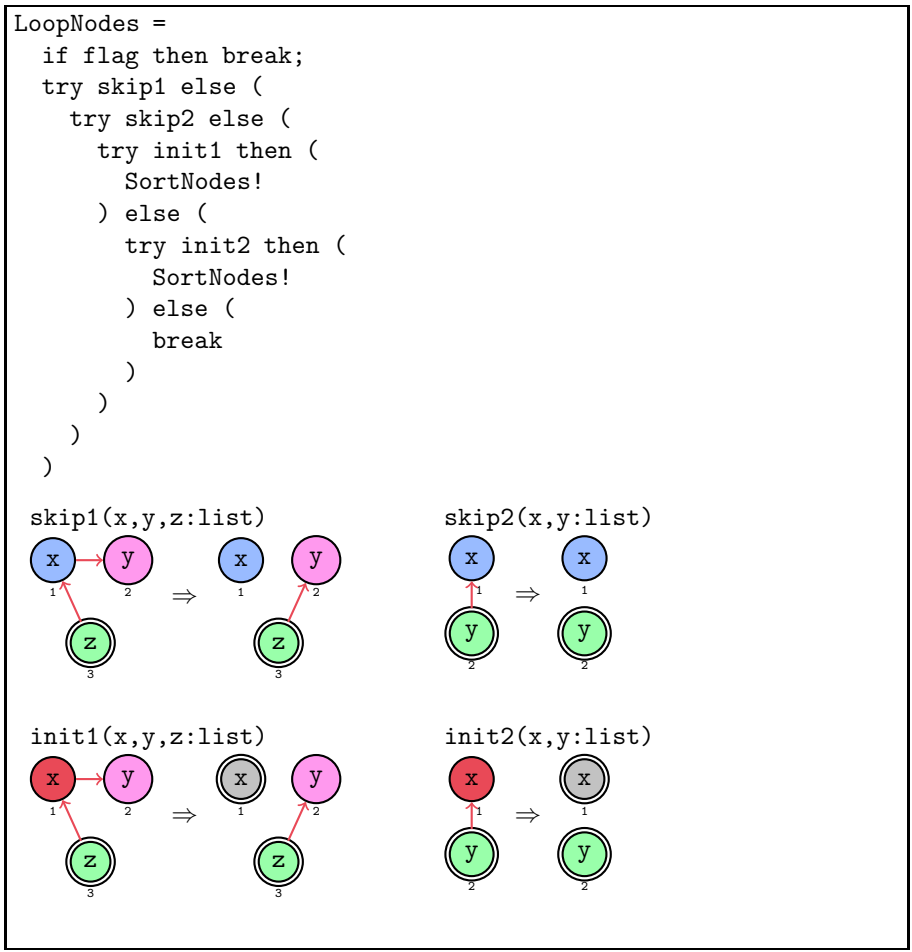
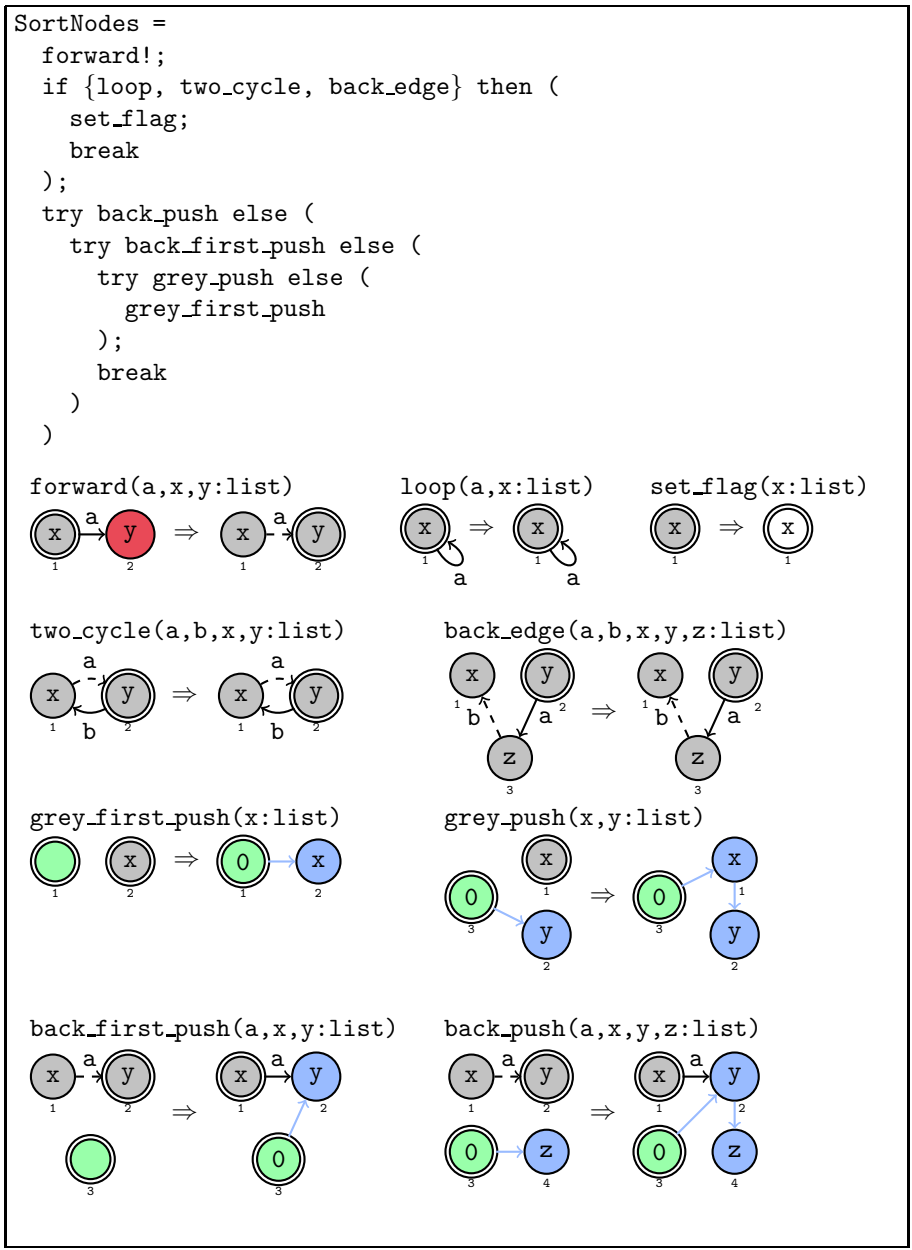
\begin{figure}[!ht]
\centering
\fbox{\begin{minipage}{11.78cm}
\begin{allintypewriter}
SortNodes =

\leavevmode\phantom{..}forward!;

\leavevmode\phantom{..}if \{loop, two\_cycle, back\_edge\} then (

\leavevmode\phantom{....}set\_flag;

\leavevmode\phantom{....}break

\leavevmode\phantom{..});

\leavevmode\phantom{..}try back\_push else (

\leavevmode\phantom{....}try back\_first\_push else (

\leavevmode\phantom{......}try grey\_push else (

\leavevmode\phantom{........}grey\_first\_push

\leavevmode\phantom{......});

\leavevmode\phantom{......}break

\leavevmode\phantom{....})

\leavevmode\phantom{..})

\medskip
\smallskip

\begin{tabular}{ p{4.5cm} p{2.9cm} p{3.10cm} }

	forward(a,x,y:list) & loop(a,x:list) & set\_flag(x:list)\\   
	
	\adjustbox{valign=t}{\begin{tikzpicture}
		\node (a) at (0,0)     [draw, circle, fill=gp2grey, thick, double, double distance=0.3mm] {x};
		\node (b) at (1,0)     [draw, circle, fill=gp2red, thick] {y};
		
		\node (c) at (1.75,0)     {$\Rightarrow$};
		
		\node (d) at (2.5,0)     [draw, circle, fill=gp2grey, thick] {x};
		\node (e) at (3.5,0)     [draw, circle, fill=gp2grey, thick, double, double distance=0.3mm] {y};
		
		\node (A) at (0,-.42)   {\tiny{1}};
		\node (B) at (1,-.42)   {\tiny{2}};
		\node (D) at (2.5,-.42)   {\tiny{1}};
		\node (E) at (3.5,-.42)   {\tiny{2}};
		
		\draw (a) edge[->, thick] node[above] {a} (b)
		(d) edge[->,dashed, thick] node[above] {a} (e);
		\end{tikzpicture}}
	
	& 
	
	\adjustbox{valign=t}{\begin{tikzpicture}
		\node (a) at (0,0) 	 [draw,circle,fill=gp2grey,thick,double,double distance=0.3mm] {x};
		
		\node (b) at (.75,0) {$\Rightarrow$};
		
		\node (c) at (1.5,0) [draw,circle,fill=gp2grey,thick,double,double distance=0.3mm] {x};
		
		\draw (a) edge[->,in=-30,out=-60,loop,thick] node[below] {a} (a)
		(c) edge[->,in=-30,out=-60,loop,thick] node[below] {a} (c);
		
		\node (d) at (0,-.42)   {\tiny{1}};
		\node (e) at (1.5,-.42) {\tiny{1}};
		\end{tikzpicture}}
	&
	
	\adjustbox{valign=t}{\begin{tikzpicture}
		\node (a) at (0,0) 	 [draw,circle,fill=gp2grey,thick,double,double distance=0.3mm] {x};
		
		\node (b) at (.75,0) {$\Rightarrow$};
		
		\node (c) at (1.5,0) [draw,circle,thick,double,double distance=0.3mm] {x};
		\node (d) at (0,-.42)   {\tiny{1}};
		\node (e) at (1.5,-.42) {\tiny{1}};
		\end{tikzpicture}}
	\\
\end{tabular}

\begin{tabular}{ p{5.09cm} p{5.09cm} }

    two\_cycle(a,b,x,y:list) & back\_edge(a,b,x,y,z:list)\\

	\adjustbox{valign=t}{\begin{tikzpicture}
		\node (a) at (0,0)      [draw, circle, fill=gp2grey, thick] {x};
		\node (b) at (1,0)     [draw, circle, fill=gp2grey, thick, double, double distance=0.3mm] {y};
		
		\node (c) at (1.75,0)     {$\Rightarrow$};
		
		\node (d) at (2.5,0)    [draw, circle, fill=gp2grey, thick] {x};
		\node (e) at (3.5,0)    [draw, circle, fill=gp2grey, thick, double, double distance=0.3mm] {y};
		
		\node (A) at (0,-.42)   {\tiny{1}};
		\node (B) at (1,-.42)   {\tiny{2}};
		\node (D) at (2.5,-.42)   {\tiny{1}};
		\node (E) at (3.5,-.42)   {\tiny{2}};
		
		\draw (a) edge[->,dashed, thick,bend left] node[above] {a} (b)
		(b) edge[->, thick,bend left] node[below] {b} (a)
		(d) edge[->,dashed, thick,bend left] node[above] {a} (e)
		(e) edge[->, thick,bend left] node[below] {b} (d);
		\end{tikzpicture}}
	
	& 
	
	\adjustbox{valign=t}{\begin{tikzpicture}
		\node (a) at (0,0)     [draw, circle, fill=gp2grey, thick] {x};
		\node (b) at (1,0)     [draw, circle, fill=gp2grey, thick, double, double distance=0.3mm] {y};
		\node (f) at (0.5,-1.1)     [draw, circle, fill=gp2grey, thick] {z};
		
		\node (c) at (1.75,-.5)     {$\Rightarrow$};
		
		\node (d) at (2.5,0)     [draw, circle, fill=gp2grey, thick] {x};
		\node (e) at (3.5,0)     [draw, circle, fill=gp2grey, thick, double, double distance=0.3mm] {y};
		\node (h) at (3,-1.1)     [draw, circle, fill=gp2grey, thick] {z};
		
		\node (A) at (-.2,-.38)   {\tiny{1}};
		\node (B) at (1.2,-.42)   {\tiny{2}};
		\node (D) at (2.3,-.38)   {\tiny{1}};
		\node (E) at (3.8,-.42)   {\tiny{2}};
		\node (F) at (0.5, -1.52)   {\tiny{3}};
		\node (H) at (3,-1.52)   {\tiny{3}};
		
		\draw 
		(b) edge[->,thick] node[right] {a} (f)
		(f) edge[->,thick,dashed] node[left] {b} (a)
		(e) edge[->,thick] node[right] {a} (h)
		(h) edge[->,thick,dashed] node[left] {b} (d)
		;
		
		\end{tikzpicture}}
    \\

	grey\_first\_push(x:list) & grey\_push(x,y:list)\\  
	
	\adjustbox{valign=t}{\begin{tikzpicture}
		\node (a) at (0,0)     [draw, circle,  fill=gp2green, thick, double, double distance=0.3mm] {\phantom{x}};
		\node (b) at (1,0)     [draw, circle, fill=gp2grey, thick, double, double distance=0.3mm] {x};
		
		\node (c) at (1.75,0)     {$\Rightarrow$};
		
		\node (d) at (2.5,0)     [draw, circle, fill=gp2green, thick, double, double distance=0.3mm] {0};
		\node (e) at (3.5,0)     [draw, circle, fill=gp2blue, thick] {x};
		
		\node (A) at (0,-.42)   {\tiny{1}};
		\node (B) at (1,-.42)   {\tiny{2}};
		\node (D) at (2.5,-.42)   {\tiny{1}};
		\node (E) at (3.5,-.42)   {\tiny{2}};
		
		\draw (d) edge[->, thick, gp2blue] (e);
		\end{tikzpicture}}
	
	& 
	
	\adjustbox{valign=t}{\begin{tikzpicture}
		\node (b) at (1,0)     [draw, circle, fill=gp2grey, thick, double, double distance=0.3mm] {x};
		\node (f) at (0,-0.5)     [draw, circle, fill=gp2green, thick, double, double distance=0.3mm] {0};
		\node (g) at (1,-1)     [draw, circle, fill=gp2blue, thick] {y};
		
		\node (c) at (1.75,-.5)     {$\Rightarrow$};
		
		\node (e) at (3.5,0)     [draw, circle, fill=gp2blue, thick] {x};
		\node (h) at (2.5,-0.5)     [draw, circle, fill=gp2green, thick, double, double distance=0.3mm] {0};
		\node (i) at (3.5,-1)     [draw, circle, fill=gp2blue, thick] {y};
		
		\node (B) at (1,-.42)   {\tiny{1}};
		\node (E) at (3.6,-.42)   {\tiny{1}};
		\node (F) at (0, -0.92)   {\tiny{3}};
		\node (G) at (1,-1.42)   {\tiny{2}};
		\node (H) at (2.5,-0.92)   {\tiny{3}};
		\node (I) at (3.5,-1.42)   {\tiny{2}};
		
		\draw (f) edge[->, thick, gp2blue] (g)
		(h) edge[->, thick, gp2blue] (e)
		(e) edge[->, thick, gp2blue] (i);
		\end{tikzpicture}}
	
	\\
	
	back\_first\_push(a,x,y:list) & back\_push(a,x,y,z:list) \\
	
	\adjustbox{valign=t}{\begin{tikzpicture}
		\node (a) at (0,0)     [draw, circle, fill=gp2grey, thick] {x};
		\node (b) at (1,0)     [draw, circle, fill=gp2grey, thick, double, double distance=0.3mm] {y};
		\node (f) at (0.5,-1.1)     [draw, circle, fill=gp2green, thick, double, double distance=0.3mm] {\phantom{x}};
		
		\node (c) at (1.75,-.5)     {$\Rightarrow$};
		
		\node (d) at (2.5,0)     [draw, circle, fill=gp2grey, thick, double, double distance=0.3mm] {x};
		\node (e) at (3.5,0)     [draw, circle, fill=gp2blue, thick] {y};
		\node (h) at (3,-1.1)     [draw, circle, fill=gp2green, thick, double, double distance=0.3mm] {0};
		
		\node (A) at (0,-.42)   {\tiny{1}};
		\node (B) at (1,-.42)   {\tiny{2}};
		\node (D) at (2.5,-.42)   {\tiny{1}};
		\node (E) at (3.6,-.42)   {\tiny{2}};
		\node (F) at (0.5, -1.52)   {\tiny{3}};
		\node (H) at (3,-1.52)   {\tiny{3}};
		
		\draw (a) edge[->,dashed, thick] node[above] {a} (b)
		(d) edge[->, thick] node[above] {a} (e)
		(h) edge[->, thick, gp2blue] (e);
		\end{tikzpicture}}
	
	&
	
	\adjustbox{valign=t}{\begin{tikzpicture}
		\node (a) at (0,0)     [draw, circle, fill=gp2grey, thick] {x};
		\node (b) at (1,0)     [draw, circle, fill=gp2grey, thick, double, double distance=0.3mm] {y};
		\node (f) at (0,-1)     [draw, circle, fill=gp2green, thick, double, double distance=0.3mm] {0};
		\node (g) at (1,-1)     [draw, circle, fill=gp2blue, thick] {z};
		
		\node (c) at (1.75,-.5)     {$\Rightarrow$};
		
		\node (d) at (2.5,0)     [draw, circle, fill=gp2grey, thick, double, double distance=0.3mm] {x};
		\node (e) at (3.5,0)     [draw, circle, fill=gp2blue, thick] {y};
		\node (h) at (2.5,-1)     [draw, circle, fill=gp2green, thick, double, double distance=0.3mm] {0};
		\node (i) at (3.5,-1)     [draw, circle, fill=gp2blue, thick] {z};
		
		\node (A) at (0,-.42)   {\tiny{1}};
		\node (B) at (1,-.42)   {\tiny{2}};
		\node (D) at (2.5,-.42)   {\tiny{1}};
		\node (E) at (3.6,-.42)   {\tiny{2}};
		\node (F) at (0, -1.42)   {\tiny{3}};
		\node (G) at (1,-1.42)   {\tiny{4}};
		\node (H) at (2.5,-1.42)   {\tiny{3}};
		\node (I) at (3.5,-1.42)   {\tiny{4}};
		
		\draw (a) edge[->,dashed, thick] node[above] {a} (b)
		(d) edge[->, thick] node[above] {a} (e)
		(f) edge[->, thick, gp2blue] (g)
		(h) edge[->, thick, gp2blue] (e)
		(e) edge[->, thick, gp2blue] (i);
		\end{tikzpicture}}
	
\end{tabular}

\end{allintypewriter}
\end{minipage}}
\caption{GP\,2 procedure \ttt{SortNodes}}
\label{fig:top-sort-program-3}
\end{figure}
\begin{figure}[!ht]
\centering
\scalebox{1.0}{\begin{tikzpicture}[scale=0.8]
	
	\node (a) at (2,1)   [draw,fill=gp2grey,circle,thick] {};
	\node (b) at (1.5,0)    [draw,fill=gp2grey,circle,thick] {};
	\node (c) at (2.5,0)    [draw,fill=gp2grey,circle,thick] {};
	\node (d) at (4,1)  [draw,fill=gp2grey,circle,thick] {};
	\node (e) at (3.5,0)    [draw,fill=gp2grey,circle,thick] {};
	\node (f) at (4.5,0)    [draw,fill=gp2grey,circle,thick] {};
	\draw
	(a) edge[->,thick] (b)
	(a) edge[->,thick] (c)
	(c) edge[->,thick] (b)
	(e) edge[->,thick] (c)
	(e) edge[->,thick] (d)
	(f) edge[->,thick] (d)
	(e) edge[->,thick] (f)
	;
	\node (t) at (3,-.75) {$\Downarrow$};
	
	\node (a) at (2,-1.5)  [draw,fill=gp2grey,circle,thick] {};
	\node (b) at (1.5,-2.5)    [draw,fill=gp2grey,circle,thick] {};
	\node (c) at (2.5,-2.5)    [draw,fill=gp2grey,circle,thick] {};
	\node (p) at (3,-1.5) [draw,circle,thick,fill=gp2green,double,double distance=0.3mm] {};
	\node (d) at (4,-1.5)  [draw,fill=gp2grey,circle,thick] {};
	\node (e) at (3.5,-2.5)    [draw,fill=gp2grey,circle,thick] {};
	\node (f) at (4.5,-2.5)    [draw,circle,thick,fill=gp2red,double,double distance=0.3mm] {};
	\draw
	(a) edge[->,thick] (b)
	(a) edge[->,thick] (c)
	(c) edge[->,thick] (b)
	(e) edge[->,thick] (c)
	(e) edge[->,thick] (d)
	(f) edge[->,thick] (d)
	(e) edge[->,thick] (f)
	(p) edge[->,thick,gp2red] (f)
	;
	\node (t) at (5.4,-1.95) {$\Rightarrow^*$};
	
	\node (a) at (6.5,-1.5)  [draw,circle,thick,fill=gp2red,double,double distance=0.3mm] {};
	\node (b) at (6,-2.5)    [draw,circle,thick,fill=gp2red] {};
	\node (c) at (7,-2.5)    [draw,circle,thick,fill=gp2red] {};
	\node (p) at (7.5,-1.5)  [draw,circle,thick,fill=gp2green,double,double distance=0.3mm] {};
	\node (d) at (8.5,-1.5) [draw,circle,thick,fill=gp2red] {};
	\node (e) at (8,-2.5)   [draw,circle,thick,fill=gp2red] {};
	\node (f) at (9,-2.5)   [draw,circle,thick,fill=gp2red] {};
	\draw
	(a) edge[->,thick,dashed] (b)
	(a) edge[->,thick] (c)
	(c) edge[->,thick,dashed] (b)
	(e) edge[->,thick,dashed] (c)
	(e) edge[->,thick,dashed] (d)
	(d) edge[->,thick,dashed] (f)
	(e) edge[->,thick] (f)
	(p) edge[->,thick,gp2red] (a)
	(a) edge[->,thick,gp2red,bend right] (b)
	(b) edge[->,thick,gp2red,bend right] (c)
	(c) edge[->,thick,gp2red,bend right] (e)
	(e) edge[->,thick,gp2red,bend left] (d)
	(d) edge[->,thick,gp2red,bend left] (f)
	;
	\node (t) at (9.9,-1.95) {$\Rightarrow^*$};
	
	\node (a) at (11,-1.5)  [draw,circle,thick,fill=gp2red] {};
	\node (b) at (10.5,-2.5)    [draw,circle,thick,fill=gp2red] {};
	\node (c) at (11.5,-2.5)    [draw,circle,thick,fill=gp2red] {};
	\node (p) at (12,-1.5)  [draw,circle,thick,fill=gp2green,double,double distance=0.3mm] {};
	\node (d) at (13,-1.5) [draw,circle,thick,fill=gp2red] {};
	\node (e) at (12.5,-2.5)   [draw,circle,thick,fill=gp2red] {};
	\node (f) at (13.5,-2.5)   [draw,circle,thick,fill=gp2red,double,double distance=0.3mm] {};
	\draw
	(a) edge[->,thick] (b)
	(a) edge[->,thick] (c)
	(c) edge[->,thick] (b)
	(e) edge[->,thick] (c)
	(e) edge[->,thick] (d)
	(f) edge[->,thick] (d)
	(e) edge[->,thick] (f)
	(p) edge[->,thick,gp2red] (a)
	(a) edge[->,thick,gp2red,bend right] (b)
	(b) edge[->,thick,gp2red,bend right] (c)
	(c) edge[->,thick,gp2red,bend right] (e)
	(e) edge[->,thick,gp2red,bend left] (d)
	(d) edge[->,thick,gp2red,bend left] (f)
	;
	\node (t) at (12,-.75) {$\Uparrow$};
	
	\node (a) at (11,1)  [draw,circle,thick,fill=gp2red] {};
	\node (b) at (10.5,0)    [draw,circle,thick,fill=gp2red] {};
	\node (c) at (11.5,0)    [draw,circle,thick,fill=gp2red] {};
	\node (p) at (12,1)  [draw,circle,thick,fill=gp2green,double,double distance=0.3mm] {};
	\node (d) at (13,1) [draw,circle,thick,fill=gp2red] {};
	\node (e) at (12.5,0)   [draw,circle,thick,fill=gp2red] {};
	\node (f) at (13.5,0)   [draw,circle,thick,fill=gp2red] {};
	\draw
	(a) edge[->,thick] (b)
	(a) edge[->,thick] (c)
	(c) edge[->,thick] (b)
	(e) edge[->,thick] (c)
	(e) edge[->,thick] (d)
	(f) edge[->,thick] (d)
	(e) edge[->,thick] (f)
	(p) edge[->,thick,gp2red] (a)
	(a) edge[->,thick,gp2red,bend right] (b)
	(b) edge[->,thick,gp2red,bend right] (c)
	(c) edge[->,thick,gp2red,bend right] (e)
	(e) edge[->,thick,gp2red,bend left] (d)
	(d) edge[->,thick,gp2red,bend left] (f)
	;
\end{tikzpicture}}
\caption{Example execution of \ttt{StackNodes!}}
\label{fig:top-sort-example-1}
\end{figure}
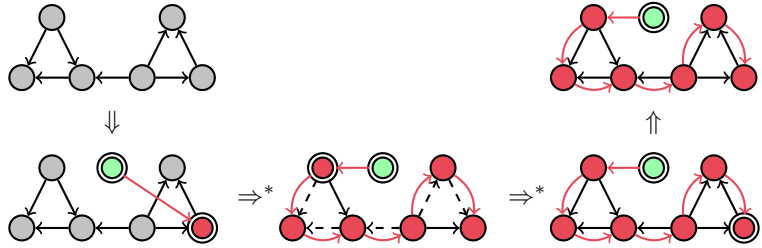
\input{dfs/fig/top-sort-example-2}

\begin{lemma}[Termination of \ttt{top-sort}]
\label{lem:top-sort-termination}
    On any host graph, the program \ttt{top-sort} terminates.
\end{lemma}

\begin{proof}
    Consider the loop \ttt{\{forward1, forward2\}!} called in the procedure \ttt{StackNodes} (Figure \ref{fig:top-sort-program-1}). In each iteration, either \ttt{forward1} is applied, \ttt{forward2} is applied, or both fail and the loop terminates. Whenever one of these rules is applied, the number of grey nodes in the host graph is reduced. Due to host graphs being finite, there are no grey nodes left eventually, and neither rule can match, terminating the loop.
    
    Next, consider the loop \ttt{StackNodes!} (Figure \ref{fig:top-sort-program-1}), a measure $\#$ consisting of the number of grey nodes of a host graph paired with the number of dashed edges, and a lexicographical ordering on said pairs. In each iteration, either \ttt{back} is applied or the loop terminates due to \ttt{break}. If \ttt{back} is applied, either the number of grey nodes remains the same and and the number of dashed edges is reduced (i.e. neither \ttt{forward1} nor \ttt{forward2} are applied), or the number of grey nodes is reduced (i.e. \ttt{forward1} or \ttt{forward2} have been applied at least once). In either case, $\#$ is reduced. Since host graphs are finite, $\#$ cannot be reduced anymore at some point, hence \ttt{back} is not applicable. Then \ttt{break} is invoked and the loop terminates.
    
    Now consider the loop \ttt{forward!} in the procedure \ttt{SortNodes} (Figure \ref{fig:top-sort-program-3}). In each iteration, either \ttt{forward} is applied, or the loop terminates. Applying \ttt{forward} reduces the number of red nodes in the host graphs. Since there are only finitely many, there will be no red nodes left for \ttt{forward} to match eventually. Hence the loop has to terminate.
    
    Next, consider the loop \ttt{SortNodes!} (Figure \ref{fig:top-sort-program-3}). In each iteration, either \ttt{back\_push} applies, \ttt{back\_first\_push} applies, or \ttt{break} is invoked. We claim that each iteration either lexicographically reduces the number of red nodes in the host graph paired with the number of grey nodes (let us call this measure $\#$), or terminates the loop. The rules \ttt{set\_flag}, \ttt{grey\_push}, and \ttt{grey\_first\_push} only get called in an iteration that terminates the loop, so we do not need to consider them for the purpose of reducing $\#$. Similarly, as rules called in the condition of an \ttt{if} statement do not modify the host graph, we do not need to consider \ttt{loop}, \ttt{two\_cycle}, and \ttt{back\_edge} for reduction purposes either. So consider an iteration where either either ttt{back\_push} or \ttt{back\_first\_push} applies. If \ttt{forward} is applied at least once this iteration, the number of red nodes is reduced, which reduces $\#$. If \ttt{forward} is not applied at all, the number of red nodes remains the same while either \ttt{back\_push} or \ttt{back\_first\_push} reduces the number of grey nodes, reducing $\#$. So eventually, $\#$ cannot be reduced any further in the finite host graph, meaning neither \ttt{back\_push} nor \ttt{back\_first\_push} can be applied, causing the loop to terminate.
    
    Finally, consider the loop \ttt{LoopNodes!} (Figure \ref{fig:top-sort-program-2}). In each iteration, either \ttt{break} is invoked, or one of the rules \ttt{skip1}, \ttt{skip2}, \ttt{init1}, and \ttt{init2} is applied. Each of these rules reduces the number of red edges in the host graph. So eventually, since host graphs are finite, there are no red edges that can be matched by these rules anymore, so all of them will fail, meaning \ttt{break} is invoked due to the structure of the nested \ttt{try} statements, and the loop terminates.
\end{proof}

The command \texttt{init; StackNodes!} creates a stack of nodes in order to navigate between strongly connected components while doing a depth-first search that travels in the direction of the edges. It is equivalent to \ttt{try init then StackNodes!} if \ttt{init} is applied successfully. We shall define stacks via their implementation.

A \emph{stack} is a finite set of red nodes connected by red edges such that the red edges form a path that does not self-intersect. The node in the path that has no incoming red edge from another path node is called the \emph{top} of the stack. Additionally, there is an unlabelled green root node called the \emph{pointer} with only one adjacent edge, namely an outgoing red edge whose target is the top of the stack.

Note that such a stack can also be defined with blue nodes and edges instead of red ones, in which case we shall call it a \emph{blue stack}. In fact, during the execution of \ttt{LoopNodes!}, a red and a blue stack coexist using the same green root.

\begin{lemma}[Correctness of \ttt{init; StackNodes!}]
\label{lem:top-sort-if-dag-corr-stacknodes}
    The command sequence \ttt{init; StackNodes!} is totally correct with respect to the specification:
    \begin{spec}
		\specinput{An \emph{input graph} $G$.}
		\specoutput{$G$ where all its nodes are in a \emph{red stack}.}
	\end{spec}
\end{lemma}

\begin{proof}
    Termination follows from Lemma \ref{lem:top-sort-termination}.
    
    The proof that all nodes of $G$ are marked red is analogous to that of Lemma \ref{lem:is-connected-output}.
    
    To show that a correctly encoded stack is formed, let us proceed by induction. The rule \ttt{init} creates a valid stack containing a single node. Now assume a valid stack is encoded in the host graph. Let us argue that after applying \ttt{StackNodes}, that is still the case. Whenever a red edge is created, the target is the top (since it is adjacent to the green root), and the source is a grey node (and hence not part of the path already, since an \emph{input graph} has grey nodes), extending the non-self-intersecting path. The green root now points toward the newly added node, making it the new top.
\end{proof}

Let us now define how to represent a topological sorting in the context of GP\,2.

\begin{definition}[Topological sorting as a graph structure]\label{def:top-sort}
    Consider a graph $G$ and the set of its blue nodes $B$. Define a binary relation on $B$ by $u \leq v$ if there is a blue edge from $u$ to $v$, or if $u=v$. $G$ contains a \emph{topological sorting} if the transitive closure of $\leq$ is a topological sorting of the subgraph of $G$ induced by the blue nodes $B$ and the unmarked edges.
\end{definition}

With such a structure, one can test whether two nodes are related with a topological sorting by checking whether there is a path of blue edges connecting them.

\begin{lemma}[Correctness of \ttt{LoopNodes!}]
\label{lem:topsort-if-dag-loopnodes}
    On a graph whose nodes are all in a red stack, and whose subgraph induced by its unmarked edges is a DAG, \ttt{LoopNodes!} outputs a graph $G$ that contains an unmarked root or a topological sorting.
\end{lemma}

\begin{proof}
    Termination follows from Lemma \ref{lem:top-sort-termination}.

    Let us assume $G$ does not contain an unmarked root, and show that it does contain a topological sorting. In fact, since no rules called in \ttt{LoopNodes!} can mark or unroot an unmarked node, we can assume that no unmarked root is introduced at any point, i.e. \ttt{set\_flag} is not applied.
    
    Consider the binary relation $\leq$ on the set of blue nodes defined by blue edges as in Definition \ref{def:top-sort}, and let us show it is a topological sorting. It is transitive since it is defined as a transitive closure.
    
    Antisymmetry follows from the fact that $\leq$ is reflexive and the fact the subgraph $H$ of $G$ induced by blue edges does not contain directed cycles. Indeed, $H$ behaves like a stack of blue nodes and edges with a green root pointing towards the top with a blue edge. When the green root is unlabelled, the stack is initialised as a single blue node, and the green root is labelled \ttt{0}. Once that label has been established, non-blue nodes are pushed. At no point does the program pop a blue node, or change the mark of a blue node. Hence no blue cycle can be introduced.
    
    Connexity follows from the fact that every node is eventually marked blue, i.e. pushed. By using arguments analogous to those in Lemma \ref{lem:is-connected-output}, we can conclude that after \ttt{SortNodes!} is applied, the grey root and all nodes reachable from it are marked blue (which we can only conclude because the \ttt{if} statement does not change the host graph since we assume \ttt{set-flag} is not applied). This difference is because the steps of the depth-first search are sensitive to edge direction. In order to sort through remaining nodes, \ttt{LoopNodes!} skips over blue, i.e. already sorted nodes in the red stack with \ttt{skip1} and \ttt{skip2}, until it reaches a red, i.e. unsorted node, which is then initialised as grey root with \ttt{init1} or \ttt{init2}. Then \ttt{StackNodes!} is called on that grey root. Since all nodes of the input graph are in the red stack by Lemma \ref{lem:top-sort-if-dag-corr-stacknodes}, all nodes are eventually marked blue.
    
    It remains to show that $\leq$ satisfies the topological property, namely that for each unmarked edge from $u$ to $v$ in the input, $u \leq v$, i.e. there is a blue path from $u$ to $v$ in the output graph. Since the blue edges form a stack of all nodes, it is enough to show that $u$ is pushed after $v$. Consider the iteration of \ttt{SortNodes!} that pushes $u$ onto the blue stack with one of the push rules (\ttt{grey\_first\_push}, \ttt{grey\_push}, \ttt{back\_first\_push}, and \ttt{back\_push}). The node $u$ has no outgoing unmarked edge with a red node as the source because then \ttt{forward!} would have had at least one more iteration, and this would not be the iteration of \ttt{SortNodes!} that pushes $u$. So $v$ is not red. It cannot be grey either because then there would be a path of dashed edges from $v$ to $u$ (since the grey nodes are in a path of dashed edges, and the root, $u$, is the final node of the path), which would mean there was a cycle of unmarked edges in the input. So $v$ must be blue, i.e. it is pushed before $u$.
\end{proof}

Now let us show the total correctness of \ttt{top-sort}. Note that we include the empty graph in the definition of DAGs. If one wishes to exclude it from the class of DAGs, it suffices to add the \ttt{else fail} to the \ttt{try} statement in \ttt{Main}, since \ttt{init} fails on the empty graph.

\begin{theorem}[Correctness of \ttt{top-sort}] \label{thm:is-dag-correctness}
	The program \ttt{top-sort} (Figures \ref{fig:top-sort-program-1}, \ref{fig:top-sort-program-2}, \ref{fig:top-sort-program-3}) is totally correct with respect to the specification:
	\begin{spec}
		\specinput{An \emph{input graph}.}
		\specoutput{Fail if the input is not a DAG, and $G$ equipped with a topological sorting otherwise.}
	\end{spec}
\end{theorem}

\begin{proof}
    If $G$ is the empty graph, \ttt{init} is not successfully applied, and the output is $G$, which defines a valid topological sorting of the empty DAG.
    
    Termination follows from Lemma \ref{lem:top-sort-termination}.
    
    If $G$ is a DAG, and no unmarked root is introduced in \ttt{LoopNodes!}, it follows from the same lemmata and the fact that no rule of \ttt{top-sort} changes the structure of the underlying graph of unmarked edges, that the output is $G$ containing a topological sorting. So we need to show that, if $G$ is a DAG, \ttt{LoopNodes!} does not introduce an unmarked root. Conversely, if $G$ is not a DAG, we need to show that an unmarked root is introduced in \ttt{LoopNodes} because matching \ttt{flag} is the only way for \ttt{Main} to fail (\ttt{unroot} always matches since \ttt{SteckNodes} leaves a red root in the host graph).
    
    The only rule that introduces an unmarked root is \ttt{set\_flag}, which is only called during the \ttt{if} statement in \ttt{SortNodes}, if the condition is satisfied. So it is enough to show that $G$ is a DAG if and only if neither \ttt{loop}, \ttt{two\_cycle}, nor \ttt{back\_edge} matches.
    
    As argued in the proof of Lemma \ref{lem:topsort-if-dag-loopnodes}, every non-pointer node is pushed onto the blue stack with one of the push rules. So the \ttt{if} statement called right before the push rules are invoked for each non-pointer node while it is a grey root.
    
    If $G$ is a DAG, \ttt{loop} and \ttt{two\_cycle} cannot match since they need a $1$-cycle or $2$-cycle respectively to be present in $G$. The rule \ttt{back\_edge} cannot match either. It contains a path from node $2$ to node $1$. As the target of a dashed edge, node $1$ is in the stack of dashed edges, so there is a path to node $1$ to node $2$, the top of the stack. This means there is a cycle.
    
    Conversely, assume that $G$ is not a DAG. If it contains a $1$- or $2$-cycle, either \ttt{loop} or \ttt{two\_cycle} matches. So assume $G$ contains a cycle of length at least $3$. Consider the first time a node of that cycle becomes a grey root due to \ttt{forward}. Eventually, \ttt{forward} is applied to make the next node in the cycle the grey root. We can repeat this argument until the last node in the cycle is the grey root (in the cycle, all edges but one are dashed, all nodes are grey, and the node with an outgoing unmarked edge is rooted). Then \ttt{back\_edge} can match.
\end{proof}

For the complexity of \ttt{top-sort}, let us return to the measures defined in Subsection \ref{subsec:is_connected}, namely $s(\ttt{r})$ (an upper bound on the number of steps of a rule \ttt{r}, i.e. the number of times \ttt{r} is called during the execution of its program), $t(\ttt{r})$ (an upper bound on the number of possible matches for a rule $r$ that have to be considered), and $K(\ttt{r}) = s(\ttt{t}) \cdot t(\ttt{r})$. We shall measure the complexity of a program or procedure \ttt{p} with $K(\ttt{p})$ (the sum of $K(\ttt{r})$ over the rules \ttt{r} called in \ttt{p}) as a function of the size of the input graph.

\begin{theorem}[Complexity of \ttt{top-sort}]
    On a class of \emph{bounded degree} \emph{input graphs}, the program \ttt{top-sort} (Figures \ref{fig:top-sort-program-1}, \ref{fig:top-sort-program-2}, \ref{fig:top-sort-program-3}) terminates in linear time with respect to the \emph{size} of its input.
\end{theorem}

\begin{proof}
    Let us first argue that there is a constant number of roots at any given time, so that we can apply Theorem \ref{thm:rooted-matching-complexity}. The rule \ttt{init} introduces a green root. In all other rules, the rootedness and mark no green nodes get modified, and no green nodes get introduced. The rule \ttt{init} also introduces a red root. In \ttt{StackNodes!}, no rules modify the number of red roots. Then with \ttt{unroot}, the one red root is removed. The rules \ttt{init1} and \ttt{init2} introduce a grey root. Whenever one of them is applied, \ttt{SortNodes!} is called. Let us show that \ttt{SortNodes!} removes the grey root. This loop can only terminate when \ttt{break} is invoked (termination itself has been shown in Lemma \ref{lem:top-sort-if-dag-corr-stacknodes}), or when \ttt{grey\_first\_push} does not find a match. In the latter case, either the grey root has already been removed (which is what we want), or the green root has a label. If the green root as a label, \ttt{grey\_push} or \ttt{back\_push} would have been applied, and \ttt{grey\_first\_push} never called. Now consider the case where \ttt{break} is invoked. This must be preceded by a successful application of \ttt{set\_flag}, \ttt{grey\_push}, or \ttt{grey\_first\_push}. In the latter two cases, the grey root is unrooted. In the former case, the grey root is unmarked, and various \ttt{break} and \ttt{fail} statements are invoked and the program terminates. In either case, the number of roots remains constant.
    
    Let us now show that $K(\ttt{top-sort})$ is linear in the size of the input graph by showing that for each rule \ttt{r} called by \ttt{top-sort}, $K(\ttt{r})$ is linear.
    
    First consider \ttt{init}. It is called once, and every node of the input graph is a valid match, so $K(\ttt{init})$ is constant. Now for each \ttt{r} rule apart from \ttt{init}, \ttt{r} is a fast rule, and hence by Theorem \ref{thm:rooted-matching-complexity}, $t(\ttt{r})$ is constant. So it remains to show that the number of calls $s(\ttt{r})$ is at most linear.
    
    The rules \ttt{unroot} and \ttt{flag} are called at most once.
    
    Note that \ttt{\{forward1, forward2\}} is only applied a linear number of times since it reduces the number of grey nodes, of which there can be only linearly many, and no other rule in \ttt{StackNodes!} introduces grey nodes. So \ttt{back} can only be applied a linear number of times. And the number of times \ttt{back} is called but not applied is once, since in that case \ttt{break} is invoked. Hence $s(\ttt{back})$ is linear. This also means that the number of iterations of \ttt{StackNodes!} is linear. So the number of times \ttt{forward1} and \ttt{forward2} are called but not applied can only be linear as well. Hence $s(\ttt{forward1})$ and $s(\ttt{forward2})$ are linear.
    
    Therefore there can only be linearly many red edges. In each successful iteration of \ttt{LoopNodes!}, one of \ttt{skip1}, \ttt{skip2}, \ttt{init1}, \ttt{init2} has to be applied, reducing the number of red edges. So there can only be linearly many iterations of \ttt{LoopNodes}, making $s(\ttt{skip1})$, $s(\ttt{skip2})$, $s(\ttt{init1})$, and $s(\ttt{init2})$ linear.
    
    The rule \ttt{forward} can only be applied a linear number of times since it reduces the number of red nodes, and no other rule in \ttt{LoopNodes!} modifies that number. So there can only be linearly many dashed edges, meaning that combined, \ttt{back\_first\_push} and \ttt{back\_push} can only be applied a linear number of times. Hence there can only be a linear number of calls of \ttt{SortNodes}. So $s(\ttt{r})$ is linear for each rule \ttt{r} called only once in \ttt{SortNodes}. That also means that the number of times \ttt{forward} is called but not applied is also linear, so $s(\ttt{forward})$ is linear.
\end{proof}

\begin{figure}[!ht]
\centering
\begin{tikzpicture}[scale=0.9]
    \begin{axis}[
    xlabel={Size of input graph},
    ylabel={Execution time (ms)},
    xmin=0,
    ymin=0,
    width=9.2cm,height=7.2cm,
    legend style={at={(1.5,0.82)}},
    ymajorgrids=true,
    grid style=dashed,
    ]
        \addplot[color=plot2, mark=square*] 
        coordinates {
        	(19040,80.37)
        	(32865,109.74)
        	(50440,153.76)
        	(71765,199.44)
        	(96840,262.41)
        	(125665,318.42)
        	(158240,399.24)
        	(194565,467.42)
        	(234640,589.15)
        	(278465,656.41)
        	(326040,802.69)
        	(377365,869.36)
        	(432440,1057.83)
        	(491265,1136.99)
        };
        \addplot[color=plot3, mark=square*] 
        coordinates {
        	(19837,74.16)
        	(30955,100.72)
        	(45601,129.39)
        	(64261,178.59)
        	(87421,216.38)
        	(115567,294.41)
        	(149185,344.49)
        	(188761,455.04)
        	(234781,528.84)
        	(287731,680.08)
        	(348097,762.40)
        	(416365,967.79)
        	(493021,1075.44)
        };
        \addplot[color=plot4, mark=square*] 
        coordinates {
        	(16381,74.87)
        	(32765,110.15)
        	(65533,179.09)
        	(131069,315.44)
        	(262141,587.88)
        	(524285,1148.18)
        };
        \addplot[color=plot5, mark=square*] 
        coordinates {
        	(20000,60.60)
        	(60000,108.83)
        	(100000,157.29)
        	(140000,206.56)
        	(180000,254.86)
        	(220000,302.12)
        	(260000,348.61)
        	(300000,399.39)
        	(340000,447.39)
        	(380000,496.80)
        	(420000,548.05)
        	(460000,596.73)
        	(500000,647.71)
        };
        \addplot[color=plot6, mark=square*] 
        coordinates {
        	(20000,58.76)
        	(60000,105.65)
        	(100000,151.27)
        	(140000,200.04)
        	(180000,246.78)
        	(220000,290.79)
        	(260000,336.25)
        	(300000,386.03)
        	(340000,431.71)
        	(380000,478.22)
        	(420000,524.70)
        	(460000,571.34)
        	(500000,618.31)
        };
        \addplot[color=plot7, mark=square*] 
        coordinates {
        	(19999,96.23)
        	(59999,192.67)
        	(99999,282.19)
        	(139999,378.29)
        	(179999,467.95)
        	(219999,563.47)
        	(259999,648.99)
        	(299999,750.58)
        	(339999,835.28)
        	(379999,935.49)
        	(419999,1035.14)
        	(459999,1118.47)
        	(499999,1222.67)
        };
        \addlegendentry{Grid Graph}
        \addlegendentry{Grid Chain}
        \addlegendentry{Binary Tree}
        \addlegendentry{Cycle Graph}
        \addlegendentry{Sun Graph}
        \addlegendentry{Linked List}
    \end{axis}
\end{tikzpicture}
\caption{Measured performance of \ttt{top-sort}}
\label{fig:top-sort-timing}
\end{figure}
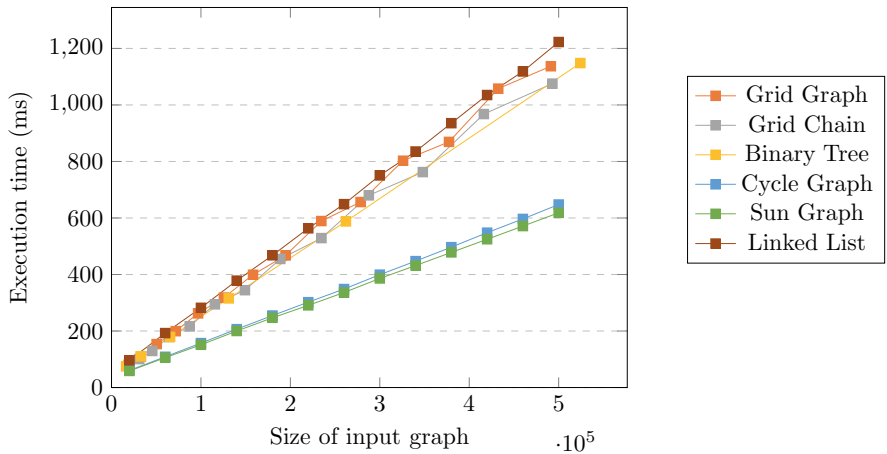

Like \ttt{2-colour}, \ttt{top-sort} is only correct on connected graphs, and can similarly be modified to work on arbitrary graphs. Finally, we have collected empirical timing results, supporting our claim that the program runs in linear time on classes of connected graphs of bounded degree (Figure \ref{fig:top-sort-timing}).

\section{Conclusion}
\label{sec:conclusion}
	
The polynomial cost of graph matching is the performance bottleneck for languages based on standard graph transformation rules. GP\,2 mitigates this problem by providing rooted rules which under mild conditions can be matched in constant time. We present rooted GP\,2 programs of two types: (1) graph reduction programs which recognise cycle graphs, trees and binary DAGs, and (2) depth-first search programs for checking connectedness and acyclicity resp.\ for producing a topological sorting. The programs are proved to be correct and to run in linear time either on arbitrary input graphs (in the case of reduction programs) or on graphs of bounded node degree (in the case of depth-first search programs). The proofs demonstrate that graph transformation rules provide a convenient and intuitive abstraction level for formal reasoning on graph programs. We also give empirical evidence for the linear run time of the programs, by presenting benchmark results for graphs of up to 500,000 nodes in various graph classes. For acyclicity checking and topological sorting, the linear behaviour is achieved by implementing depth-first search strategies based on an encoding of stacks in graphs.	
	
In future work, we intend to investigate for more graph algorithms whether and under what conditions their time complexity in conventional programming languages can be reached in GP\,2. The more involved the data structures of those algorithms are, the more challenging will be the implementation task. This is because in GP\,2, the internal graph data structure is (intentionally) hidden from the programmer and hence any data structures used by an algorithm need to be encoded in host graphs. A simple example of this is the encoding of stacks as linked lists in the programs for acyclicity checking and topological sorting.

Additional future work is the refinement of unrooted programs into more efficient rooted programs. It is not obvious how to do this in general, or what refinement tactics could be used.
	
The programs using depth-first search need host graphs of bounded node degree in order to run in linear time. A topic for future work is therefore to find a mechanism that allows us to overcome this restriction. Clearly, such a mechanism will require to modify GP\,2 and its implementation.

\bibliography{ms}

\begin{thebibliography}{10}
\expandafter\ifx\csname url\endcsname\relax
  \def\url#1{\texttt{#1}}\fi
\expandafter\ifx\csname urlprefix\endcsname\relax\def\urlprefix{URL }\fi
\expandafter\ifx\csname href\endcsname\relax
  \def\href#1#2{#2} \def\path#1{#1}\fi

\bibitem{Ehrig-Ermel-Golas-Hermann15a}
H.~Ehrig, C.~Ermel, U.~Golas, F.~Hermann, Graph and Model Transformation,
  Monographs in Theoretical Computer Science, Springer, 2015.
\newblock \href {https://doi.org/10.1007/978-3-662-47980-3}
  {\path{doi:10.1007/978-3-662-47980-3}}.

\bibitem{Runge-Ermel-Taentzer11a}
O.~Runge, C.~Ermel, G.~Taentzer, {AGG 2.0} -- new features for specifying and
  analyzing algebraic graph transformations, in: Proc.\ 4th International
  Symposium on Applications of Graph Transformations with Industrial Relevance
  ({AGTIVE 2011}), Vol. 7233 of Lecture Notes in Computer Science, Springer,
  2011, pp. 81--88.
\newblock \href {https://doi.org/10.1007/978-3-642-34176-2_8}
  {\path{doi:10.1007/978-3-642-34176-2_8}}.

\bibitem{Agrawal-Karsai-Neema-Shi-Vizhanyo06a}
A.~Agrawal, G.~Karsai, S.~Neema, F.~Shi, A.~Vizhanyo, The design of a language
  for model transformations, Software {\&} Systems Modeling 5~(3) (2006)
  261--288.
\newblock \href {https://doi.org/10.1007/s10270-006-0027-7}
  {\path{doi:10.1007/s10270-006-0027-7}}.

\bibitem{Ghamarian-deMol-Rensink-Zambon-Zimakova12a}
A.~Ghamarian, M.~de~Mol, A.~Rensink, E.~Zambon, M.~Zimakova, Modelling and
  analysis using {GROOVE}, International Journal on Software Tools for
  Technology Transfer 14~(1) (2012) 15--40.
\newblock \href {https://doi.org/10.1007/s10009-011-0186-x}
  {\path{doi:10.1007/s10009-011-0186-x}}.

\bibitem{Jakumeit-Buchwald-Kroll10a}
E.~Jakumeit, S.~Buchwald, M.~Kroll, {GrGen.NET} -- the expressive, convenient
  and fast graph rewrite system, International Journal on Software Tools for
  Technology Transfer 12~(3--4) (2010) 263--271.
\newblock \href {https://doi.org/10.1007/s10009-010-0148-8}
  {\path{doi:10.1007/s10009-010-0148-8}}.

\bibitem{Arendt-Biermann-Jurack-Krause-Taentzer10a}
T.~Arendt, E.~Biermann, S.~Jurack, C.~Krause, G.~Taentzer, Henshin: Advanced
  concepts and tools for in-place {EMF} model transformations, in: Proc.\ 13th
  International Conference on Model Driven Engineering Languages and Systems
  ({MODELS 2010}), Vol. 6394 of Lecture Notes in Computer Science, Springer,
  2010, pp. 121--135.
\newblock \href {https://doi.org/10.1007/978-3-642-16145-2_9}
  {\path{doi:10.1007/978-3-642-16145-2_9}}.

\bibitem{Fernandez-Kirchner-Mackie-Pinaud14a}
M.~Fernandez, H.~Kirchner, I.~Mackie, B.~Pinaud, Visual modelling of complex
  systems: Towards an abstract machine for {PORGY}, in: Proc.\ 10th Conference
  on Computability in Europe ({CiE 2014}), Vol. 8493 of Lecture Notes in
  Computer Science, Springer, 2014, pp. 183--193.
\newblock \href {https://doi.org/10.1007/978-3-319-08019-2_19}
  {\path{doi:10.1007/978-3-319-08019-2_19}}.

\bibitem{Plump12a}
D.~Plump, The design of {GP\,2}, in: Proc.\ 10th International Workshop on
  Reduction Strategies in Rewriting and Programming ({WRS 2011}), Vol.~82 of
  Electronic Proceedings in Theoretical Computer Science, 2012, pp. 1--16.
\newblock \href {https://doi.org/10.4204/EPTCS.82.1}
  {\path{doi:10.4204/EPTCS.82.1}}.

\bibitem{Plump17a}
D.~Plump, From imperative to rule-based graph programs, Journal of Logical and
  Algebraic Methods in Programming 88 (2017) 154--173.
\newblock \href {https://doi.org/10.1016/j.jlamp.2016.12.001}
  {\path{doi:10.1016/j.jlamp.2016.12.001}}.

\bibitem{Poskitt-Plump12a}
C.~Poskitt, D.~Plump, {Hoare}-style verification of graph programs, Fundamenta
  Informaticae 118~(1--2) (2012) 135--175.
\newblock \href {https://doi.org/10.3233/FI-2012-708}
  {\path{doi:10.3233/FI-2012-708}}.

\bibitem{Poskitt-Plump14a}
C.~Poskitt, D.~Plump, Verifying monadic second-order properties of graph
  programs, in: Proc.\ 7th International Conference on Graph Transformation
  ({ICGT 2014}), Vol. 8571 of Lecture Notes in Computer Science, Springer,
  2014, pp. 33--48.
\newblock \href {https://doi.org/10.1007/978-3-319-09108-2_3}
  {\path{doi:10.1007/978-3-319-09108-2_3}}.

\bibitem{Hristakiev-Plump18a}
I.~Hristakiev, D.~Plump, Checking graph programs for confluence, in: Software
  Technologies: Applications and Foundations -- {STAF} 2017 Collocated
  Workshops, Revised Selected Papers, Vol. 10748 of Lecture Notes in Computer
  Science, Springer, 2018, pp. 92--108.
\newblock \href {https://doi.org/10.1007/978-3-319-74730-9_8}
  {\path{doi:10.1007/978-3-319-74730-9_8}}.

\bibitem{Doerr95b}
H.~D{\"o}rr, Efficient Graph Rewriting and its Implementation, Vol. 922 of
  Lecture Notes in Computer Science, Springer, 1995.
\newblock \href {https://doi.org/10.1007/BFb0031909}
  {\path{doi:10.1007/BFb0031909}}.

\bibitem{Bak-Plump12a}
C.~Bak, D.~Plump, Rooted graph programs, in: Proc.\ 7th International Workshop
  on Graph Based Tools ({GraBaTs 2012}), Vol.~54 of Electronic Communications
  of the {EASST}, 2012.
\newblock \href {https://doi.org/10.14279/tuj.eceasst.54.780}
  {\path{doi:10.14279/tuj.eceasst.54.780}}.

\bibitem{Bak-Plump16a}
C.~Bak, D.~Plump, Compiling graph programs to {C}, in: Proc.\ 9th International
  Conference on Graph Transformation ({ICGT 2016}), Vol. 9761 of Lecture Notes
  in Computer Science, Springer, 2016, pp. 102--117.
\newblock \href {https://doi.org/10.1007/978-3-319-40530-8_7}
  {\path{doi:10.1007/978-3-319-40530-8_7}}.

\bibitem{Sedgewick02a}
R.~Sedgewick, Algorithms in {C}. Part 5: Graph Algorithms, 3rd Edition,
  Addison-Wesley, 2002.

\bibitem{Campbell-Romo-Plump20d}
G.~Campbell, J.~Rom{\"o}, D.~Plump, \href{https://arxiv.org/abs/2010.03993}{The
  improved {GP\,2} compiler}, Tech. rep., Department of Computer Science,
  University of York, {UK} (2020).
\newline\urlprefix\url{https://arxiv.org/abs/2010.03993}

\bibitem{Campbell-Courtehoute-Plump19b}
G.~Campbell, B.~Courtehoute, D.~Plump, Linear-time graph algorithms in {GP\,2},
  in: Proc.\ 8th Conference on Algebra and Coalgebra in Computer Science
  ({CALCO} 2019), Vol. 139 of Leibniz International Proceedings in Informatics
  ({LIPIcs}), Schloss Dagstuhl--Leibniz-Zentrum f{\"u}r Informatik, 2019, pp.
  16:1--16:23.
\newblock \href {https://doi.org/10.4230/LIPIcs.CALCO.2019.16}
  {\path{doi:10.4230/LIPIcs.CALCO.2019.16}}.

\bibitem{Bak15a}
C.~Bak, \href{https://etheses.whiterose.ac.uk/12586/}{{GP\,2}: Efficient
  implementation of a graph programming language}, Ph.D. thesis, Department of
  Computer Science, University of York, {UK} (2015).
\newline\urlprefix\url{https://etheses.whiterose.ac.uk/12586/}

\bibitem{Habel-Plump02c}
A.~Habel, D.~Plump, Relabelling in graph transformation, in: Proc.\ First
  International Conference on Graph Transformation ({ICGT 2002}), Vol. 2505 of
  Lecture Notes in Computer Science, Springer, 2002, pp. 135--147.
\newblock \href {https://doi.org/10.1007/3-540-45832-8_12}
  {\path{doi:10.1007/3-540-45832-8_12}}.

\bibitem{Plotkin04a}
G.~D. Plotkin, A structural approach to operational semantics, Journal of Logic
  and Algebraic Programming 60--61 (2004) 17--139.
\newblock \href {https://doi.org/10.1016/j.jlap.2004.05.001}
  {\path{doi:10.1016/j.jlap.2004.05.001}}.

\bibitem{Aho-Hopcroft-Ullman74a}
A.~Aho, J.~Hopcroft, J.~Ullman, The Design and Analysis of Computer Algorithms,
  Addison-Wesley, 1974.

\bibitem{Skiena08a}
S.~Skiena, The Algorithm Design Manual, 2nd Edition, Springer, 2008.
\newblock \href {https://doi.org/10.1007/978-1-84800-070-4}
  {\path{doi:10.1007/978-1-84800-070-4}}.

\bibitem{Campbell19a}
G.~Campbell, \href{https://arxiv.org/abs/1906.05170}{Efficient graph
  rewriting}, Bsc thesis, Department of Computer Science, University of York,
  {UK} (2019).
\newline\urlprefix\url{https://arxiv.org/abs/1906.05170}

\end{thebibliography}

\end{document}